\documentclass[12pt]{article}
\usepackage{amsmath}
\usepackage{graphicx,psfrag,epsf}
\usepackage{enumerate}
\usepackage{caption}
\usepackage{natbib}
\usepackage{url}
\usepackage{amsmath}
\usepackage{graphicx}
\usepackage{enumerate}
\usepackage{natbib}
\usepackage{url}
\usepackage{tabularx}
\usepackage{rotating}
\usepackage{algorithm}
\usepackage{float}
\usepackage{color}
\usepackage{pgfplots}
\usepackage{tikz}
\usepackage{algpseudocode}
\usetikzlibrary{positioning, shapes.geometric}
\usetikzlibrary{shapes,snakes}
\usepackage{amsfonts}
\usepackage[bookmarks=true]{hyperref}
\usepackage{ragged2e,subfigure}
\usepackage{verbatim}
\usepackage{extarrows}
\usepackage{epstopdf}

\newtheorem{theorem}{Theorem}

\newtheorem{lemma}{Lemma}
\newtheorem{proposition}{Proposition}
\newtheorem{corollary}{Corollary}
\def\beq{\begin{equation}}
\def\eeq{\end{equation}}
\def\beqr{\begin{eqnarray}}
\def\eeqr{\end{eqnarray}}
\def\beqrs{\begin{eqnarray*}}
\def\eeqrs{\end{eqnarray*}}
\def\bet{\begin{theorem}}
\def\eet{\end{theorem}}
\def\bel{\begin{lemma}}
\def\eel{\end{lemma}}
\def\bep{\begin{proposition}}
\def\eep{\end{proposition}}
\def\bec{\begin{corollary}}
\def\eec{\end{corollary}}
\def\n{\nonumber}

\def\mR{\mathbb{R}}
\def\mE{\mathbb{E}}

\def\cov{\mbox{cov}}
\def\vec{\mbox{vec}}
\def\tr{\mbox{tr}}
\def\blue{\color{blue}}

\newcommand{\blind}{1}

\begin{document}

\def\spacingset#1{
\renewcommand{\baselinestretch}
{#1}\small\normalsize}
\spacingset{1}
\date{}


\if1\blind
{
\title{\bf Modeling and Learning on High-Dimensional Matrix-Variate Sequences}

\author{Xu Zhang$^1$, Catherine C. Liu$^2$, Jianhua Guo$^3$, K. C. Yuen$^4$ and A. H. Welsh$^5$\\
\it{$^1$South China Normal University,
$^2$The Hong Kong Polytechnic University,}\\
\it{$^3$Beijing Technology and Business University, $^4$The University of Hong Kong}\\
\it{and $^5$The Australian National University}
}
  \maketitle
} \fi

\if0\blind
{
  \bigskip
  \bigskip
  \bigskip
  \begin{center}
{\LARGE\bf Modeling and Learning on High-Dimensional Matrix-Variate Sequences}
\end{center}
  \medskip
} \fi

\bigskip
\begin{abstract}

We propose a new matrix factor model, named RaDFaM, which is strictly derived based on the general rank decomposition and assumes a structure of a high-dimensional vector factor model for each basis vector.
RaDFaM contributes a novel class of low-rank latent structure that makes tradeoff between signal intensity and dimension reduction from the perspective of tensor subspace.
Based on the intrinsic separable covariance structure of RaDFaM, for a collection of matrix-valued observations, we derive a new class of PCA variants for estimating loading matrices, and sequentially the latent factor matrices.
The peak signal-to-noise ratio of RaDFaM is proved to be superior in the category of PCA-type estimations.
We also establish the asymptotic theory including the consistency, convergence rates, and asymptotic distributions for components in the signal part.
Numerically, we demonstrate the performance of RaDFaM in applications such as matrix reconstruction, supervised learning, and clustering, on uncorrelated and correlated data, respectively.

\end{abstract}

\noindent%
\vfill

\newpage
\spacingset{1.4} 

\section{Introduction}
\label{sec:introduction}

High-dimensional matrix objects arise in a broad range of applications, occurring as the slices of computed tomography (CT) or as magnetic resonance imaging (MRI) data in medical imaging, as multinational macroeconomic indices data in economics, to name a few \citep{Zhang2017matrix, GuptaNagar2018matrix, FanLiZhangZou2020statistical}.
Low-rank approximation of a matrix-variate is critical for discovery based on matrix-valued observations.
In this paper, we aim to model a new class of latent low-rank signal, and establish the inferential methodology and theory under a collection of matrix-variates.

\subsection{Rank-Decomposition-Based Factor Modeling}
\label{subsec:modeling}

Recall that a matrix $\mathbf{X} \in \mR^{p_{1}\times p_{2}}$ of rank-$l$ has the general rank decomposition expressible as $\mathbf{X} =\mathbf{U}\mathbf{V}^{\top}$, where $\mathbf{U}=(\mathbf{U}_{1},\ldots,\mathbf{U}_{l})\in\mR^{p_1\times l}$ and
$\mathbf{V}=(\mathbf{V}_{1},\ldots,\mathbf{V}_{l})\in\mR^{p_2\times l}$ are \textit{full column rank matrices}.
Let $\text{span}(\mathbf{M})$ be the \textit{column space} spanned by columns of the placeholder matrix $\mathbf{M}$.
Then $\{\mathbf{U}_{i}\}_{i=1}^l$ and $\{\mathbf{V}_{i}\}_{i=1}^l$ are \textit{bases} of $\text{span}(\mathbf{U})$ and $\text{span}(\mathbf{V})$, respectively.

For a high-dimensional matrix-variate $\mathbf{X}$, where $p_1$ and $p_2$ may tend to infinity,
by enduing all basis vectors $\mathbf{U}_{i}$ and $\mathbf{V}_{i}$ with the structure of the high-dimensional vector factor model,
we postulate the following rank-decomposition-based matrix factor model (RaDFaM)
\beqr
\label{RaDFaM_rank}
\left\{\begin{array}{l}
\mathbf{X}=\mathbf{U}\mathbf{V}^{\top}=\sum\limits_{i=1}^{l}\mathbf{U}_{i}\mathbf{V}_{i}^{\top},~~~~~~~~~~~~~~~~(\text{rank-$l$}~\text{decomposition})\\
    \mathbf{U}_{i} = \mathbf{R}\mathbf{A}_{i\cdot}+\boldsymbol{\xi}_{i},~~    \mathbf{V}_{i}=\mathbf{C}\mathbf{B}_{i\cdot}+\boldsymbol{\eta}_{i},~~(\text{vector factor models})
\end{array}\right.
\eeqr
where $\mathbf{R}\in\mR^{p_1\times k_1}$ ($k_1\ll p_1$) and $\mathbf{C}\in\mR^{p_2\times k_2}$ ($k_2\ll p_2$) are called loading matrices, $\mathbf{A}_{i\cdot}\in\mR^{k_1}$ and $\mathbf{B}_{i\cdot}\in\mR^{k_2}$ are vectors of latent factors, and $\boldsymbol{\xi}_{i}\in\mR^{p_1}$ and $\boldsymbol{\eta}_{i}\in\mR^{p_2}$ are vectors of idiosyncratic errors.

To brief the hierarchical structure in expression \eqref{RaDFaM_rank}, we adopt matrix notation to denote $\mathbf{A}=(\mathbf{A}_{1\cdot},\ldots,\mathbf{A}_{l\cdot})^{\top}$ $\in\mR^{l\times k_1}$, $\mathbf{B}=(\mathbf{B}_{1\cdot},\ldots,\mathbf{B}_{l\cdot})^{\top} \in\mR^{l\times k_2}$,
$\boldsymbol{\xi}=(\boldsymbol{\xi}_{1},\ldots,\boldsymbol{\xi}_{l})\in\mR^{p_1\times l}$, and $\boldsymbol{\eta}=(\boldsymbol{\eta}_{1},\ldots,\boldsymbol{\eta}_{l})\in\mR^{p_2\times l}$.
Let $\mathbf{G}_{i\cdot}\in\mR^{n}$ and $\mathbf{G}_{j}\in\mR^{m}$ be the $i$th row and the $j$th column of a placeholder matrix $\mathbf{G}\in\mR^{m\times n}$.
Denote $\mathbf{Z}=\mathbf{A}^{\top}\mathbf{B}\in\mR^{k_1\times k_2}$, $\mathbf{E}=\boldsymbol{\eta}\mathbf{A}\in\mR^{p_2\times k_1}$, $\mathbf{F}=\boldsymbol{\xi}\mathbf{B}\in\mR^{p_1\times k_2}$, and $\mathbf{e}=\boldsymbol{\xi}\boldsymbol{\eta}^{\top}\in\mR^{p_1\times p_2}$.
Then in pure matrix notation, RaDFaM has the following equivalent expression
\beqr\label{RaDFaM}
\mathbf{X} = \underbrace{\mathbf{R} \mathbf{Z} \mathbf{C}^\top + \mathbf{R} \mathbf{E}^\top + \mathbf{F} \mathbf{C}^\top}_{\text{low-rank signal}} + \underbrace{\mathbf{e}}_{\text{noise}}
\equiv \mathbf{S} + \mathbf{e}.
\eeqr

Different insights of assignments of parts of signal and noise in model \eqref{RaDFaM} may lead to
different low-rank approximations of the matrix-variate $\mathbf{X}$.
Specifically,
\beqr\label{BiMFaM}
\mathbf{X}
&=&\underbrace{\mathbf{R}\mathbf{Z}\mathbf{C}^{\top}}_{\text{latent structure}}
+\underbrace{\mathbf{R}\mathbf{E}^{\top}
+\mathbf{F}\mathbf{C}^{\top}
+\mathbf{e}}_{\text{error}}
\equiv\underbrace{\mathbf{R}\mathbf{Z}\mathbf{C}^{\top}}_{\text{low-rank signal}}+\text{noise},~~~~(\text{BiMFaM})
\eeqr
\beqr\label{2w-DFM}
\mathbf{X}
&=&\underbrace{\mathbf{R}\mathbf{E}^{\top}
+\mathbf{F}\mathbf{C}^{\top}}_{\text{latent structure}}
+\underbrace{\mathbf{R}\mathbf{Z}\mathbf{C}^{\top}+\mathbf{e}}_{\text{error}}
\equiv\underbrace{\mathbf{R} \mathbf{E}^\top + \mathbf{F} \mathbf{C}^\top}_{\text{low-rank signal}}+\text{noise}.~~~~~~~~(\text{2w-DFM})
\eeqr
In fact, models \eqref{BiMFaM} and \eqref{2w-DFM} throw specific quantities from the individual latent structure part into the error part, respectively.
That is, if one exposes only the bilinear-form product involving the low-dimensional interactive latent factor ($\mathbf{Z}$) or the linear addition involving the two low-dimensional main latent factors (row-wise $\mathbf{F}$ and column-wise $\mathbf{E}$) as signals, respectively,
then model \eqref{RaDFaM_rank} generates two classes of matrix factor models in the existing literature, the popular bilinear-form matrix factor model (BiMFaM) and a newly presented two-way dynamic factor model (2w-DFM), respectively \citep{WangLiuChen2019JOE-factor,
YuHeKongZhang2022JOE-projected, ChenFan2023JASA-statistical, YuanGaoHeHuangGuo2023JRSSB-two}.

\subsection{Tensor Subspace and Dimension Reduction}
\label{subsec:subspace}

In this subsection, we aim to elucidate advantages and disadvantages of RaDFaM from the tensor subspace perspective.
Recall that a tensor space $\mR^{p_1\times \ldots\times p_D}$ is conceptually consistent with a vector space, the operations of which satisfy the axioms of a real vector space \citep[Section 3.1]{Shores2007applied}.
A tensor subspace  $\mathcal{S}\subset\mR^{p_1\times \ldots\times p_D}$ includes the zero tensor and is closed under finite tensor addition and scalar multiplication;
the dimension of $\mathcal{S}$, denoted as $\text{dim}(\mathcal{S})$, is the number of elements in its basis \citep{ ZhouLuFengLinYan2019TPAMI-tensor}.

For different tensor decompositions, the deduced low-rank signal structures correspond to different tensor subspaces \citep{LuPlataniotisVenetsanopoulos2013multilinear, WangAggarwalAeron2019PRL-principal, ZhangLiLiuGuo2022arXiv-Tucker}.
For the case of $D=2$ in our study, the signal in BiMFaM is essentially in the form of the $2$nd-order Tucker decomposition, and corresponds to the tensor subspace \citep[Section IV-B]{Zare2018IEEE-extension}
{\footnotesize
\begin{align*}
\label{BiMFaM_space}
&\mathcal{S}_{BiM}=\left\{\sum_{i_{1}=1}^{k_{1}} \sum_{i_{2}=1}^{k_{2}} \mathbf{Z}_{i_{1}i_{2}}\left(\mathbf{R}_{i_{1}} \circ \mathbf{C}_{ i_{2}}\right)|\mathbf{Z}=(\mathbf{Z}_{i_1 i_2})\in\mR^{k_1\times k_2}\right\},
\end{align*}}\noindent
where $\{\mathbf{R}_{i_1} \circ \mathbf{C}_{i_2}\in \mR^{p_1 \times p_2}\}$ represents the set of rank-$1$ matrices spanning $\mathcal{S}_{BiM}$ with $\circ$ being the vector outer product, and the interactive latent factor $\mathbf{Z}$ becomes the matrix of coefficients of the linear combination.
Denote the vectorized $\mathcal{S}_{BiM}$ as $\widetilde{\mathcal{S}}_{BiM}=\{\sum_{i_{1}=1}^{k_{1}} \sum_{i_{2}=1}^{k_{2}} \mathbf{Z}_{i_{1}i_{2}}(\mathbf{C}_{ i_{2}}\otimes \mathbf{R}_{i_{1}})|\mathbf{Z}=(\mathbf{Z}_{i_1 i_2})\in\mR^{k_1\times k_2}\}$, where $\mathbf{C}_{i_{2}}\otimes \mathbf{R}_{i_{1}}=\vec(\mathbf{R}_{i_{1}} \circ \mathbf{C}_{i_{2}})$, and $\otimes$ and $\vec(\cdot)$ are the Kronecker product and the vectorization operation, respectively.
The isomorphism of $\mathcal{S}_{BiM}$ and $\widetilde{\mathcal{S}}_{BiM}$ yields their identical dimensionality \citep[Subsection 7.3]{Nicholson2020linear}.
It is easy to see that $\{\mathbf{C}_{ i_{2}}\otimes \mathbf{R}_{i_{1}}\}$ is the basis of $\widetilde{\mathcal{S}}_{BiM}$ due to the column orthogonality of $\mathbf{R}$ and $\mathbf{C}$ (see Assumption 3).
Hence,
$$\text{dim}(\mathcal{S}_{BiM})=\text{dim}(\widetilde{\mathcal{S}}_{BiM})= k_1k_2\ll p_1p_2.$$

In the same spirit, some algebra relates the following tensor subspaces to RaDFaM and 2w-DFM:
{\footnotesize
\begin{align*}
&\mathcal{S}_{RaD}=\Bigg\{\sum_{i_{1}=1}^{k_{1}} \sum_{i_{2}=1}^{k_{2}} \mathbf{Z}_{i_{1}i_{2}}\left(\mathbf{R}_{ i_{1}} \circ \mathbf{C}_{i_{2}}\right)+
\sum_{i_{1}=1}^{k_{1}} \sum_{i_{2}=1}^{p_{2}} \mathbf{E}_{i_{2}i_{1}}\left(\mathbf{R}_{ i_{1}} \circ \mathbf{I}_{p_2, i_{2}}\right)
+\sum_{i_{1}=1}^{p_{1}} \sum_{i_{2}=1}^{k_{2}} \mathbf{F}_{i_{1}i_{2}}\left(\mathbf{I}_{p_1, i_{1}} \circ \mathbf{C}_{ i_{2}}\right)\\\n
&~~~~~~~~~~~~~~
|\mathbf{Z}=(\mathbf{Z}_{i_1 i_2})\in\mR^{k_1\times k_2},\mathbf{E}=(\mathbf{E}_{i_2 i_1})\in\mR^{p_2\times k_1},\mathbf{F}=(\mathbf{F}_{i_1 i_2})\in\mR^{p_1\times k_2}\Bigg\},\\
&\mathcal{S}_{2w}=\Bigg\{
\sum_{i_{1}=1}^{k_{1}} \sum_{i_{2}=1}^{p_{2}} \mathbf{E}_{i_{2}i_{1}}\left(\mathbf{R}_{ i_{1}} \circ \mathbf{I}_{p_2, i_{2}}\right)
+\sum_{i_{1}=1}^{p_{1}} \sum_{i_{2}=1}^{k_{2}} \mathbf{F}_{i_{1}i_{2}}\left(\mathbf{I}_{p_1, i_{1}} \circ \mathbf{C}_{ i_{2}}\right)
|\mathbf{E}\in\mR^{p_2\times k_1},\mathbf{F}\in\mR^{p_1\times k_2}\Bigg\},
\end{align*}}\noindent
where $\mathbf{I}_{d}$ is the $d\times d$ identity matrix.
Note that $\vec(\mathbf{R}_{i_{1}} \circ \mathbf{I}_{p_2, i_{2}})=\mathbf{I}_{p_2, i_{2}}\otimes \mathbf{R}_{i_{1}}$ and $\vec(\mathbf{I}_{p_1,i_{1}} \circ \mathbf{C}_{ i_{2}})=\mathbf{C}_{ i_{2}}\otimes\mathbf{I}_{p_1,i_{1}}$.
Then the sets of vectors that span the \textit{vectorized isomorphic subspaces} under RaDFaM and 2w-DFM are $\{\mathbf{C}_{i_{2}}\otimes \mathbf{R}_{i_{1}},\mathbf{I}_{p_2,i_{2}}\otimes \mathbf{R}_{i_{1}},\mathbf{C}_{i_{2}}\otimes\mathbf{I}_{p_1,i_{1}}\}$ and $\{\mathbf{I}_{p_2,i_{2}}\otimes \mathbf{R}_{i_{1}},\mathbf{C}_{i_{2}}\otimes\mathbf{I}_{p_1,i_{1}}\}$, respectively.
Due to the fact that $\text{span}(\mathbf{C}\otimes\mathbf{R})\subset\text{span}(\mathbf{I}_{p_2}\otimes\mathbf{R})\cup\text{span}(\mathbf{C}\otimes \mathbf{I}_{p_1})$, we have
$$\text{dim}(\mathcal{S}_{RaD})=\text{dim}(\mathcal{S}_{2w})=p_1k_2+k_1p_2-\text{dim}\{\text{span}(\mathbf{I}_{p_2}\otimes \mathbf{R})\cap\text{span}(\mathbf{C}\otimes \mathbf{I}_{p_1})\}\ll p_1p_2.$$

Based on the above discussion of tensor subspaces and companion dimensions, we can see that:
i) RaDFaM possesses the strongest low-rank signal strength.
Though RaDFaM reduces the dimension less than BiMFaM ($\text{dim}(\mathcal{S}_{RaD}) > \text{dim}(\mathcal{S}_{BiM})$), RaDFaM is able to extract additional mode-wise information ($\mathbf{E}$ and $\mathbf{F}$); though RaDFaM and 2w-DFM have the same tensor subspace dimension ($\text{dim}(\mathcal{S}_{RaD}) = \text{dim}(\mathcal{S}_{2w})$), RaDFaM is able to extract additional interaction information  ($\mathbf{Z}$).
ii) RaDFaM incurs no extra computational burden in the sense that the basis parameters $\mathbf{R}$ and $\mathbf{C}$ are identical for all three factor models.

\subsection{Separable Covariance Structure}
\label{subsec:separable}

Recall that a matrix-variate is said to have a separable covariance structure if the variance-covariance matrix of its vectorization is the Kronecker product of column-wise and row-wise variance-covariance matrices \citep{Dawid1981Biometrika-some, GuptaNagar2018matrix}.
Indeed, the rank-$l$ matrix-variate $\mathbf{X}$ under RaDFaM possesses the separable covariance structure, where its vectorization is expressible as
\begin{align*}\vec(\mathbf{X})
=\sum\limits_{i=1}^l\mathbf{V}_i\otimes \mathbf{U}_i.
\end{align*}
Let $\mathbf{\Psi}_{G}$ be the variance-covariance matrix of the rows or columns of the placeholder matrix $\mathbf{G}$.
Under regularity assumptions of zero mean, mutual uncorrelatedness of $\mathbf{A}$, $\mathbf{B}$, $\boldsymbol{\xi}$ and $\boldsymbol{\eta}$, and uncorrelatedness among rows or columns of the factor matrix $\mathbf{A}$ ($\mathbf{B}$) and the error matrix $\boldsymbol{\xi}$ ($\boldsymbol{\eta}$),
by some algebra, one has
$\cov(\mathbf{V}_i)=\mathbf{C}\mathbf{\Psi}_B \mathbf{C}^{\top}+\mathbf{\Psi}_{\eta}$, $\cov(\mathbf{U}_i)=\mathbf{R}\mathbf{\Psi}_A \mathbf{R}^{\top}+\mathbf{\Psi}_{\xi}$; and
\beqr
\label{separable_covariance}
\cov\{\vec(\mathbf{X})\}
= l\{\cov(\mathbf{V}_i)\otimes\cov(\mathbf{U}_i)\}
\equiv l(\mathbf{\Sigma}_C\otimes\mathbf{\Sigma}_R),
\eeqr
where $\mathbf{\Sigma}_R$ and $\mathbf{\Sigma}_C$ are called row-wise and column-wise variance-covariance matrices of $\mathbf{X}$ under RaDFaM, respectively, and their elements control the correlation of different rows and columns in $\mathbf{X}$, respectively.

Separable covariance structure brings convenience for inference based on covariance decompositions \citep{Hoff2011BA-separable, FosdickHoff2014AOAS-separable, FanLiLiao2021AR-recent}.
The proposed RaDFaM enjoys this property, which guarantees our PCA-type estimation of the loading matrices is doable.
Thus later we name our proposed approach of PCA-type estimation as sPCA.

\subsection{Related Work and Organization}

The bilinear-form low-rank structure has been extensively researched in engineering and computer vision from the algorithmic perspective \citep[among others]{YangZhangFrangiYang2004PAMI-two, Ye2004ICML-generalized,LiuChenZhouTan2010IEEE-generalized,AhmadiRezghi2020PR-generalized}.
The corresponding BiMFaM was first explored for matrix-variate time series \citep{WangLiuChen2019JOE-factor}, then extended to uncorrelated or weakly correlated observations
\citep{ChenFan2023JASA-statistical}, and its convergence rates were enhanced by the projection technique \citep{YuHeKongZhang2022JOE-projected};
their estimation approaches were PCA based.
Recently, 2w-DFM was presented and a refined quasi maximum likelihood estimation (Q-MLE) enhanced the PCA-type initial estimation \citep{YuanGaoHeHuangGuo2023JRSSB-two};
a similar latent structure was separately investigated from the spiked covariance perspective \citep{TangYuanZhang2023arXiv-mode}.
Our proposed RaDFaM with sPCA estimation shows superior signal intensity both theoretically and numerically in the category of PCA-type estimation approaches, albeit with some sacrifice of theoretical convergence rates;
and sPCA has a much less computational cost than Q-MLE, though they are comparable in reconstruction accuracy.

The remainder of this article is organized as follows.
Section 2 develops PCA-type estimation based on the separable covariance structure, and states the theoretical evidence for the strong signal intensity of RaDFaM.
Section 3 establishes the asymptotic theory, and Section 4 presents simulation studies for the finite sample performance evaluation.
Section 5 conducts real data analysis on uncorrelated and correlated matrix-variate observations, respectively, followed by concluding remarks in Section 6.
All the technical proofs are given in the online supplement.

\section{Estimation for Matrix-Valued Observations}
\label{sec:estimation}

Let $\mathbf{X}_t\in\mR^{p_1\times p_2}$, $t\in[T]\equiv\{1,\ldots,T\}$, be a sequence of high-dimensional matrix-valued observations that might be weakly correlated or uncorrelated, where the mode-wise dimensions $p_1$ and $p_2$, and the number of observations $T$, can all tend to infinity.
The matrix-valued observations under RaDFaM are expressible as
\beqr\label{RaDFaM_t}
\begin{array}{l}
\mathbf{X}_t = \mathbf{S}_t + \mathbf{e}_t,~~
\mathbf{S}_t = \mathbf{R} \mathbf{Z}_t \mathbf{C}^\top + \mathbf{R} \mathbf{E}_t^\top + \mathbf{F}_t \mathbf{C}^\top.
\end{array}
\eeqr
In this section, under model$+$data in (\ref{RaDFaM_t}),
we propose two-step estimation for the components in the signal part, including so-called sPCA estimators (based on the property of separable covariance structure) for the loading matrices in Subsection \ref{subsec:estimation_loadings} and least squares estimators for the factor matrices in Subsection \ref{subsec:estimation_factor}.
Furthermore, we show the superiority of matrix reconstruction using RaDFaM compared to BiMFaM and 2w-DFM based on PCA-type estimators in Subsection \ref{subsec:reconstruction_error}.
We consider the estimation of factor numbers in Subsection \ref{subsec:estimation_factornumber}.

\subsection{Separable PCA Estimators for Loading Matrices}
\label{subsec:estimation_loadings}

It is known that estimators of loading matrices $\mathbf{R}$ and $\mathbf{C}$ are uniquely determined up to a rotation \citep{WangLiuChen2019JOE-factor, ChenFan2023JASA-statistical}.
Thus we restrict the loading matrices by setting $p_{1}^{-1}\mathbf{R}^{\top}\mathbf{R}=\mathbf{I}_{k_{1}}$
and $p_{2}^{-1} \mathbf{C}^{\top} \mathbf{C}=\mathbf{I}_{k_{2}}${\blue,} known as the strong factor assumption in the literature of factor models \citep{StockWatson2002JASA-forecasting, LamYao2011Biometrika-estimation}.
In Assumption 3 of Section \ref{sec:theory}, we use the strict limit form of the strong factor assumption.

The spectral or PCA-type methods that are most commonly used for estimating loading matrices are based on the eigen-decomposition or singular value decomposition of some moment statistics \citep[Remark 2]{ChenFan2023JASA-statistical}.
One of the rationales behind these methods is that, under the strong factor assumption, top eigenspaces (spaces spanned by the top eigenvectors) of population moment matrices are consistent estimators of the column spaces of loading matrices \citep{DavisKahan1970rotation}.

Nonetheless, spectral methods based on second moments may not work under model (\ref{RaDFaM_t}) for matrix-valued observations.
Take the row-wise second moment for instance
\beqrs
\mE\left(\mathbf{X}_t\mathbf{X}_t^{\top}\right)
=  p_2\mathbf{R} \mE\left(\mathbf{Z}_t\mathbf{Z}_t^{\top}\right) \mathbf{R}^{\top}
+\left\{\mathbf{R} \mE\left(\mathbf{E}_t^{\top}\mathbf{E}_t\right)\mathbf{R}^{\top}
+ p_2\mE\left(\mathbf{F}_t\mathbf{F}_t^{\top}\right)\right\}
+\mE\left(\mathbf{e}_t\mathbf{e}_t^{\top}\right),
\eeqrs
where we assume $\mathbf{Z}_t$, $\mathbf{E}_t$, $\mathbf{F}_t$ and $\mathbf{e}_t$ are uncorrelated.
Compared to BiMFaM \citep[Subsection 2.2]{ChenFan2023JASA-statistical}, the two terms in braces on the right-hand side, which are exactly the main terms in $\mE(\mathbf{X}_t\mathbf{X}_t^{\top})$ under 2w-DFM, are not negligible.
Fortunately, from equation  \eqref{separable_covariance}, we have the following empowerment.

\bep
\label{second-moment} (Row-wise second moment)
Under RaDFaM, based on the separable covariance structure in equation \eqref{separable_covariance}, we have
\beqrs
\mE(\mathbf{X}_t\mathbf{X}_t^{\top})=\{l\tr(\mathbf{\Sigma}_C)\}\mathbf{\Sigma}_R.
\eeqrs
\eep
Proposition \ref{second-moment} shows that $\mE(\mathbf{X}_t\mathbf{X}_t^{\top})$ and $\mathbf{\Sigma}_R$ are equivalent up to a constant $l\tr(\mathbf{\Sigma}_C)$, so have identical top eigenspaces, that is
\beqr
\label{span_R_1}
\text{span}[\text{eig}\{\mE(\mathbf{X}_t\mathbf{X}_t^{\top}),k_1\}]=\text{span}\{\text{eig}(\mathbf{\Sigma}_R,k_1)\},
\eeqr
where $\mbox{eig}({\mathbf{M}},r)$ denotes a matrix with columns given by the top $r$ eigenvectors of the placeholder matrix $\mathbf{M}$.
Recall that $\mathbf{\Sigma}_R=\mathbf{R}\mathbf{\Psi}_A \mathbf{R}^{\top}+\mathbf{\Psi}_{\xi}$.
Using the general pervasiveness assumption and applying the Davis-Kahan Theorem \citep{FanLiZhangZou2020statistical,DavisKahan1970rotation}, we can show the approximate equivalence between the top eigenspace of $\mathbf{\Sigma}_R$ and the column space of $\mathbf{R}$
\beqr
\label{span_R_2}
\frac{1}{p_1}\left\|\text{eig}(\mathbf{\Sigma}_R,k_1)\text{eig}^{\top}(\mathbf{\Sigma}_R,k_1) - \mathbf{R}\mathbf{R}^{\top}\right\|_F=o(1),
\eeqr
where $\|\cdot\|_F$ is the Frobenius norm.
Hence, based on (\ref{span_R_1}) and (\ref{span_R_2}), we have the approximate equivalence of the top eigenspace of $\mE(\mathbf{X}_t\mathbf{X}_t^{\top})$ and the column space of $\mathbf{R}$
\beqr
\label{span_R}
\text{span}[\text{eig}\{\mE(\mathbf{X}_t\mathbf{X}_t^{\top}),k_1\}]\approx\text{span}(\mathbf{R}),
\eeqr
and we can derive the sPCA estimator of $\mathbf{R}$ as
\beqr
\label{estimator_R}
\widehat{\mathbf{R}}=\sqrt{p_1}\mbox{eig}(\widehat{\mathbf{M}}_{1},k_1)~~\text{with}~~\widehat{\mathbf{M}}_1= \frac{1}{Tp_1p_2}\sum_{t=1}^T \mathbf{X}_t \mathbf{X}_t^\top.
\eeqr
By a similar argument, the sPCA estimator of $\mathbf{C}$ is
\beqr
\label{estimator_C}
\widehat{\mathbf{C}}=\sqrt{p_2}\mbox{eig}(\widehat{\mathbf{M}}_{2},k_2)~~\text{with}~~
\widehat{\mathbf{M}}_2 =\frac{1}{Tp_1p_2}\sum_{t=1}^T \mathbf{X}_t^\top \mathbf{X}_t.
\eeqr

\subsection{Least Squares Estimators for Factor Matrices}
\label{subsec:estimation_factor}

Given $\mathbf{R}$ and $\mathbf{C}$, estimators of factor matrices $\mathbf{Z}_t$, $\mathbf{E}_t$ and $\mathbf{F}_t$ can be obtained by minimizing the matrix reconstruction error
\beqrs
\min_{\substack{p_{1}^{-1} \mathbf{R}^{\top}\mathbf{R}=\mathbf{I}_{k_{1}}, p_{2}^{-1} \mathbf{C}^{\top}\mathbf{C}=\mathbf{I}_{k_{2}}\\
\{\mathbf{Z}_t\}_{t=1}^T,\{\mathbf{E}_t\}_{t=1}^T,\{\mathbf{F}_t\}_{t=1}^T}}
\frac{1}{T p_1 p_2}\sum\limits_{t=1}^T\left\| \mathbf{X}_t - \mathbf{R}\mathbf{Z}_t \mathbf{C}^\top - \mathbf{R}\mathbf{E}_t^\top - \mathbf{F}_t \mathbf{C}^\top \right\|_F^2.
\eeqrs
Using the fact that $\|\mathbf{A}\|_{F}^2=\tr(\mathbf{A}\mathbf{A}^{\top})$, we expand the objective function, and then take partial derivatives with respect to $\mathbf{Z}_t$, $\mathbf{E}_t$ and $\mathbf{F}_t$, respectively, to obtain the normal equations.
After some algebra, we obtain the system of equations
\beqrs
\left\{\begin{array}{l}
  \mathbf{Z}_t = \frac{\mathbf{R}^\top \mathbf{X}_t \mathbf{C}}{p_1p_2}-\frac{\mathbf{E}_t^\top \mathbf{C}}{p_2}-\frac{\mathbf{R}^\top \mathbf{F}_t}{p_1},\\
      (\mathbf{I}_{p_2}-\frac{\mathbf{C}\mathbf{C}^\top}{p_2})(\mathbf{E}_t-\frac{\mathbf{X}_t^\top\mathbf{R}}{p_1})=\mathbf{0}_{p_2\times k_1},\\
    (\mathbf{I}_{p_1}-\frac{\mathbf{R}\mathbf{R}^\top}{p_1})(\mathbf{F}_t-\frac{\mathbf{X}_t\mathbf{C}}{p_2})=\mathbf{0}_{p_1\times k_2},
\end{array}\right.
\eeqrs
where $\mathbf{0}_{p\times q}$ is the $p\times q$ matrix of zeros.
For any linear equations $\mathbf{A}\mathbf{x}=\mathbf{b}$, $\mathbf{A}^{\dagger}\mathbf{b}$ is the one of minimum-norm among the least-squares solutions, where $\mathbf{A}^{\dagger}$ represents the pseudoinverse of $\mathbf{A}$ \citep[Corollary 3, Chapter 3]{Ben-IsraelGreville2003generalized}.
Note that $\mathbf{I}_{p_1}-p_1^{-1}\mathbf{R}\mathbf{R}^\top$ is an orthogonal projection matrix, and its pseudoinverse is itself.
Hence, for the linear equations
$(\mathbf{I}_{p_1}-p_1^{-1}\mathbf{R}\mathbf{R}^\top)({\mathbf{F}}_t-p_2^{-1}\mathbf{X}_t{\mathbf{C}})=\mathbf{0}$,
the minimum-norm solutions of the columns of ${\mathbf{F}}_t-p_2^{-1}\mathbf{X}_t{\mathbf{C}}$ are the columns of $(\mathbf{I}_{p_1}-p_1^{-1}\mathbf{R}\mathbf{R}^\top)\mathbf{0}$ ($=\mathbf{0}$).
That is, $\mathbf{F}_t-p_2^{-1}\mathbf{X}_t\mathbf{C}=\mathbf{0}$.
Similarly, ${\mathbf{E}}_t - p_1^{-1}\mathbf{X}_t^{\top}{\mathbf{R}}=\mathbf{0}$.
Then we have $\mathbf{E}_t = p_1^{-1}\mathbf{X}_t^{\top}{\mathbf{R}}$ and $\mathbf{F}_t = p_2^{-1}\mathbf{X}_t{\mathbf{C}}$.
Substituting these expressions into the first equation of the system, we have $\mathbf{Z}_t=-(p_1p_2)^{-1}{\mathbf{R}}^{\top}\mathbf{X}_t{\mathbf{C}}$.
Using $\widehat{\mathbf{R}}$ and $\widehat{\mathbf{C}}$ in equations \eqref{estimator_R} and \eqref{estimator_C}, the factor matrices can be estimated by
\beqr
\label{factormatrix_RaDFaM}
\widehat{\mathbf{E}}_t = \frac{\mathbf{X}_t^{\top}\widehat{\mathbf{R}}}{p_1},\quad
\widehat{\mathbf{F}}_t = \frac{\mathbf{X}_t\widehat{\mathbf{C}}}{p_2},\quad
\widehat{\mathbf{Z}}_t = -\frac{\widehat{\mathbf{R}}^{\top}\mathbf{X}_t\widehat{\mathbf{C}}}{p_1p_2}.
\eeqr
Accordingly, the estimated signal part of RaDFaM is
\beqr
\label{signal_RaDFaM}
\widehat{\mathbf{S}}_t=-\frac{\widehat{\mathbf{R}}\widehat{\mathbf{R}}^{\top}\mathbf{X}_t \widehat{\mathbf{C}}\widehat{\mathbf{C}}^{\top}}{p_1p_2}
+\frac{\widehat{\mathbf{R}}\widehat{\mathbf{R}}^{\top}\mathbf{X}_t}{p_1}
+\frac{\mathbf{X}_t\widehat{\mathbf{C}}\widehat{\mathbf{C}}^{\top}}{p_2}.
\eeqr
Hereafter, we refer to the two-step estimation procedure which is summarized in Algorithm \ref{algorithm} as sPCA for ease of reference.

\begin{algorithm}
\caption{sPCA}
\begin{algorithmic}[1]
\State \textbf{Input}: matrix observations $\{\mathbf{X}_t\}_{t=1}^T$, factor numbers $k_1$ and $k_2$.
\State Estimate loading matrices by equations \eqref{estimator_R} and \eqref{estimator_C}.
\State Estimate factor matrices and the signal part by equations \eqref{factormatrix_RaDFaM} and \eqref{signal_RaDFaM} for $t\in[T]$.
\State \textbf{Output}: $\widehat{\mathbf{R}}$, $\widehat{\mathbf{C}}$, $\{\widehat{\mathbf{Z}}_t\}_{t=1}^T$,
$\{\widehat{\mathbf{F}}_t\}_{t=1}^T$, $\{\widehat{\mathbf{E}}_t\}_{t=1}^T$, and $\{\widehat{\mathbf{S}}_t\}_{t=1}^T$.
\end{algorithmic}
\label{algorithm}
\end{algorithm}

\subsection{Reconstruction Error Comparison}
\label{subsec:reconstruction_error}

A grayscale digital image can be represented by a matrix in which each element represents a pixel value (i.e., brightness) ranging from 0 to 255 \citep{Suetens2017fundamentals}.
We compare the matrix reconstruction performance of RaDFaM, BiMFaM and 2w-DFM in terms of the peak signal-to-noise ratio (PSNR), which is a standard metric to measure the quality of reconstructed images compared with the originals.
The PSNR is defined as
\beqrs
\label{psnr}
\text{PSNR}=10 \log_{10}\left\{ \frac{\|\mathbf{X}\|_{max}^2}{\|\mathbf{X}-\widehat{\mathbf{X}}\|_F^2/(p_1p_2)}\right\},
\eeqrs
where $\mathbf{X}=(\mathbf{X}_{ij})\in\mR^{p_1\times p_2}$ is an original image, $\widehat{\mathbf{X}}=(\widehat{\mathbf{X}}_{ij})\in\mR^{p_1\times p_2}$ is a reconstructed image, and $\|\mathbf{X}\|_{max}$ is the maximum of the absolute pixel values of $\mathbf{X}$;
the denominator inside the logarithm is the mean square error (MSE) between $\mathbf{X}$ and $\widehat{\mathbf{X}}$, that is $\text{MSE}=(p_1p_2)^{-1}\|\mathbf{X}-\widehat{\mathbf{X}}\|_F^2$, the average of the square of pixel differences of the two images, i.e., $({p_1p_2})^{-1}\sum_{i=1}^{p_1}\sum_{j=1}^{p_2}(\mathbf{X}_{ij}-\widehat{\mathbf{X}}_{ij})^2$.
Therefore, the more the reconstructed image resembles the original image, the smaller MSE will be and the larger PSNR will be \citep{Salomon2004data}.
The PSNR is preferred to the MSE because it is dimensionless and the logarithm in the PSNR reduces the sensitivity to small variations in the reconstructed image.

Before conducting the comparison,
we show that $\widehat{\mathbf{R}}$ and $\widehat{\mathbf{C}}$ (equations \eqref{estimator_R} and \eqref{estimator_C}) can also act as PCA-type estimators for loading matrices $\mathbf{R}$ and $\mathbf{C}$
under BiMFaM and 2w-DFM.
For BiMFaM, $\widehat{\mathbf{R}}$ and $\widehat{\mathbf{C}}$ are exactly $\alpha$-PCA estimators with $\alpha=0$ in \cite{ChenFan2023JASA-statistical} or the initial estimators in \cite{YuHeKongZhang2022JOE-projected}.
The corresponding estimators of the latent factor matrix $\mathbf{Z}_t$ and the signal part are straightforwardly
\beqr
\label{signal_bilinear}
\widetilde{\mathbf{Z}}_t=\frac{\widehat{\mathbf{R}}^{\top}\mathbf{X}_t\widehat{\mathbf{C}}}{p_1p_2}
\quad\text{and}\quad
\widetilde{\mathbf{S}}_t=\frac{\widehat{\mathbf{R}}\widehat{\mathbf{R}}^{\top} \mathbf{X}_t\widehat{\mathbf{C}}\widehat{\mathbf{C}}^{\top}}{p_1p_2}.
\eeqr
For 2w-DFM, let $\mathbf{\Psi}_E$ and $\mathbf{\Psi}_F$ be the variance-covariance matrices of the rows of $\mathbf{E}_t$ and $\mathbf{F}_t$, respectively, and $\sigma^2$ be the average of the element-wise variances of the noise matrix.
Based on the working variance-covariance matrix ($\mathbf{\Sigma}$) of $\vec(\mathbf{X}_t)$ \citep[Section 3.1]{YuanGaoHeHuangGuo2023JRSSB-two}, one may show
$\mE(\mathbf{X}_t\mathbf{X}_t^{\top})
= p_2\mathbf{R}\mathbf{\Psi}_E\mathbf{R}^{\top} + p_2\tr(\mathbf{\Psi}_{F}) \mathbf{I}_{p_1} + p_2\sigma^2\mathbf{I}_{p_1}$;
the order of eigenvalues of $p_2\tr(\mathbf{\Psi}_{F}) \mathbf{I}_{p_1} + p_2\sigma^2\mathbf{I}_{p_1}$ is smaller than that of $p_2\mathbf{R}\mathbf{\Psi}_E\mathbf{R}^{\top}$.
Then, equation \eqref{span_R} holds under the strong factor assumption and yields $\widehat{\mathbf{R}}$, and so as its column-wise counterpart and $\widehat{\mathbf{C}}$.
Analog to the minimization of reconstruction error in Subsection \ref{subsec:estimation_factor}, the estimators of the latent factor matrices and the signal part are
\beqr
\label{signal_2w-DFM}
\breve{\mathbf{F}}_t = \frac{\mathbf{X}_t\widehat{\mathbf{C}}}{p_2},\quad
\breve{\mathbf{E}}_t = \frac{\mathbf{X}_t^{\top}\widehat{\mathbf{R}}}{p_1},\quad
\breve{\mathbf{S}}_t=\frac{\widehat{\mathbf{R}}\widehat{\mathbf{R}}^{\top}\mathbf{X}_t}{p_1}
+\frac{\mathbf{X}_t\widehat{\mathbf{C}}\widehat{\mathbf{C}}^{\top}}{p_2}.
\eeqr

For the $t$th image or matrix,
$\widehat{\mathbf{S}}_t$, $\widetilde{\mathbf{S}}_t$, and $\breve{\mathbf{S}}_t$ in equations \eqref{signal_RaDFaM},  \eqref{signal_bilinear}, and
\eqref{signal_2w-DFM}, which are the specific estimates $\widehat{\mathbf{X}}_t$ under unified PCA estimates $\widehat{\mathbf{R}}$ and $\widehat{\mathbf{C}}$, will yield their levels of the PSNR or the MSE under the three matrix factor models.
The following proposition shows that RaDFaM is superior to BiMFaM and 2w-DFM, respectively, in terms of the PSNR or the MSE.

\bep
\label{PSNR}
For each matrix observation $\mathbf{X}_t$, $t\in [T]$, we have
\beqrs
\text{PSNR}_{t,RaD} \geq \text{PSNR}_{t,BiM}~\text{and}~
\text{PSNR}_{t,RaD} \geq \text{PSNR}_{t,2w};
\eeqrs
or equivalently,
\beqrs
\text{MSE}_{t,RaD} \leq \text{MSE}_{t,BiM}~\text{and}~
\text{MSE}_{t,RaD} \leq \text{MSE}_{t,2w}.
\eeqrs
\eep
In addition, our numerical studies show that, even based on other PCA-type estimators, i.e., autoPCA \citep{WangLiuChen2019JOE-factor} and proPCA \citep{YuHeKongZhang2022JOE-projected} for BiMFaM, and Step-App for 2w-DFM \citep{YuanGaoHeHuangGuo2023JRSSB-two}, RaDFaM still has better reconstruction performance.

\subsection{Factor Number Estimation}
\label{subsec:estimation_factornumber}

The above subsections all assume known $k_1$ and $k_2$. When $k_1$ and $k_2$ are unknown, we need to determine them before estimation.
We adopt general ratio-type estimators for factor numbers.
Let $\widehat{\lambda}_{1}(\widehat{\mathbf{M}}_1) \geq \widehat{\lambda}_{2}(\widehat{\mathbf{M}}_1) \geq \ldots \geq \widehat{\lambda}_{p_1}(\widehat{\mathbf{M}}_1) \geq 0$ denote the ordered eigenvalues of $\widehat{\mathbf{M}}_1$ and let $k_{\text {max}}$ be a given upper bound for $k_1$. The number of row-wise factors can be estimated by
\beqr\label{number_k}
\widehat{k}_1=\underset{1 \leq j \leq k_{\max }}{\arg \max }\,\, \frac{\widehat{\lambda}_{j}(\widehat{\mathbf{M}}_1)}{\widehat{\lambda}_{j+1}(\widehat{\mathbf{M}}_1)},
\eeqr
and the column-wise $\widehat{k}_2$ can be defined similarly.
This ratio-type estimator is widely used in factor models including the vector case and the matrix case \citep{LamYao2012AOS-factor, YuHeKongZhang2022JOE-projected}.
We will prove its consistency in the next section.

\section{Asymptotic Properties}
\label{sec:theory}

In this section, we state necessary assumptions in Subsection \ref{subsec:assumptions} and provide asymptotic properties including the rate of convergence for estimators of loading matrices and the signal parts, the asymptotic distributions of estimators of the loading matrices, and the consistency of estimators of factor numbers, in Subsection \ref{subsec:asymptotic_properties}.

\subsection{Assumptions}
\label{subsec:assumptions}

\noindent
\textbf{Assumption 1.} \textit{$\alpha$-mixing: the vectorized factor processes $\{\vec(\mathbf{Z}_t)\}$, $\{\vec(\mathbf{F}_t)\}$, $\{\vec(\mathbf{E}_t)\}$, and the noise process $\{\vec(\mathbf{e}_t)\}$ are $\alpha$-mixing.}

A vector process $\{\boldsymbol{u}_t\}$ is $\alpha$-mixing,
if $\sum_{h=1}^{\infty} \alpha(h)^{1-2 / \gamma}<\infty$ for some $\gamma > 2$, where
$\alpha(h)=\sup _{i} \sup _{A \in \mathcal{C}_{-\infty}^{i}, B \in \mathcal{C}_{i+h}^{\infty}}|P(A \cap B)-P(A) P(B)|$ with $\mathcal{C}_{i}^{j}$ the $\sigma$-field generated by
$\left\{\boldsymbol{u}_{t}: i \leq t \leq j\right\}$.
The $\alpha$-mixing condition means that variables that are sufficiently far apart are asymptotic independent \citep[Appendix A.3]{FrancqZakoian2019garch}.

\noindent
{\bf{Assumption 2}} \textit{Common factors: assume that $k_1$ and $k_2$ are fixed. For any $t\in[T], i\in[p_1], j\in[p_2], u\in[k_1]$ and $v\in[k_2]$, there exists a positive constant $m$ such that $\mE(\mathbf{Z}^4_{t,uv})\leq m$, $\mE(\mathbf{F}^8_{t,iv})\leq m$, $\mE(\mathbf{E}^8_{t,ju})\leq m$ and
{\small$$\frac{1}{T}\sum\limits_{t=1}^T\mathbf{Z}_t\mathbf{Z}_t^\top \stackrel{a.s.}{\rightarrow}\mathbf{\Sigma}_{Z1},\quad \frac{1}{T}\sum\limits_{t=1}^T\mathbf{Z}_t^\top \mathbf{Z}_t \stackrel{a.s.}{\rightarrow}\mathbf{\Sigma}_{Z2},\quad
\frac{1}{Tp_1}\sum\limits_{t=1}^T \mathbf{F}_t^\top \mathbf{F}_t\stackrel{a.s.}{\rightarrow}\mathbf{\Sigma}_{F},\quad \frac{1}{Tp_2}\sum\limits_{t=1}^T \mathbf{E}_t^\top \mathbf{E}_t\stackrel{a.s.}{\rightarrow}\mathbf{\Sigma}_{E},$$}\noindent
where $\mathbf{\Sigma}_{Z1}\in\mR^{k_1\times k_1}$, $\mathbf{\Sigma}_{Z2}\in\mR^{k_2\times k_2}$, $\mathbf{\Sigma}_{F}\in\mR^{k_2\times k_2}$ and $\mathbf{\Sigma}_{E}\in\mR^{k_1\times k_1}$ are positive definite matrices.
Let $\mathbf{\Sigma}_1=\mathbf{\Sigma}_E+\mathbf{\Sigma}_{Z1}$ and $\mathbf{\Sigma}_2=\mathbf{\Sigma}_F+\mathbf{\Sigma}_{Z2}$. Assume that $\mathbf{\Sigma}_1$ and $\mathbf{\Sigma}_2$ have spectral decompositions $\mathbf{\Sigma}_1=\mathbf{\Gamma}_1\mathbf{\Lambda}_1\mathbf{\Gamma}_1^\top$ and $\mathbf{\Sigma}_2=\mathbf{\Gamma}_2\mathbf{\Lambda}_2\mathbf{\Gamma}_2^\top$, where the diagonal elements of $\mathbf{\Lambda}_1$ and $\mathbf{\Lambda}_2$ are distinct and arranged in decreasing order.}

Assumption 2 requires the interactive factor matrix $\mathbf{Z}_t$ to have bounded fourth moments, and the row-wise and column-wise factor matrices $\mathbf{F}_t$ and $\mathbf{E}_t$ to have bounded eighth moments.
It also requires the second sample moments of the factor matrices to converge to positive definite matrices (This can be derived under Assumption 1, see \citep[Chapter 16]{AthreyaLahiri2006measure} and \citep[Appendix A.3.]{FrancqZakoian2019garch} for more details).
The assumption of distinct and decreasing eigenvalues of $\mathbf{\Sigma}_1$ and $\mathbf{\Sigma}_2$ results in unique eigen-decompositions and identifiable eigenvectors.
Assumption 2 is an extension of the assumption in \cite{YuHeKongZhang2022JOE-projected} with extra row-wise and column-wise factors.

\noindent
{\bf{Assumption 3}} \textit{Loading matrices: there exist positive constants $\bar{r}$ and $\bar{c}$ such that
$\|\mathbf{R}\|_{max} \leq \bar{r}$ and $\|\mathbf{C}\|_{max} \leq \bar{c}$.
As $\min \left\{p_1,p_2\right\} \rightarrow \infty$,
we have $\|p_{1}^{-1} \mathbf{R}^{\top}\mathbf{R}-\mathbf{I}_{k_{1}}\| \rightarrow 0$ and
$\|p_{2}^{-1} \mathbf{C}^{\top}\mathbf{C}-$ $\mathbf{I}_{k_{2}}\| \rightarrow 0$,
where $\|\cdot\|$ is the $L_2$-norm of a matrix.}

Assumption 3 is a model identification condition.
It guarantees that our model belongs to the strong factor regime, which means that the factors are pervasively shared by elements of observations, and the signal component has spiked eigenvalues relative to the noise component \citep{StockWatson2002JASA-forecasting, LamYao2011Biometrika-estimation, FanLiaoMincheva2013JRSSB-large, ChenFanWang2020SCM-high}.

\noindent
{{\bf{Assumption 4}} \textit{Correlation: for any $s, t, t_1\in[T], i, i_1, i_2\in[p_1], j, j_1, j_2\in[p_2], u_1, u_2, u_3, u_4\in[k_1]$ and $v_1, v_2, v_3, v_4\in[k_2]$, there exists a positive constant $m$ such that}
\begin{enumerate}[(1)]
  \item[(4.1)] $\sum\limits_{t_2=1}^{T} |\mE(\mathbf{Z}_{t_1,u_1v_1}\mathbf{Z}_{t_2,u_2v_2})|\leq m$.

  \item [(4.2)]
  $\mE(\mathbf{e}_{t,ij}^8)\leq m$;
  $\sum\limits_{t_2=1}^{T} \sum\limits_{i_2=1}^{p_1} \sum\limits_{j_2=1}^{p_2} |\mE(\mathbf{e}_{t_1,i_1j_1}\mathbf{e}_{t_2,i_2j_2})|\leq m$;\\
  $\sum\limits_{t_2=1}^{T}
  \sum\limits_{i_3,i_4=1}^{p_1}
  \sum\limits_{j_3,j_4=1}^{p_2}
  |\cov(\mathbf{e}_{t_1,i_1j_1} \mathbf{e}_{t_1,i_2j_2},  \mathbf{e}_{t_2,i_3j_3}\mathbf{e}_{t_2,i_4j_4})|\leq m$;\\   $\sum\limits_{t_2=1}^{T}
  \sum\limits_{i_3,i_4=1}^{p_1}
  \sum\limits_{j_3,j_4=1}^{p_2}
  |\cov(\mathbf{e}_{s,i_1j_1} \mathbf{e}_{t_1,i_2j_2},  \mathbf{e}_{s,i_3j_3}\mathbf{e}_{t_2,i_4j_4})|\leq m$.

  \item[(4.3)]
  $\sum\limits_{t_2=1}^{T} \sum\limits_{i_2=1}^{p_1} |\mE(\mathbf{F}_{t_1,i_1v_1}\mathbf{F}_{t_2,i_2v_2})|\leq m$;   $\sum\limits_{t_2=1}^{T}\sum\limits_{i_3,i_4=1}^{p_1}
  |\cov(\mathbf{F}_{t_1,i_1v_1} \mathbf{F}_{t_1,i_2v_2},  \mathbf{F}_{t_2,i_3v_3}\mathbf{F}_{t_2,i_4v_4})|\leq m$;\\
  $\sum\limits_{t_2=1}^{T}
  \sum\limits_{i_3,i_4=1}^{p_1}
  |\cov(\mathbf{F}_{s,i_1v_1} \mathbf{F}_{t_1,i_2v_2},  \mathbf{F}_{s,i_3v_3}\mathbf{F}_{t_2,i_4v_4})|\leq m$.

  \item[(4.4)]
  $\sum\limits_{t_2=1}^{T} \sum\limits_{j_2=1}^{p_2} |\mE(\mathbf{E}_{t_1,j_1u_1}\mathbf{E}_{t_2,j_2u_2})|\leq m$;
  $\sum\limits_{t_2=1}^{T}
  \sum\limits_{j_3,j_4=1}^{p_2}
  |\cov(\mathbf{E}_{t_1,j_1u_1} \mathbf{E}_{t_1,j_2u_2},  \mathbf{E}_{t_2,j_3u_3}\mathbf{E}_{t_2,j_4u_4})|\leq m$;\\
  $\sum\limits_{t_2=1}^{T}
  \sum\limits_{j_3,j_4=1}^{p_2}
  |\cov(\mathbf{E}_{s,j_1u_1} \mathbf{E}_{t_1,j_2u_2},  \mathbf{E}_{s,j_3u_3}\mathbf{E}_{t_2,j_4u_4})|\leq m$.
\end{enumerate}

Assumption 4 focuses on the cross-sectional and temporal correlation of the noise matrices $\mathbf{e}_t$ and the factor matrices $\mathbf{Z}_t$, $\mathbf{F}_t$ and $\mathbf{E}_t$.
Assumption 4.1 requires the temporal correlation of $\mathbf{Z}_t$ to be weak, but no constraint is put on the cross-sectional correlation $\mathbf{Z}_t$ due to the finiteness of $k_1$ and $k_2$.  {Similar to Assumption 4.1,} Assumptions 4.2-4.4 are made for $\mathbf{e}_t$, $\mathbf{F}_t$ and $\mathbf{E}_t$, respectively.
Taking Assumption 4.2 of the matrix noise as an example, it requires a bounded eighth moment for each element, the cross-sectional and temporal correlation to be weak, and
the cross-sectional and temporal correlation of $\mathbf{e}_t\circ\mathbf{e}_t\in\mR^{p_1\times p_2\times p_1\times p_2}$ and $\mathbf{e}_s\circ\mathbf{e}_t\in\mR^{p_1\times p_2\times p_1\times p_2}$ to be weak.
In summary, we assume the correlations in the noise part and the factors are weak, and the independent matrix observations satisfy Assumption 4.}

\noindent
{\bf{Assumption 5}} \textit{Central limit theorems}:
for $i\in[p_1]$, $j\in[p_2]$, $s, t\in[T]$,
$$\frac{1}{\sqrt{T}}\sum\limits_{t=1}^T \mathbf{Z}_t \mathbf{F}_{t,i\cdot} \stackrel{d}{\rightarrow}N(\mathbf{0},\mathbf{V}_{1i}),~~\frac{1}{\sqrt{T}}\sum\limits_{t=1}^T \mathbf{Z}_t^\top \mathbf{E}_{t,j\cdot} \stackrel{d}{\rightarrow}N(\mathbf{0},\mathbf{V}_{2j}),$$
where $\mathbf{V}_{1i} = \mathop{\text{lim}}\limits_{T{\rightarrow}\infty}{T}^{-1}\sum_{s,t}
\mE(\mathbf{Z}_t\mathbf{F}_{t,i\cdot}\mathbf{F}_{s,i\cdot}^\top \mathbf{Z}_s^\top)$ and $\mathbf{V}_{2j} = \mathop{\text{lim}}\limits_{T{\rightarrow}\infty}{T}^{-1}\sum_{s,t}
  \mE(\mathbf{Z}_t^\top \mathbf{E}_{t,j\cdot}\mathbf{E}_{s,j\cdot}^\top \mathbf{Z}_s^\top).$

Assumption 5 is useful when deriving the asymptotic distributions of the estimated loading matrices.
Under Assumptions 1-4, Assumption 5 can be derived by applying the central limit theorem for $\alpha$-mixing processes \citep{AthreyaLahiri2006measure, FrancqZakoian2019garch}. Assumption 5 is different from the corresponding condition for BiMFaM in that the main terms differ due to the incorporation of row-wise and column-wise factors.

\subsection{Asymptotic Properties}
\label{subsec:asymptotic_properties}

Let $\widehat{\mathbf{\Lambda}}_1$ and $\widehat{\mathbf{\Lambda}}_2$ be the diagonal matrices with elements the top $k_1$ and $k_2$ eigenvalues of $\widehat{\mathbf{M}}_1$ and $\widehat{\mathbf{M}}_2$, respectively.
Based on $\widehat{\mathbf{M}}_1 \widehat{\mathbf{R}} = \widehat{\mathbf{R}} \widehat{\mathbf{\Lambda}}_1$ and $\widehat{\mathbf{M}}_2 \widehat{\mathbf{C}} = \widehat{\mathbf{C}} \widehat{\mathbf{\Lambda}}_2$, one may define the following asymptotic orthogonal matrices (see Appendix D in the online supplement for more details),
\beqrs
\mathbf{H}_1 &=& \frac{1}{T p_{1}} \sum\limits_{t=1}^{T}\mathbf{Z}_t\mathbf{Z}_t^{\top}\mathbf{\mathbf{R}}^{\top}\widehat{\mathbf{\mathbf{R}}}\widehat{\mathbf{\mathbf{\Lambda}}}_1^{-1}
+\frac{1}{T p_{1} p_{2}} \sum\limits_{t=1}^{T}\mathbf{E}_t^{\top}\mathbf{E}_t\mathbf{\mathbf{R}}^{\top}\widehat{\mathbf{\mathbf{R}}}\widehat{\mathbf{\mathbf{\Lambda}}}_1^{-1},\\
\mathbf{H}_2 &=& \frac{1}{T p_{2}} \sum\limits_{t=1}^{T}\mathbf{Z}_t^{\top}\mathbf{Z}_t\mathbf{\mathbf{C}}^{\top}\widehat{\mathbf{\mathbf{C}}}\widehat{\mathbf{\mathbf{\Lambda}}}_2^{-1}
+\frac{1}{T p_{1} p_{2}} \sum\limits_{t=1}^{T}\mathbf{F}_t^{\top}\mathbf{F}_t\mathbf{\mathbf{C}}^{\top}\widehat{\mathbf{\mathbf{C}}}\widehat{\mathbf{\mathbf{\Lambda}}}_2^{-1}.
\eeqrs
We first present the following convergence rates of $\widehat{\mathbf{R}}$ and $\widehat{\mathbf{C}}$ in the Frobenius norm.

\bet\label{consistency_loading}
Suppose that $T$, $p_1$ and $p_2$ tend to infinity, and that $k_1$ and $k_2$ are fixed.
If Assumptions 1-4 hold, then there exist asymptotic orthogonal matrices ${\mathbf{H}}_1$ and ${\mathbf{H}}_2$ such that
\[\frac{1}{p_1}||\widehat{\mathbf{R}}-\mathbf{R}{\mathbf{H}}_1||_F^2=O_p(\frac{1}{T}+\frac{1}{p_1^2}),\quad\frac{1}{p_2}||\widehat{\mathbf{C}}-\mathbf{C}{\mathbf{H}}_2||_F^2=O_p(\frac{1}{T}+\frac{1}{p_2^2}).\]
\eet
Our convergence rates for $\widehat{\mathbf{R}}$ and $\widehat{\mathbf{C}}$ are slower than those for PCA-type estimators under BiMFaM, though they achieve those in the high-dimensional vector factor model \citep{Bai2003Econometrica-inferential};
the convergence rates of proPCA \citep{YuHeKongZhang2022JOE-projected} are faster than those of other known methods.
This result reflects our strategy of sacrificing the theoretical rate of convergence to enhance the signal strength.

We next derive the asymptotic distributions of the estimated loading matrices.
As the row sizes of $\mathbf{R}$ and $\mathbf{C}$ tend to infinity, we establish the row-wise asymptotic distribution in the following Theorem \ref{asymptotic_normal}.

\bet\label{asymptotic_normal}
Suppose that $T$, $p_1$ and $p_2$ tend to infinity, and that $k_1$ and $k_2$ are fixed.
If Assumptions 1-5 hold, then
\begin{enumerate}
 \item for $i\in[p_1]$, we have
  $$\left\{\begin{array}{lr}
  \sqrt{T}(\widehat{\mathbf{R}}_{i\cdot}-{\mathbf{H}}_1^\top \mathbf{R}_{i\cdot}) \stackrel{d}{\rightarrow}
  N(\mathbf{0},\mathbf{\Lambda}_1^{-1}\mathbf{\Gamma}_1^\top \mathbf{V}_{1i}\mathbf{\Gamma}_1\mathbf{\Lambda}_1^{-1}),~~~ T=o(p_1^2),\\
  \widehat{\mathbf{R}}_{i\cdot}-{\mathbf{H}}_1^\top \mathbf{R}_{i\cdot} = O_p(\frac{1}{p_1}),~~~ p_1^2=O(T);
\end{array}\right.$$
  \item for $j\in[p_2]$, we have
  $$\left\{\begin{array}{lr}
  \sqrt{T}(\widehat{\mathbf{C}}_{j\cdot}-{\mathbf{H}}_2^\top \mathbf{C}_{j\cdot}) \stackrel{d}{\rightarrow}
  N(\mathbf{0},\mathbf{\Lambda}_2^{-1}\mathbf{\Gamma}_2^\top \mathbf{V_{2j}}\mathbf{\Gamma}_2\mathbf{\Lambda}_2^{-1}),~~~ T=o(p_2^2),\\
  \widehat{\mathbf{C}}_{j\cdot}-{\mathbf{H}}_2^\top \mathbf{C}_{j\cdot} = O_p(\frac{1}{p_2}),~~~ p_2^2=O(T).
\end{array}\right.$$
\end{enumerate}
\eet
Theorem \ref{asymptotic_normal} shows that the loading matrix estimator can be asymptotically normal when $p_1$ and $p_2$ are sufficiently large.
Even if the central limit theorem does not hold, the rows of estimated loading matrices are still consistent with no restriction on the limiting relationship between $T$ and $\{p_1, p_2\}$.

We now turn to the convergence rate of the estimated signal component $\widehat{\mathbf{S}}_t$ in equation \eqref{signal_RaDFaM}.
As the size of $\widehat{\mathbf{S}}_t$ tends to infinity, we will prove its element-wise consistency.

\bet\label{consistency_signal}
Suppose that $T$, $p_1$ and $p_2$ tend to infinity, and that $k_1$ and $k_2$ are fixed.
If Assumptions 1-4 hold, then we have
$$|\widehat{\mathbf{S}}_{t,ij}-\mathbf{S}_{t,ij}|=O_p(\frac{1}{\sqrt{T}}+\frac{1}{\sqrt{p_1}}+\frac{1}{\sqrt{p_2}}),~\text{for any}~i\in[p_1]~\text{and}~j\in[p_2].$$
\eet
Similar to the convergence rates of the estimators of the loading matrices, the convergence rate of the estimator of the signal component in RaDFaM is slower than that in BiMFaM.
However, as shown in Proposition \ref{PSNR}, RaDFaM has better reconstruction performance.

Theorems \ref{consistency_loading}-\ref{consistency_signal} are proved with fixed factor numbers $k_1$ and $k_2$.
If the factor numbers are unknown, we show that the ratio-type estimator (\ref{number_k}) provides consistent estimators.

\bet\label{convergence_k}
Suppose that $T$, $p_1$ and $p_2$ tend to infinity, and $k_{max}$ is not smaller than $\max\{k_1,k_2\}$.
If Assumptions 1-4 hold, then we have
$$
{P}(\widehat{k}_1 \neq k_1) \rightarrow 0 \quad\text{and}\quad
{P}(\widehat{k}_2 \neq k_2) \rightarrow 0.
$$
\eet
Theorem \ref{convergence_k} shows that with large enough $k_{max}$, the factor numbers can be estimated consistently.
The proof follows arguments similar to those used in the proof of Lemma 1 given in the online supplement.
As $\widehat{\mathbf{R}}$ and $\widehat{\mathbf{C}}$ of sPCA are identical to the estimator of $\mathbf{R}$ and $\mathbf{C}$ in $\alpha$-PCA with $\alpha=0$, they can use the same ratio-type estimators.

\section{Simulation Studies}

In this section, we present simulations to demonstrate the finite sample performance of the proposed RaDFaM.
The simulation settings for six different scenarios and measures of performance are introduced in Subsection \ref{subsec:simulation_settings}.
The simulation results assessing the performance of loading matrix estimators, signal part estimators and matrix reconstruction,
the time cost of computation,
the estimation error of factor numbers,
and the asymptotic normality of loading matrix estimators are reported in Subsection \ref{subsec:simulation_results2}.

\subsection{Simulation Settings and Performance Measures}
\label{subsec:simulation_settings}

We first introduce the simulation settings.
The matrix observations are generated with factor numbers $(k_1,k_2)=(3,3)$, and 6 sets of $(T,p_1,p_2)$, which are $(20,50,100)$, $(20,100,100)$, $(50,20,50)$, $(50,100,100)$, $(100,20,50)$ and $(100,50,100)$, respectively.
The loading matrix $\mathbf{R}$ is taken as $\sqrt{p_1}$ times the matrix of the top $k_1$ left singular vectors of the SVD decomposition of a $p_1\times k_1$-dimensional matrix with independent standard normal elements, and the loading matrix $\mathbf{C}$ is generated similarly.
Then $\mathbf{Z}_t$, $\mathbf{E}_t$ and $\mathbf{F}_t$ are generated from the following vector auto-regression models
\beqrs
\vec(\mathbf{Z}_t)=\phi\vec(\mathbf{Z}_{t-1})+\sqrt{1-\phi^2}\mathbf{u}_t,\\
\vec(\mathbf{F}_t)=\psi \vec(\mathbf{F}_{t-1})+\sqrt{1-\psi^2}\boldsymbol{\xi}_t,\\
\vec(\mathbf{E}_t)=\gamma vec(\mathbf{E}_{t-1})+\sqrt{1-\gamma^2}\boldsymbol{\eta}_t,
\eeqrs
where $\mathbf{u}_t\sim N(\mathbf{0},\mathbf{I}_{k_1k_2})$,
$\boldsymbol{\xi}_t\sim N(\mathbf{0},\mathbf{I}_{p_1k_2})$,
$\boldsymbol{\eta}_t\sim N(\mathbf{0},\mathbf{I}_{p_2k_1})$, and the entries of the initial values of $\vec(\mathbf{Z}_t)$, $\vec(\mathbf{F}_t)$ and $\vec(\mathbf{E}_t)$ are standard normal.
The coefficients $\phi$, $\psi$ and $\gamma$ control the temporal correlation of the factor matrices.
We generate $\mathbf{e}_t \sim MN\left(\mathbf{0};\mathbf{\Omega}_{e}, \mathbf{\Delta}_{e}\right)$, where $\mathbf{\Omega}_{e}$ and $\mathbf{\Delta}_{e}$
both have $1$'s on the diagonal and constant off-diagonal elements $1/p_1$ and $1/p_2$, respectively.
We consider six scenarios in the following table, the first three are the uncorrelated cases and the last three are the correlated cases.

\vspace{-0.8cm}
\setcounter{table}{-1}
\begin{table}[H]
  \caption{Simulation scenarios}
  \label{scenarios}
  \centering
\begin{tabular}{{ll|ccccc}}
\hline\hline
&   & $\phi$ & $\psi$ & $\gamma$ & Generating Model\\
\hline
&Scenario I   & 0 & 0 & 0 & RaDFaM\\
Uncorrelated&Scenario II  & 0 & - & - & BiMFaM\\
&Scenario III & - & 0 & 0 & 2w-DFM\\
\hline
&Scenario IV  & 0.6 & 0.8 & 0.8 & RaDFaM\\
Correlated&Scenario V   & 0.6 & - & - & BiMFaM\\
&Scenario VI  & - & 0.8 & 0.8 & 2w-DFM\\
\hline\hline
\end{tabular}
\end{table}

We compare sPCA under RaDFaM; autoPCA \citep{WangLiuChen2019JOE-factor}, $\alpha$-PCA with $\alpha=0$ \citep{ChenFan2023JASA-statistical}, and proPCA \citep{YuHeKongZhang2022JOE-projected} under BiMFaM; 2w-PCA (estimators in equation \eqref{signal_2w-DFM}), and Step-App and Q-MLE, the initial estimators and quasi-likelihood estimators in \cite{YuanGaoHeHuangGuo2023JRSSB-two}, respectively, under 2w-DFM.

Next, we introduce the measures of performance.
For the loading matrices, we use the following column space distance to characterize the performance of factor loading estimators
\beqrs
\mathcal{D}(\widehat{\mathbf{R}},\mathbf{R})
=\frac{1}{p_1}\left\|\widehat{\mathbf{R}}\widehat{\mathbf{R}}^{\top}-\mathbf{R}\mathbf{R}^{\top}\right\|_2,\quad
\mathcal{D}(\widehat{\mathbf{C}},\mathbf{C})
=\frac{1}{p_2}\left\|\widehat{\mathbf{C}}\widehat{\mathbf{C}}^{\top}-\mathbf{C}{\mathbf{C}}^{\top}\right\|_2.
\eeqrs
For the signal part, we measure the distance between the estimator and the true signal using
$$
\mathcal{D}_{signal} = \sqrt{\frac{1}{Tp_1p_2}\sum\limits_{t=1}^T\|\widehat{\mathbf{S}}_t-\mathbf{S}_t\|_F^2}.
$$
For assessing the reconstruction performance, we use
$$\text{RMSE} = \sqrt{\frac{1}{T}\sum\limits_{t=1}^T\text{MSE}_t},\quad
\overline{\text{PSNR}} = \frac{1}{T}\sum\limits_{t=1}^T\text{PSNR}_t,
$$
where the $\text{MSE}_t$ and the $\text{PSNR}_t$ are defined in Subsection \ref{subsec:reconstruction_error}.
For the estimated factor numbers, we assess the accuracy by
\beqrs
\mbox{acc}=\mathbf{I}(\widehat{k}_1=k_1,\widehat{k}_2=k_2)*100\%.
\eeqrs

\subsection{Simulation Results}
\label{subsec:simulation_results2}

\begin{figure}[H]
  \centering
\includegraphics[width=5in]{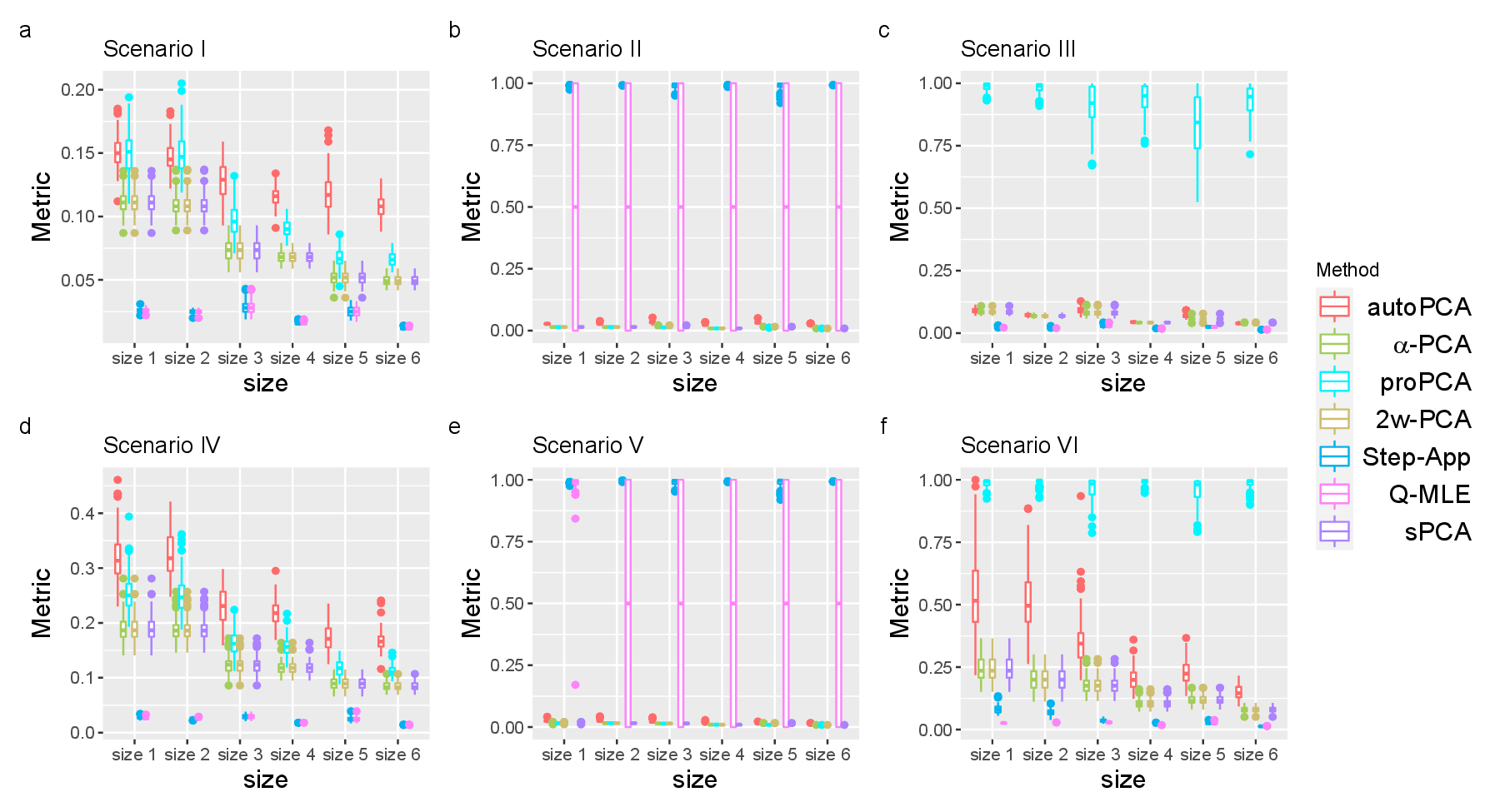}
\caption{Boxplots of $\mathcal{D}(\widehat{\mathbf{R}},\mathbf{R})$ for seven estimation methods under six size settings in six scenarios.}
  \label{simu_R}
\end{figure}

First, we focus on the performance of row-wise loading matrix estimators and the similar results for the column-wise counterpart are left in Appendix H of the online supplement.
Boxplots of $\mathcal{D}(\widehat{\mathbf{R}}, \mathbf{R})$ for the seven comparison methods over 100 replications are reported in Figure \ref{simu_R}.
Note that $\alpha$-PCA, 2w-PCA and sPCA provide identical estimators regardless of the generating model, so they have identical results under all scenarios.
For Scenarios I and IV,
autoPCA performs the worst, proPCA performs poorly, and Step-App and Q-MLE perform the best, followed by $\alpha$-PCA/2w-PCA/sPCA;
for Scenarios II and V, Q-MLE fails most (abnormal upright bars in Subfigures b and e) and performs poorly when it works, Step-App performs worst, and other methods have similar results;
for Scenarios III and VI, proPCA performs the worst, autoPCA performs poorly for correlated observations, and Step-App and Q-MLE perform the best, followed by $\alpha$-PCA/2w-PCA/sPCA.

\begin{figure}[H]
\centering
\includegraphics[height=2.5in,width=5in]{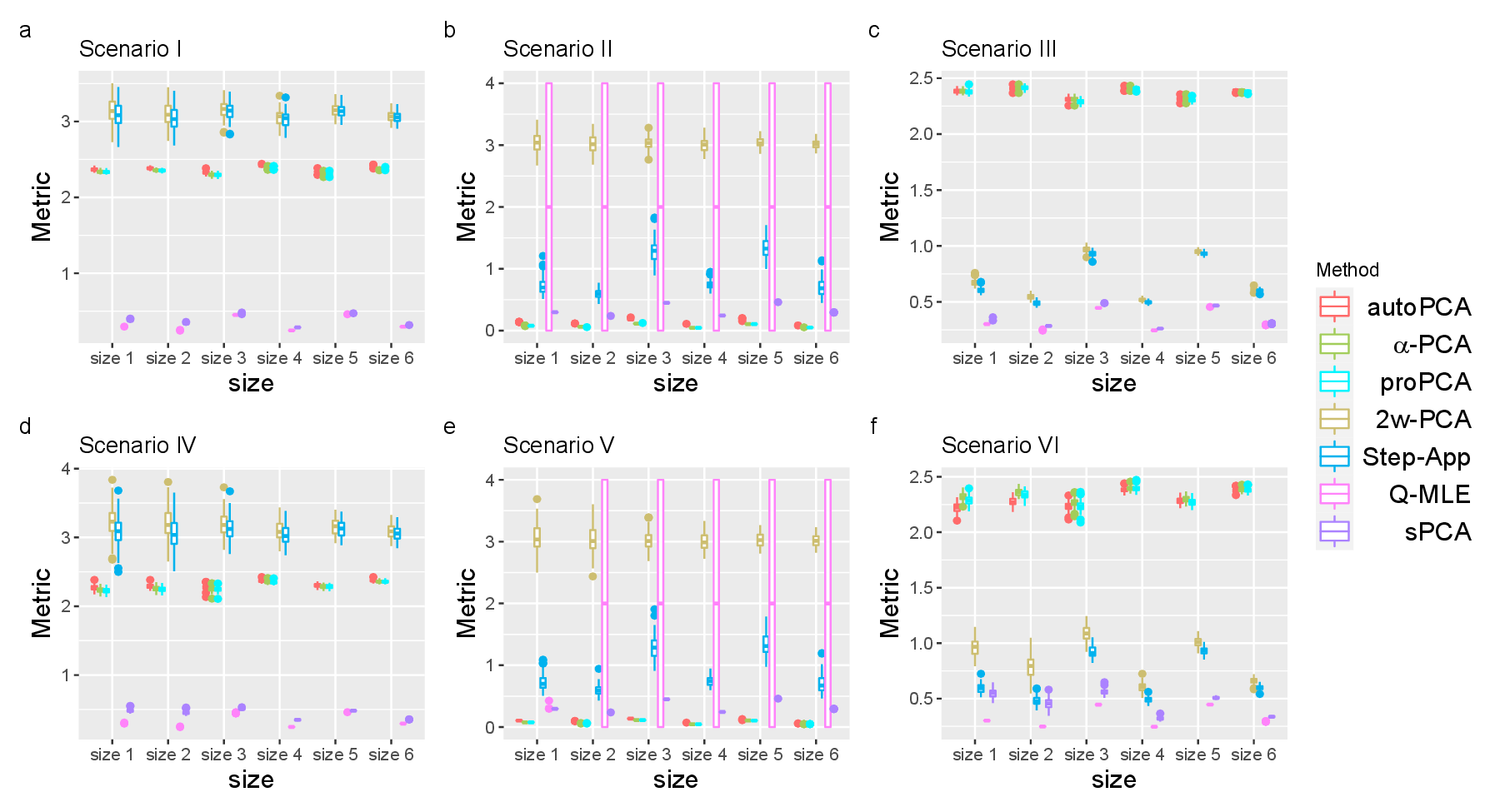}
\caption{Boxplots of $\mathcal{D}_{signal}$ for seven estimation methods under six size settings in six scenarios.}
\label{simu_S}
\end{figure}

\begin{figure}[H]
\centering
\includegraphics[height=2.5in,width=5in]{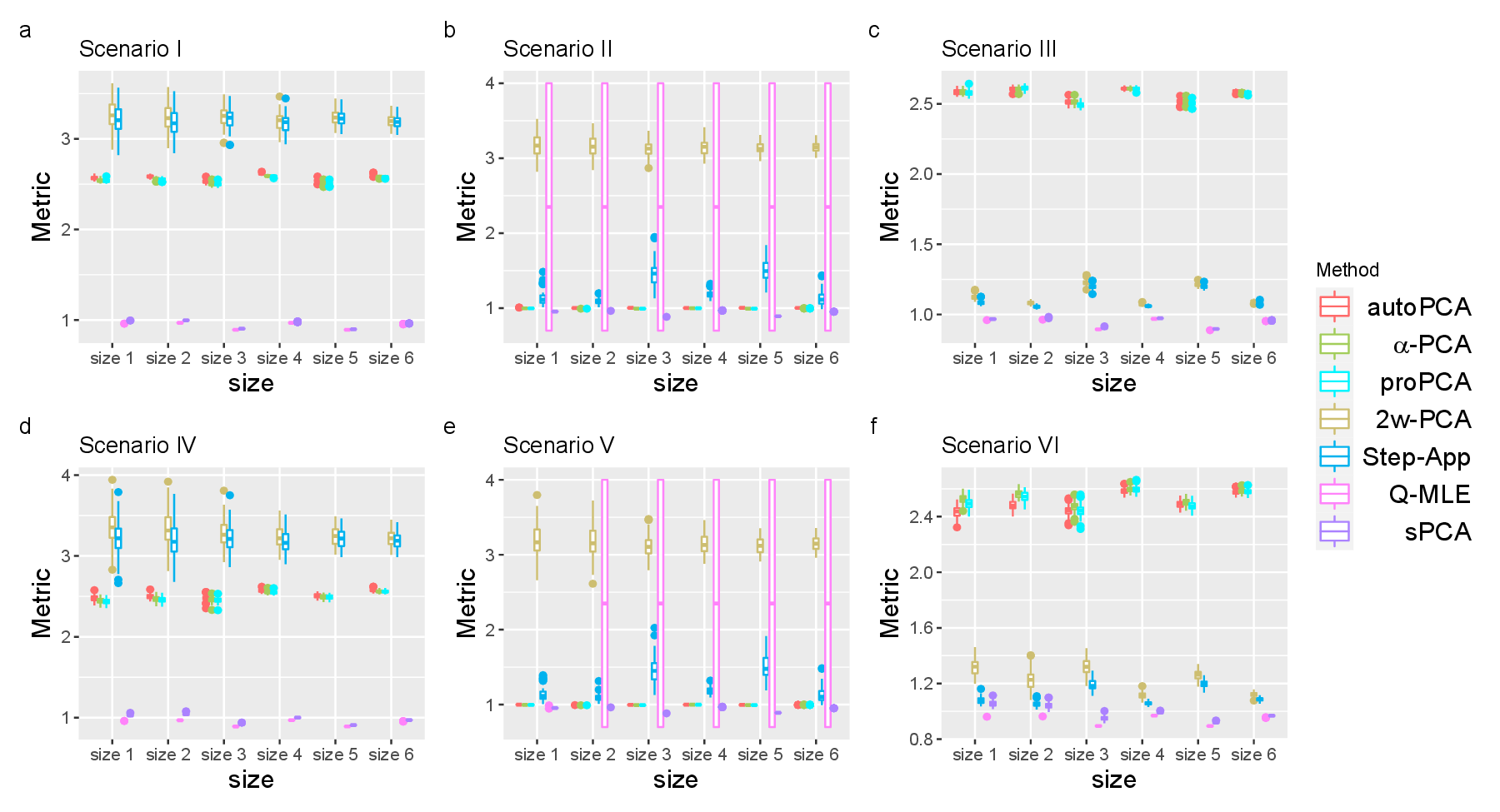}
\caption{Boxplots of the $\text{RMSE}$ for seven estimation methods under six size settings in six scenarios.}
\label{simu_X1}
\end{figure}

\begin{figure}
\centering
\includegraphics[height=2.5in,width=5in]{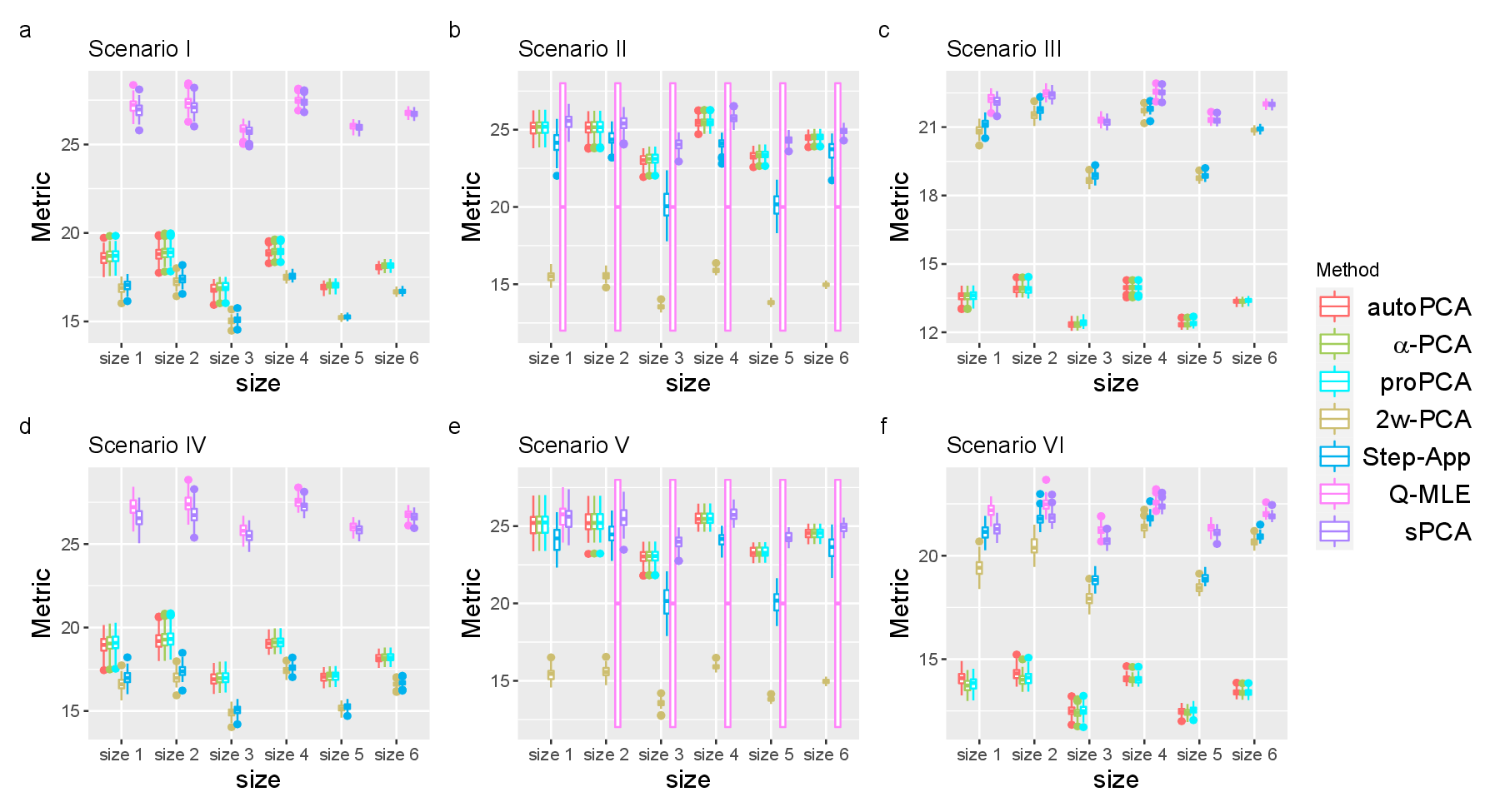}
\caption{Boxplots of the $\overline{\text{PSNR}}$ for seven estimation methods under six size settings in six scenarios.}
\label{simu_X2}
\end{figure}

Second, for the performance of the signal parts estimators and matrix reconstruction, boxplots of the $\mathcal{D}_{signal}$, the $\text{RMSE}$ and the $\overline{\text{PSNR}}$ from 100 replications are shown in Figures \ref{simu_S} - \ref{simu_X2}.
For Scenarios I and IV, 2w-PCA and Step-App perform the worst, autoPCA, $\alpha$-PCA and proPCA perform poorly, and sPCA and Q-MLE perform the best;
for Scenarios II and V, Q-MLE fails at most cases, 2w-PCA performs worst, Step-App performs poorly, and other methods have similar results;
for Scenarios III and VI, autoPCA, $\alpha$-PCA and proPCA perform the worst, 2w-PCA and Step-App perform poorly, and sPCA and Q-MLE perform the best.

\begin{figure}[H]
\centering
\includegraphics[height=2.5in,width=5in]{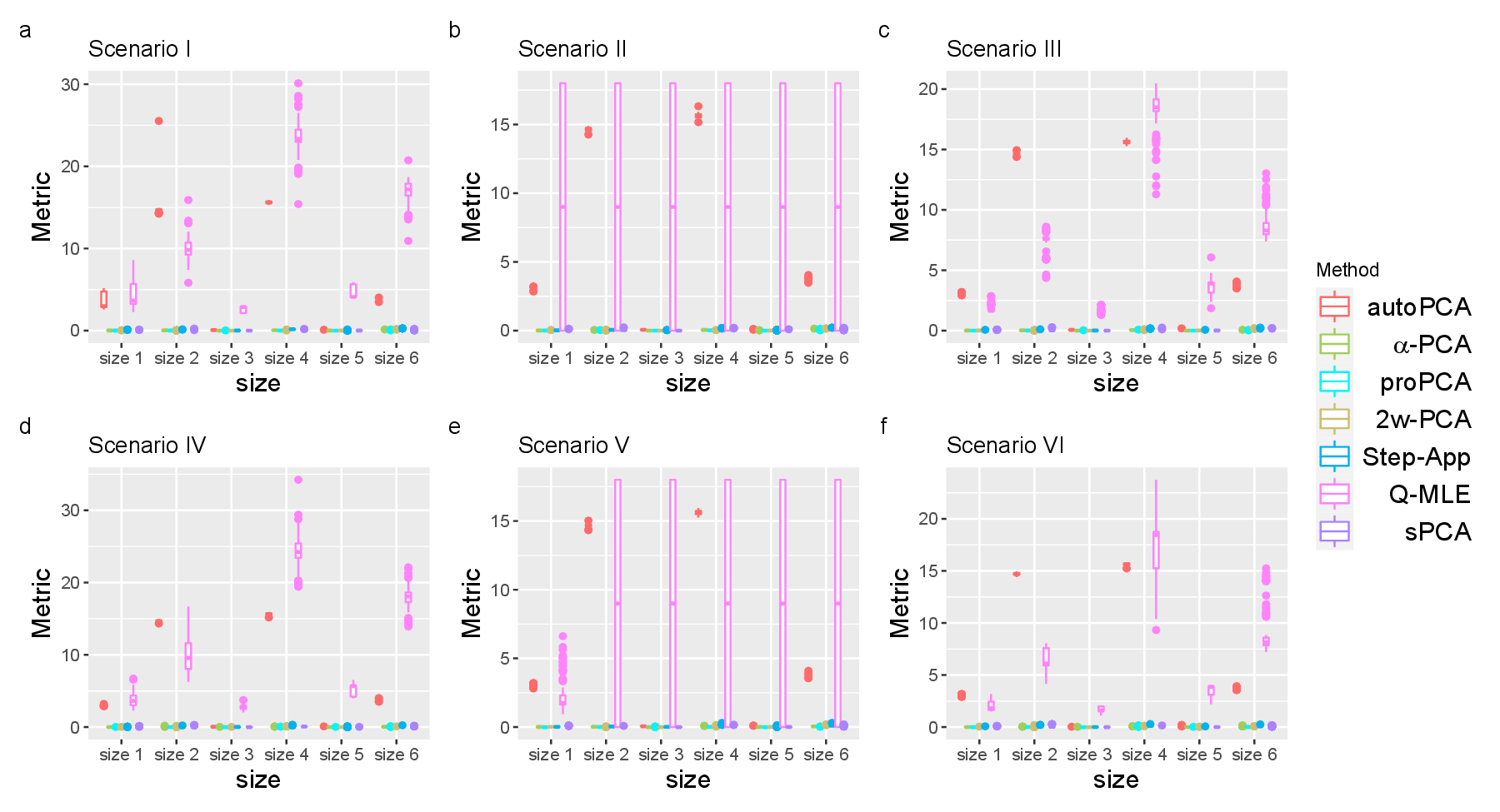}
\caption{Boxplots of the computational time for seven estimation methods under six size settings in six scenarios.}
\label{simu_TIME}
\end{figure}

Third, to evaluate the computational time required for different methods, boxplots of the run time in seconds from 100 replications are shown in Figure \ref{simu_TIME}.
It is evident that the Q-MLE incurs the highest computational time cost, particularly for large matrices and sample sizes. Following Q-MLE, autoPCA exhibits a relatively higher time cost, while the remaining methods show similar time requirements.

Fourth, the results of the estimation error of factor numbers are shown in Table \ref{FacNum}, from which it can be seen that $\alpha$-PCA/2w-PCA/sPCA estimate the factor numbers almost correctly under all scenarios, autoPCA does not work under RaDFaM and 2w-DFM, proPCA does not work under 2w-DFM, and Step-App and Q-MLE do not work under BiMFaM.

\begin{table}
\caption{The result for estimated factor numbers}
\label{FacNum}
\begin{center}
\scalebox{0.8}{
\resizebox{1.2\columnwidth}{!}{
\begin{tabular}{|c|cccccc|cccccc|cccccc|}
\hline
& \multicolumn{6}{c|}{scenario I} & \multicolumn{6}{c|}{scenario II} & \multicolumn{6}{c|}{scenario III}\\
\hline
methods & size 1 & size 2 & size 3 & size 4 & size 5 & size 6
& size 1 & size 2 & size 3 & size 4 & size 5 & size 6
& size 1 & size 2 & size 3 & size 4 & size 5 & size 6\\
\hline
autoPCA       & 2   & 10  & 6   &  99 & 61  & 100
              & 100 & 100 & 100 & 100 & 100 & 100
              &   0 &   0 &   0 &   0 &   0 &   0\\
$\alpha$-PCA  & 100 & 100 & 100 & 100 & 100 & 100
              & 100 & 100 & 100 & 100 & 100 & 100
              & 100 & 100 & 100 & 100 & 100 & 100\\
proPCA        & 100 & 100 & 100 & 100 & 100 & 100
              & 100 & 100 & 100 & 100 & 100 & 100
              & 2   & 1   & 13  & 2   & 18  & 18 \\
2w-PCA        & 100 & 100 & 100 & 100 & 100 & 100
              & 100 & 100 & 100 & 100 & 100 & 100
              & 100 & 100 & 100 & 100 & 100 & 100\\
Step-App      & 100 & 100 & 100 & 100 & 100 & 100
              &   0 &   0 &   0 &   0 &   0 &   0
              & 100 & 100 & 100 & 100 & 100 & 100\\
Q-MLE         & 100 & 100 & 100 & 100 & 100 & 100
              &   - &   - &   - &   - &   - &   -
              & 100 & 100 & 100 & 100 & 100 & 100\\
sPCA          & 100 & 100 & 100 & 100 & 100 & 100
              & 100 & 100 & 100 & 100 & 100 & 100
              & 100 & 100 & 100 & 100 & 100 & 100\\
\hline
& \multicolumn{6}{c|}{scenario IV} & \multicolumn{6}{c|}{scenario V} & \multicolumn{6}{c|}{scenario VI}\\
\hline
methods & size 1 & size 2 & size 3 & size 4 & size 5 & size 6
& size 1 & size 2 & size 3 & size 4 & size 5 & size 6
& size 1 & size 2 & size 3 & size 4 & size 5 & size 6\\
\hline
autoPCA       &   0 &   0 &   0 &   0 &   0 &   0
              & 100 & 100 & 100 & 100 & 100 & 100
              &  33 &  57 &   1 &  77 &  15 &   4\\
$\alpha$-PCA  & 100 & 100 & 100 & 100 & 100 & 100
              & 100 & 100 & 100 & 100 & 100 & 100
              &  94 & 100 & 100 & 100 & 100 & 100\\
proPCA        &  99 &  99 & 100 & 100 & 100 & 100
              & 100 & 100 & 100 & 100 & 100 & 100
              &   4 &   6 &   0 &   2 &   4 &   1\\
2w-PCA        & 100 & 100 & 100 & 100 & 100 & 100
              & 100 & 100 & 100 & 100 & 100 & 100
              &  94 & 100 & 100 & 100 & 100 & 100\\
Step-App      & 100 & 100 & 100 & 100 & 100 & 100
              &   0 &   0 &   0 &   0 &   0 &   0
              & 100 & 100 & 100 & 100 & 100 & 100\\
Q-MLE         & 100 & 100 & 100 & 100 & 100 & 100
              &   0 &   - &   - &   - &   - &   -
              & 100 & 100 & 100 & 100 & 100 & 100\\
sPCA          & 100 & 100 & 100 & 100 & 100 & 100
              & 100 & 100 & 100 & 100 & 100 & 100
              &  94 & 100 & 100 & 100 & 100 & 100\\
\hline
\end{tabular}}}
\end{center}
\end{table}

\begin{figure}[H]
\centering
\includegraphics[height=2in,width=3.5in]{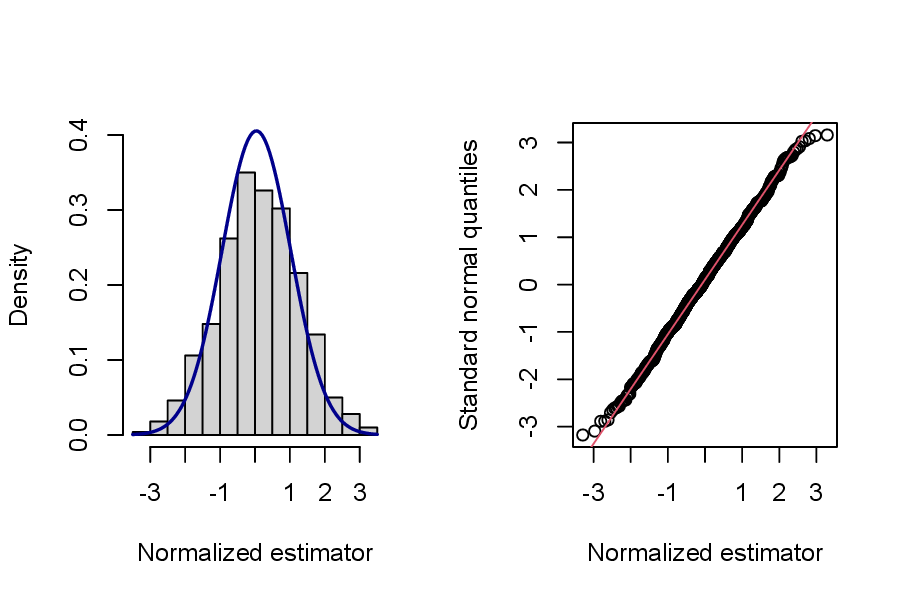}
\caption{Histogram (with superimposed standard normal density) and QQ-plot of the normalised estimator $\widehat{\mathbf{R}}_{13}$.}
\label{distribution_R}
\end{figure}

Finally, we illustrate the asymptotic normality of the sPCA estimator $\widehat{\mathbf{R}}$ under Scenario I with $(T,p_1,p_2)=(100,150,150)$.
In this case, the asymptotic variance-covariance matrix of $\widehat{\mathbf{R}}_{i\cdot}$ is $k_2/(k_2+1)^2\mathbf{I}_{k_1}$, so we present
the histogram and QQ plot of the third element of $(k_2+1)\sqrt{T/k_2}(\widehat{\mathbf{R}}_{1\cdot}-{\mathbf{H}}_1^{\top}\mathbf{R}_{1\cdot})$ over 1000 replications in Figure \ref{distribution_R}.
The smooth curve in the left plot is the probability density function of the standard normal distribution, and the QQ plot supports the normal approximation (the $p$-value of the Shapiro-Wilk test for normality is 0.4243).

In summary, based on PCA-type estimation, sPCA demonstrates excellent performance in estimating loading matrices and factor numbers, and outperforms other methods in terms of matrix reconstruction, even when dealing with data generated from BiMFaM and 2w-DFM.
Compared to Q-MLE, which refines Step-App, sPCA's performance in estimating loading matrices, factor numbers, and matrix reconstruction is comparable with much less computational time, and is robust for dealing with data generated from BiMFaM when Q-MLE usually fails.

\begin{figure}[H]
\centering
\subfigure[Original]
{\captionsetup{font=footnotesize}
\includegraphics[height=1.2in,width=1.2in]{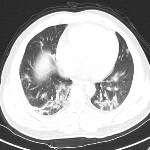}}
\quad
\subfigure[$k_1=k_2=5$]
{\includegraphics[height=1.2in,width=1.2in]{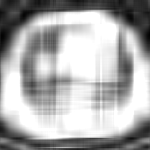}}
\quad
\subfigure[$k_1=k_2=15$]
{\includegraphics[height=1.2in,width=1.2in]{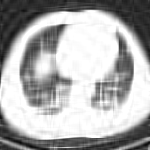}}
\\
\subfigure[$k_1=k_2=25$]
{\includegraphics[height=1.2in,width=1.2in]{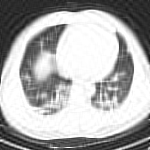}}
\quad
\subfigure[$k_1=k_2=35$]
{\includegraphics[height=1.2in,width=1.2in]{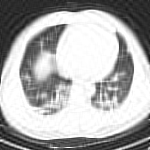}}
\quad
\subfigure[$k_1=k_2=45$]
{\includegraphics[height=1.2in,width=1.2in]{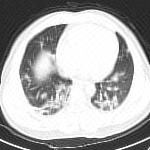}}
\caption{Reconstructed images for different factor numbers under RaDFaM.}
\label{COVID-FIGURE}
\end{figure}

\section{Real Data Analysis}

\subsection{Uncorrelated Data: CT Images for COVID-19}
\label{subsec:CTimage}

In this subsection, we apply the proposed approach to a real open-source chest CT data called COVID-CT, which was one of the largest publicly available COVID-19 CT scan dataset in the early pandemic \citep{YangHeZhaoZhangXie2020arXiv-covid,Zhang3Liu2-2023SIM-low}.
It contains 349 COVID-19 positive CT scans manually selected from the embedded figures out of 760 preprints on medRxiv2 and bioRxiv3, and 397 negative CT scans from four open-access databases.
Due to figure format problems, we finally have 317 positive scans and 397 negative scans.
As discussed in Subsection \ref{subsec:reconstruction_error}, grayscale image data can be regarded as matrix data, we have $T=714$, $p_1=150$ and $p_2=150$ by resizing each image to $150\times 150$.

\textbf{\textit{Image reconstruction}}
RaDFaM, BiMFaM and 2w-DFM can be used to extract low-dimensional features from the matrix-valued observations, and the estimated low-rank signal parts can be regarded as the reconstruction of the original image.
We conduct a comparison of the reconstruction performance of the seven methods used in simulation studies.
The factor numbers $k_1$ and $k_2$ are set to be equal and take on the values of $5$, $15$, $25$, $35$ and $45$.
Furthermore, we also compare with the autoencoder (AE) methods: i) AE-1 and AE-5 with $1$ and $5$ hidden layers  are used, and the number of neurons are set to be  $k_1$ and $(256,128,k_1,128,256)$, respectively \citep{HintonSalakhutdinov2006Science-reducing, FanMaZhong2021SS-selective}.
We use the whole sample as the batch, and carry out 500 epochs;
ii) a convolutional autoencoder (CAE) of $8$ hidden layers is used to make full use of the matrix structural information, where the number of neurons in the central fully connected bottleneck layer is set to be $k_1$ \citep{GuoLiuZhuYin2017ICONIP-deep}.
We take 32 as the batch size, and carry out 50 epochs.

The results of the $\text{RMSE}$ and the $\overline{\text{PSNR}}$ for all methods are shown in Tables \ref{COVID1} and \ref{COVID2}, respectively.
We can see that:
i) except 2w-PCA, Step-App and AE methods, the performance of other methods improves as the number of factors increases; see the reconstructed results of sPCA for different factor numbers in Figure \ref{COVID-FIGURE} for visulization;
ii) the relatively small sample size ($T=714$) makes the performance of AE methods worse than the statistical methods;
iii) sPCA outperforms other PCA-type estimations;
iv) sPCA uses much less computational time than Q-MLE and has comparable performance; refer to the computational time (in seconds) in Table \ref{COVID3}.

\begin{table}[H]
  \caption{\label{COVID1}RMSEs of different estimation methods for different choices of $(k_1,\,k_2)$ applied to the COVID-CT data.}
  \centering
\begin{tabular}{*{7}{c}}
\hline\hline
Model & Method/ $(k_1,k_2)$ & $(5,5)$ & $(15,15)$ & $(25,25)$ & $(35,35)$ & $(45,45)$ \\
\hline\hline
BiMFaM & autoPCA      & 0.2009 & 0.1347 & 0.1097 & 0.0949 & 0.0836\\
       & $\alpha$-PCA & 0.1907 & 0.1238 & 0.0975 & 0.0814 & 0.0698\\
       & proPCA       & 0.1899 & 0.1230 & 0.0969 & 0.0810 & 0.0695\\
\hline
2w-DFM & 2w-PCA       & 0.6623 & 0.6710 & 0.6732 & 0.6744 & 0.6752\\
       & Step-App     & 0.6370 & 0.6558 & 0.6567 & 0.6580 & 0.6594\\
       &   Q-MLE      & \textbf{0.1169} & \textbf{0.0708} & \textbf{0.0528} & 0.0488 & \textbf{0.0368}\\
       \hline
RaDFaM & sPCA         & 0.1246 & 0.0782 & 0.0579 & \textbf{0.0456} & 0.0369\\
\hline
       & AE-1         & 0.3403 & 0.3403 & 0.3403 & 0.3404 & 0.3404\\
AE     & AE-5         & 0.3403 & 0.3403 & 0.3403 & 0.3403 & 0.3403\\
       & CAE          & 0.3383 & 0.3383 & 0.3383 & 0.3384 & 0.3384\\
\hline\hline
\end{tabular}
\end{table}

\begin{table}[H]
\caption{\label{COVID2}$\overline{\text{PSNR}}$s of different estimation methods for different choices of $(k_1,\,k_2)$ applied to the COVID-CT data.}
\centering
\begin{tabular}{*{7}{c}}
\hline\hline
Model &  Method/ $(k_1,k_2)$ & $(5,5)$ & $(15,15)$ & $(25,25)$ & $(35,35)$ & $(45,45)$ \\
\hline\hline
BiMFaM & autoPCA      & 14.1498 & 17.6611 & 19.4595 & 20.7566 & 21.9110\\
       & $\alpha$-PCA & 14.6213 & 18.3889 & 20.5233 & 22.1957 & 23.6577\\
       & proPCA       & 14.6597 & 18.4334 & 20.5756 & 22.2437 & 23.6945\\
       \hline
2w-DFM & 2w-PCA       &  3.9152 &  3.7891 &  3.7549 &  3.7362 &  3.7242\\
       & Step-App     &  4.2524 &  4.0008 &  3.9877 &  3.9683 &  3.9478\\
       & Q-MLE        & \textbf{18.8588} & \textbf{23.3783} & \textbf{26.1569} & 26.9265 & 29.7442\\
       \hline
RaDFaM & sPCA         & 18.3016 & 22.4788 & 25.3164 & \textbf{27.6773} & \textbf{29.8249}\\
\hline
       & AE-1         & 9.5018 & 9.5010 & 9.5004 & 9.4994 & 9.4976\\
AE     & AE-5         & 9.5023 & 9.5022 & 9.5019 & 9.5020 & 9.5020\\
       & CAE          & 9.5534 & 9.5537 & 9.5518 & 9.5516 & 9.5499\\
\hline\hline
\end{tabular}
\end{table}

\begin{table}[H]
\caption{\label{COVID3}Computational time (s) of different estimation methods for different choices of $(k_1,\,k_2)$ applied to the COVID-CT data.}
\centering
\begin{tabular}{*{7}{c}}
\hline\hline
Model &  Method/ $(k_1,k_2)$ & $(5,5)$ & $(15,15)$ & $(25,25)$ & $(35,35)$ & $(45,45)$ \\
\hline\hline
BiMFaM & autoPCA      & 287.64 & 291.38 & 288.88 & 289.68 & 293.33\\
       & $\alpha$-PCA &   3.33 &   3.27 &   3.23 &   3.47 &   3.66\\
       & proPCA       &   0.47 &   0.80 &   1.27 &   2.73 &   2.35\\
       \hline
2w-DFM & 2w-PCA       &    3.88 &   4.39 &   4.82 &   4.83 &   5.50\\
       & Step-App     &    6.42 &   8.67 &  10.68 &  15.14 &  18.45\\
       & Q-MLE        &  435.47 &3374.95 &1247.70 &1723.92 &3894.55\\
       \hline
RaDFaM & sPCA         &    4.99 &   4.75 &   6.17 &   5.88 &   9.01\\
\hline\hline
\end{tabular}
\end{table}

\textbf{\textit{Image classification: RaDFaM vs BiMFaM vs 2w-DFM}}
Once unsupervised learning on high-dimensional matrix-variate objects is built up under a specific latent structure, one may conduct supervised learning based on the extracted low-dimensional features.
Take RaDFaM on the aforementioned CT scanning data set $\{(Y_t, \mathbf{X}_t) , t \in [714]\}$ for instance, where each CT scan $\mathbf{X}_t$ is a biomedical image biomarker to diagnose the disease label $Y_t \in \{0,1\}$ with 0 being negative and 1 otherwise.
Let
 $\langle\mathbf{A},\mathbf{B}\rangle=\vec^{\top}(\mathbf{A})\vec(\mathbf{B})$ be the matrix inner product for the placeholder matrices $\mathbf{A}$ and $\mathbf{B}$.
Then a two-step latent logistic regression model can be practically deployed, named \textit{LLR$_{RaD}$},
\beqrs
\label{hlr}
\left\{\begin{array}{l}
\mbox{logit}P(Y_t=1|\mathbf{Z}_t,\mathbf{F}_t,\mathbf{E}_t)=\gamma + \langle\mathbf{B}_1,\mathbf{Z}_t\rangle + \langle\mathbf{B}_2,\mathbf{F}_t\rangle + \langle\mathbf{B}_3,\mathbf{E}_t\rangle,\\
\mathbf{X}_t=\mathbf{R} \mathbf{Z}_t \mathbf{C}^\top + \mathbf{R} \mathbf{E}_t^\top + \mathbf{F}_t \mathbf{C}^\top+\mathbf{e}_t,
\end{array}\right.
\eeqrs
where $\gamma$ is the intercept, and
$\mathbf{B}_1\in\mR^{k_1\times k_2}$, $\mathbf{B}_2\in\mR^{p_1\times k_2}$, and $\mathbf{B}_3\in\mR^{p_2\times k_1}$ are the regression coefficients accompanied with the latent-factor regressors
$\mathbf{Z}_t$, $\mathbf{F}_t$, and $\mathbf{E}_t$, respectively.
Analogous latent models on the same CT scan dataset are denoted as
\textit{LLR$_{BiM}$} and \textit{LLR$_{2w}$} with
the latent factor regressor $\mathbf{Z}_t$ and latent factor regressors
$\mathbf{E}_t$, $\mathbf{F}_t$, based on the unsupervised learning of the matrix factor models BiMFaM and 2w-DFM, respectively.


We randomly split samples into a training set ($80\%$) and a testing set ($20\%$).
In the training or fitting process, we estimate loadings and latent factors by sPCA (\textit{LLR$_{RaD}$}), $\alpha$-PCA and proPCA (\textit{LLR$_{BiM}$}), and 2w-PCA and Step-App (\textit{LLR$_{2w}$}), respectively, followed by estimating the matrix regression coefficients (considering the computational time cost, autoPCA and Q-MLE are not used).
In the testing or prediction process, we obtain the new regressors by the factor score equations (such as \eqref{factormatrix_RaDFaM}, \eqref{signal_bilinear} and \eqref{signal_2w-DFM}) based on the estimated loadings, followed by predicting labels.
The procedure is randomly repeated for 100 times.

The factor numbers $k_1$ and $k_2$ are set to be equal, and take values of 1, 5, 9, 13, 17, 21, and 25, respectively.
Note that LASSO type thresholding is applied to sPCA, 2w-PCA, and Step-App due to the high dimensionality of $\vec(\mathbf{F}_t)$ and $\vec(\mathbf{E}_t)$, and it is also used for $\alpha$-PCA and proPCA when the factor numbers exceed 5 due to the increasing dimensionality of $\vec(\mathbf{Z}_t)$ with the growing factor numbers.

\begin{figure}[H]
\centering
\includegraphics[height=2.3in,width=3.3in]{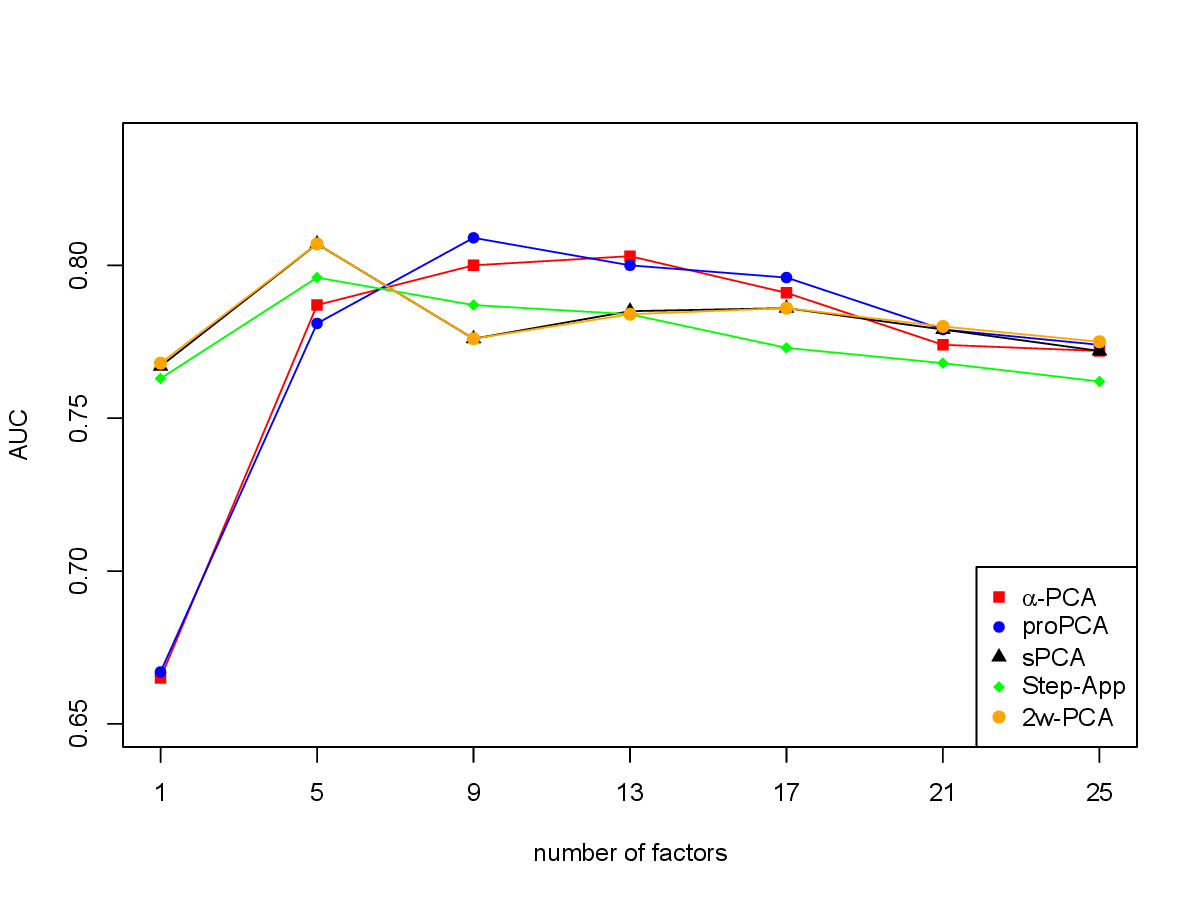}
\caption{Average AUC values for different choice of $(k_1,k_2)$ for COVID-CT data.}
\label{AUC}
\end{figure}

Figure \ref{AUC} is a scatter plot of the average AUC values against different factor numbers.
We may observe that,
i) the performance of sPCA (black) and 2w-PCA (orange) are almost identical.
This may be owing to their identical estimations of $\mathbf{E}_t$ and $\mathbf{F}_t$ in equations \eqref{factormatrix_RaDFaM} and \eqref{signal_2w-DFM}, and the shrinkage function of LASSO that eliminates the interactive effect $\mathbf{Z}_t$ automatically;
ii) the number of factors may affect the prediction accuracy of the latent logistic regression models, particular for the supervised learning derived based on BiMFaM.

\subsection{Correlated Data: Multinational Macroeconomic Indices}
\label{subsec:MultiMacroIndeice}

We analyze the quarterly multinational macroeconomic indices data collected from the Organisation of Economic Cooperation and Development (OECD) containing $10$ quarterly macroeconomic indices from $14$ countries for $74$ quarters from 2000.Q1 to 2018.Q2.
Samples of each quarter can be treated as a matrix-variate with rows, columns, and each element representing different countries, macroeconomic indices, and the value of the index of the corresponding country, respectively.

The $10$ indices consist of four groups: the production group (P:TIEC, P:TM, GDP), the consumer price group (CPI:FOOD, CPI:ENER, CPI:TOT), the money market group (IR:Long, IR:SHORT), and the international trade group (IT:EXP, IT:IM).
The countries and corresponding abbreviations are given in Appendix I of the online supplement.
After transformations of each univariate time series along the index mode as in \cite{ChenFan2023JASA-statistical} (see Table 10 in the supplement),
we center the matrix-variate series to eliminate mean effects, and obtain observations with $T = 72$, $p_1=14$ and $p_2=10$.

\begin{figure}
\centering
\subfigure[autoPCA.]
{\includegraphics[height=1.3in,width=1.8in]{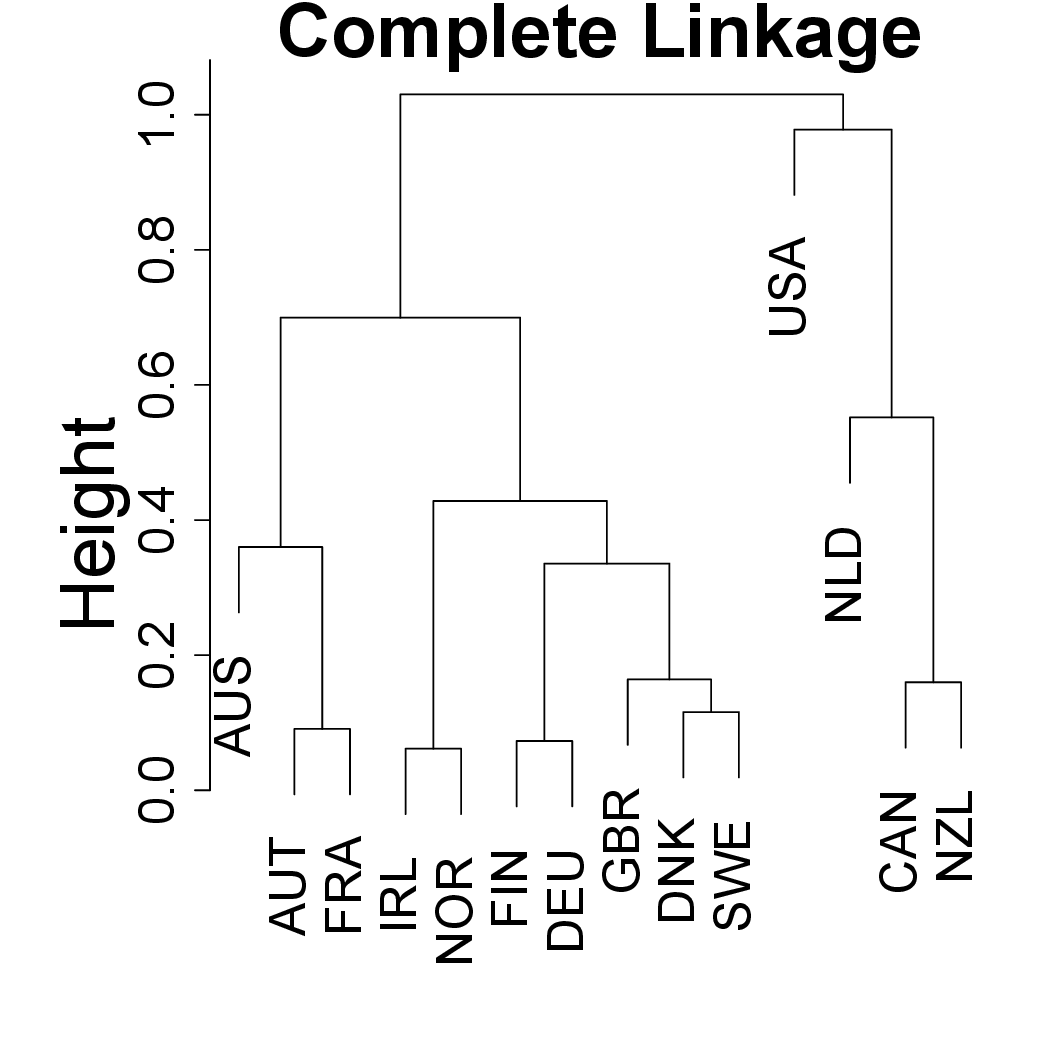}}
\quad
\subfigure[$\alpha$-PCA.]
{\includegraphics[height=1.3in,width=1.8in]{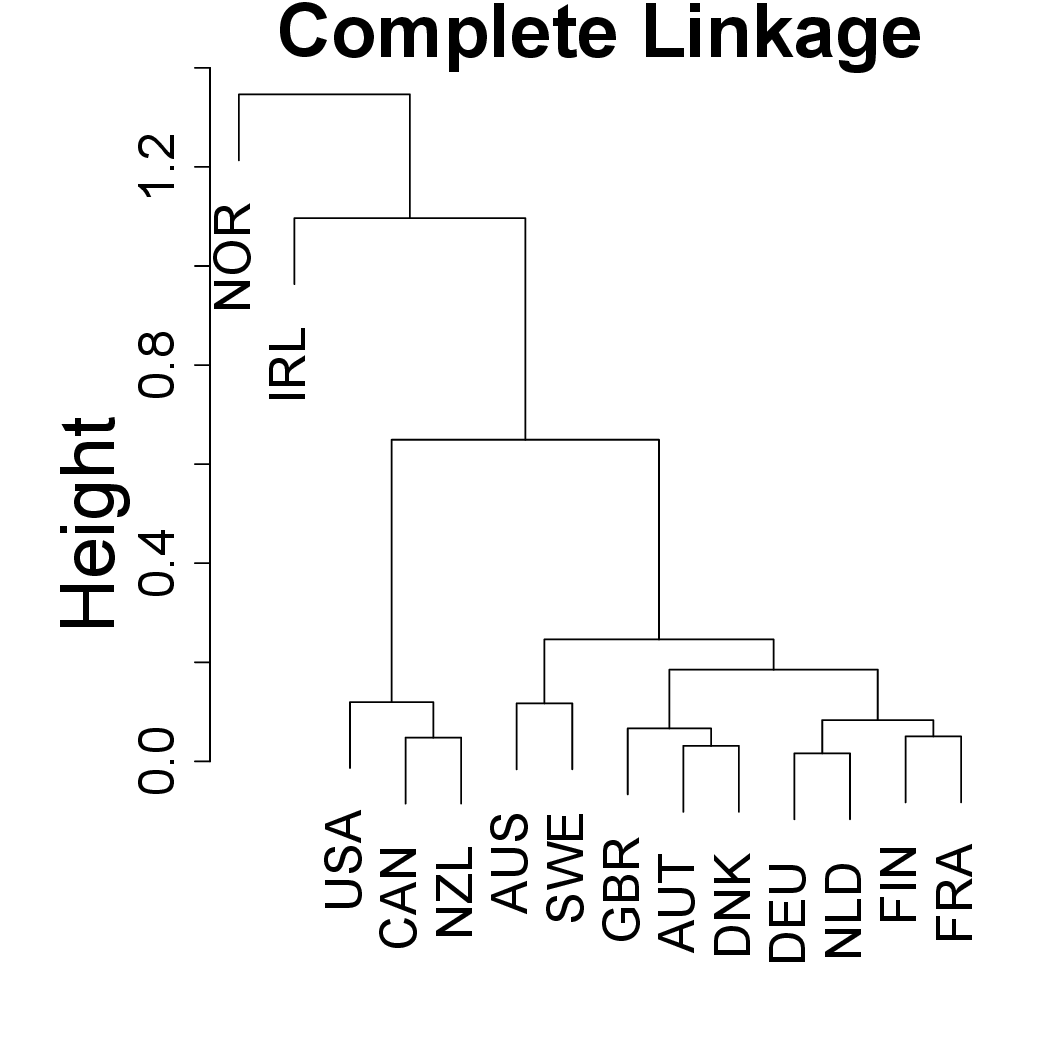}}
\quad
\subfigure[proPCA.]
{\includegraphics[height=1.3in,width=1.8in]{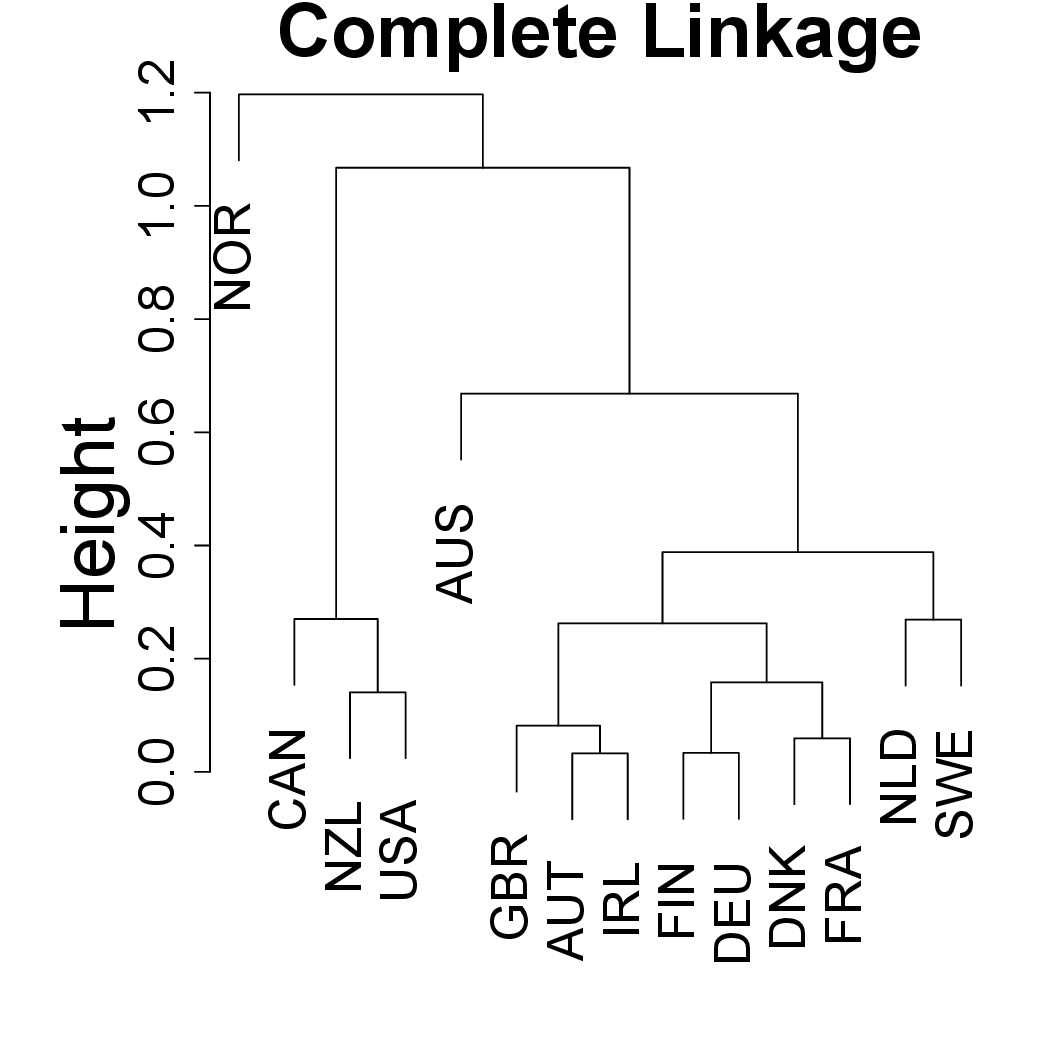}}
\quad
\subfigure[Step-App.]
{\includegraphics[height=1.3in,width=1.8in]{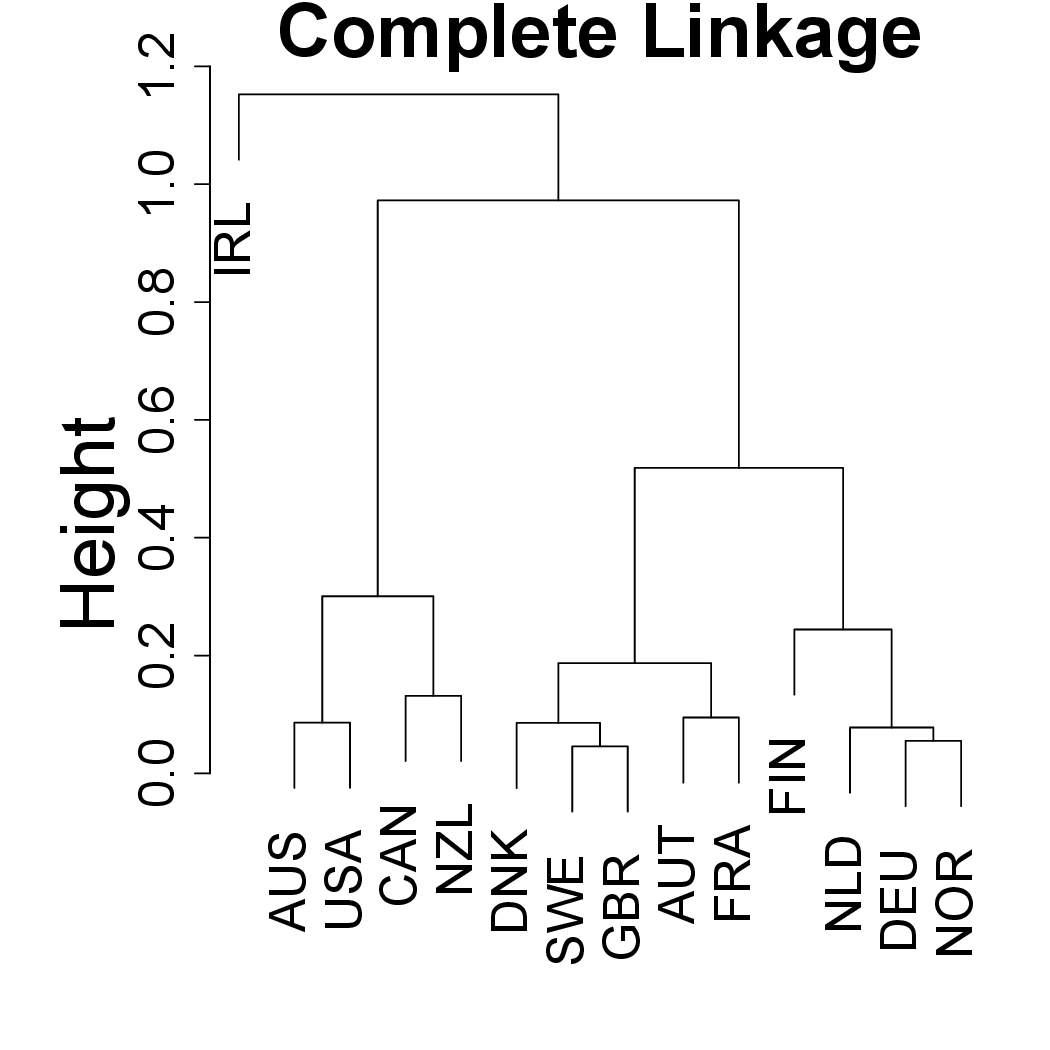}}
\quad
\subfigure[Q-MLE.]
{\includegraphics[height=1.3in,width=1.8in]{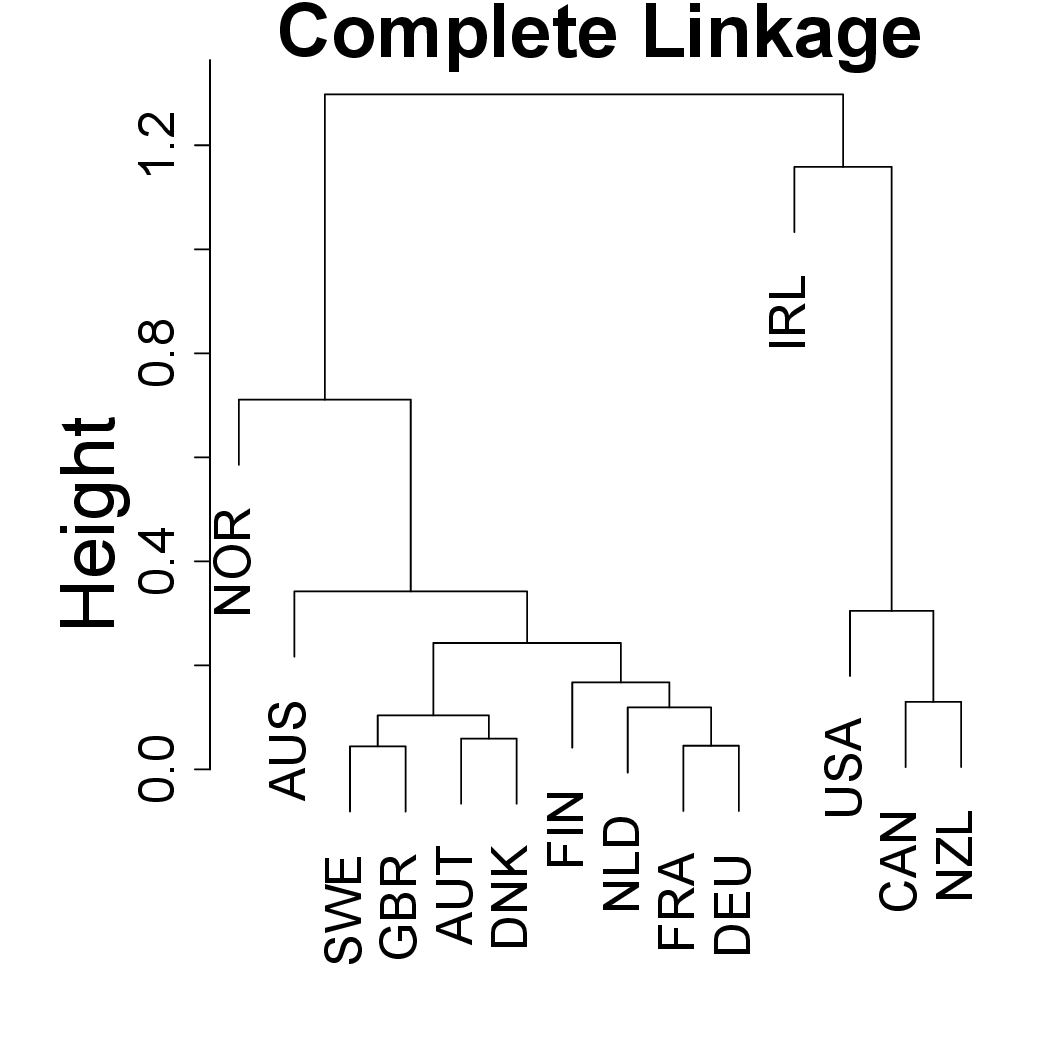}}
\quad
\subfigure[sPCA.]
{\includegraphics[height=1.3in,width=1.8in]{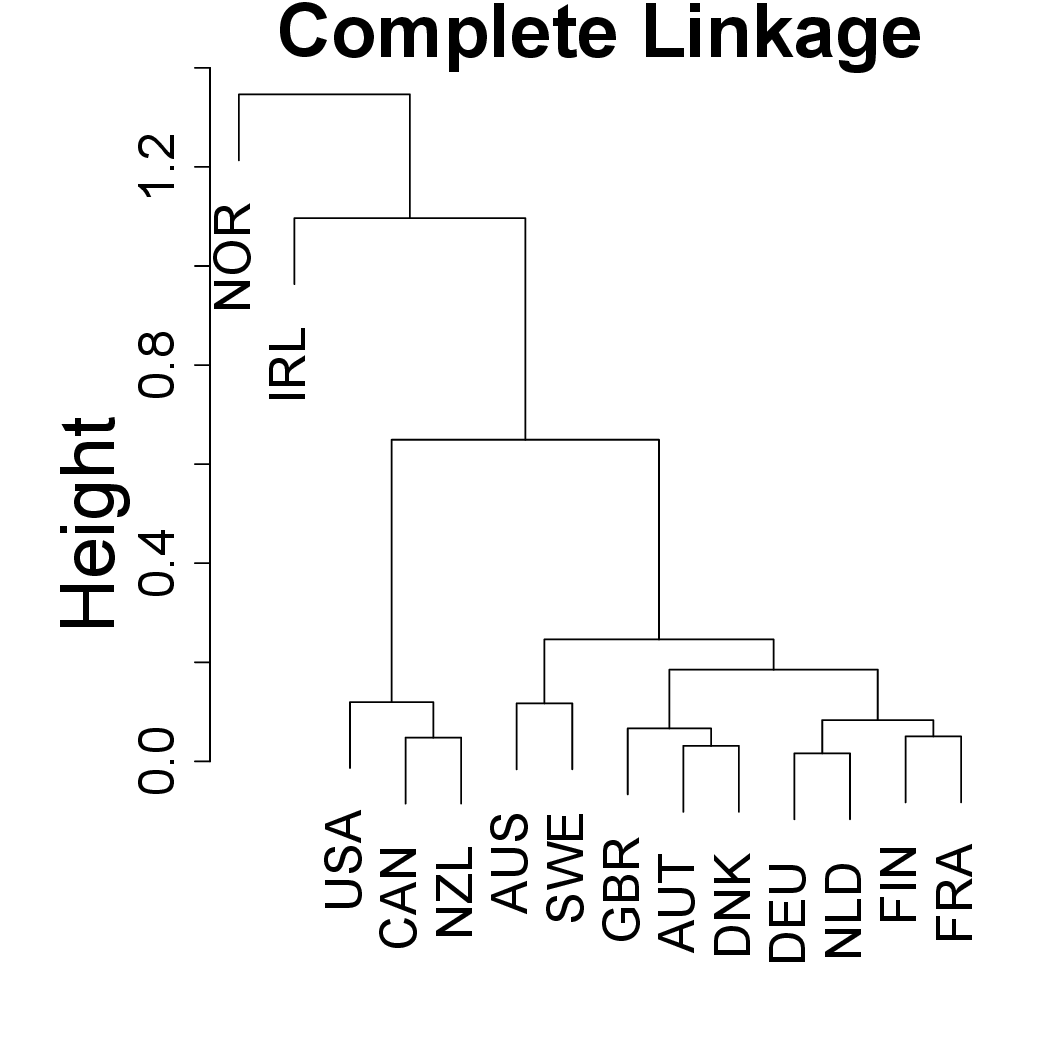}}
\caption{Hierarchical clustering results from the estimated row-wise loading matrix for different methods of estimation applied to the multinational, macroeconomic indices data.}
\label{hclustering_R}
\end{figure}

\begin{figure}
\centering
\subfigure[autoPCA.]
{\includegraphics[height=1.3in,width=1.8in]{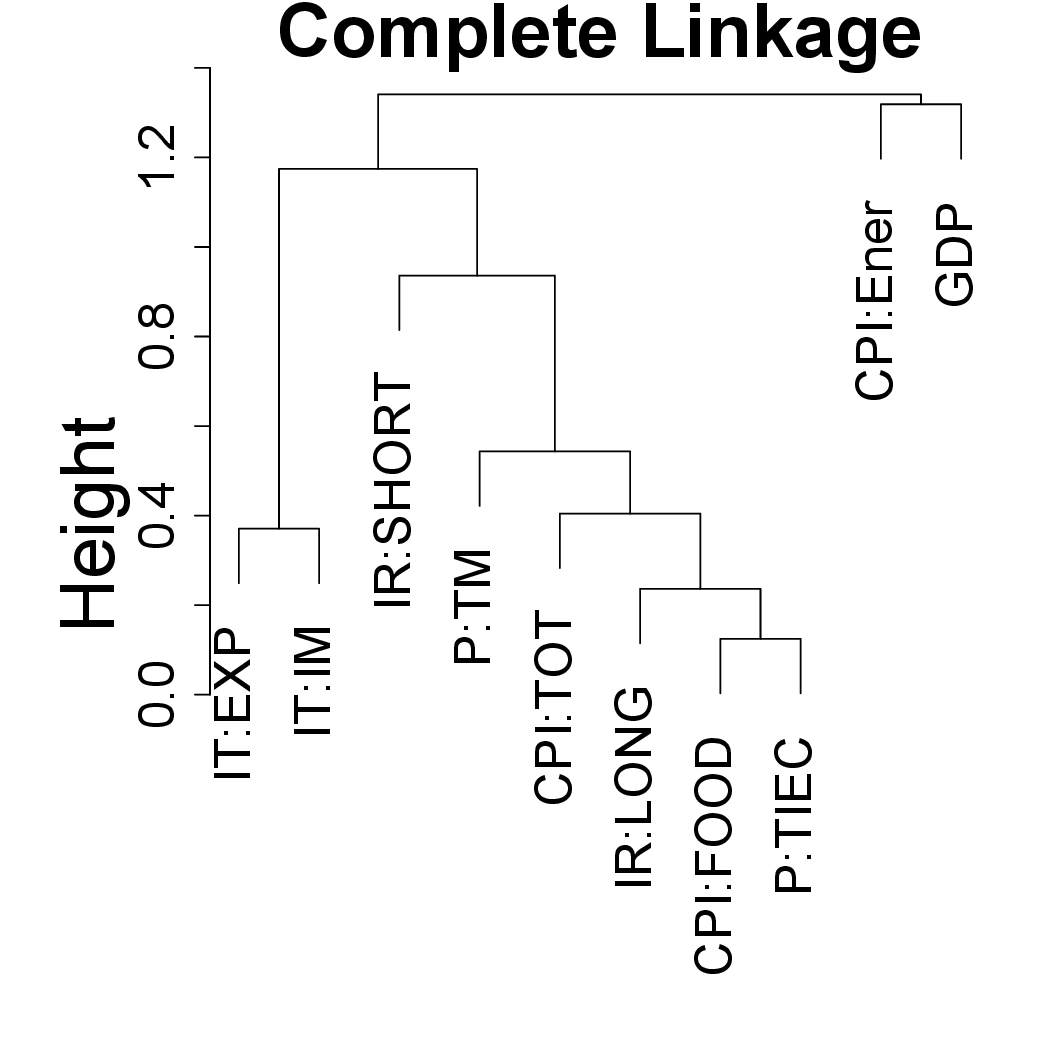}}
\quad
\subfigure[$\alpha$-PCA.]
{\includegraphics[height=1.3in,width=1.8in]{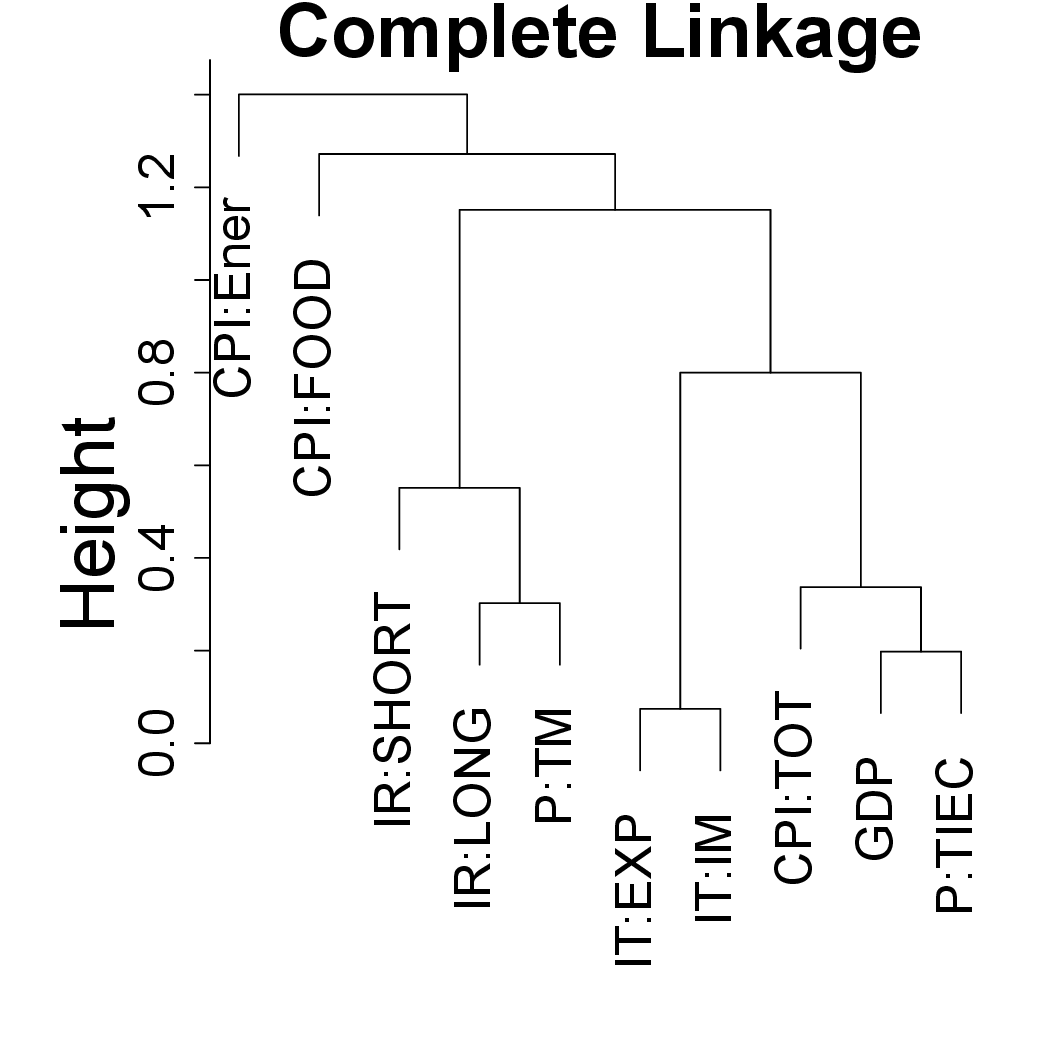}}
\quad
\subfigure[proPCA.]
{\includegraphics[height=1.3in,width=1.8in]{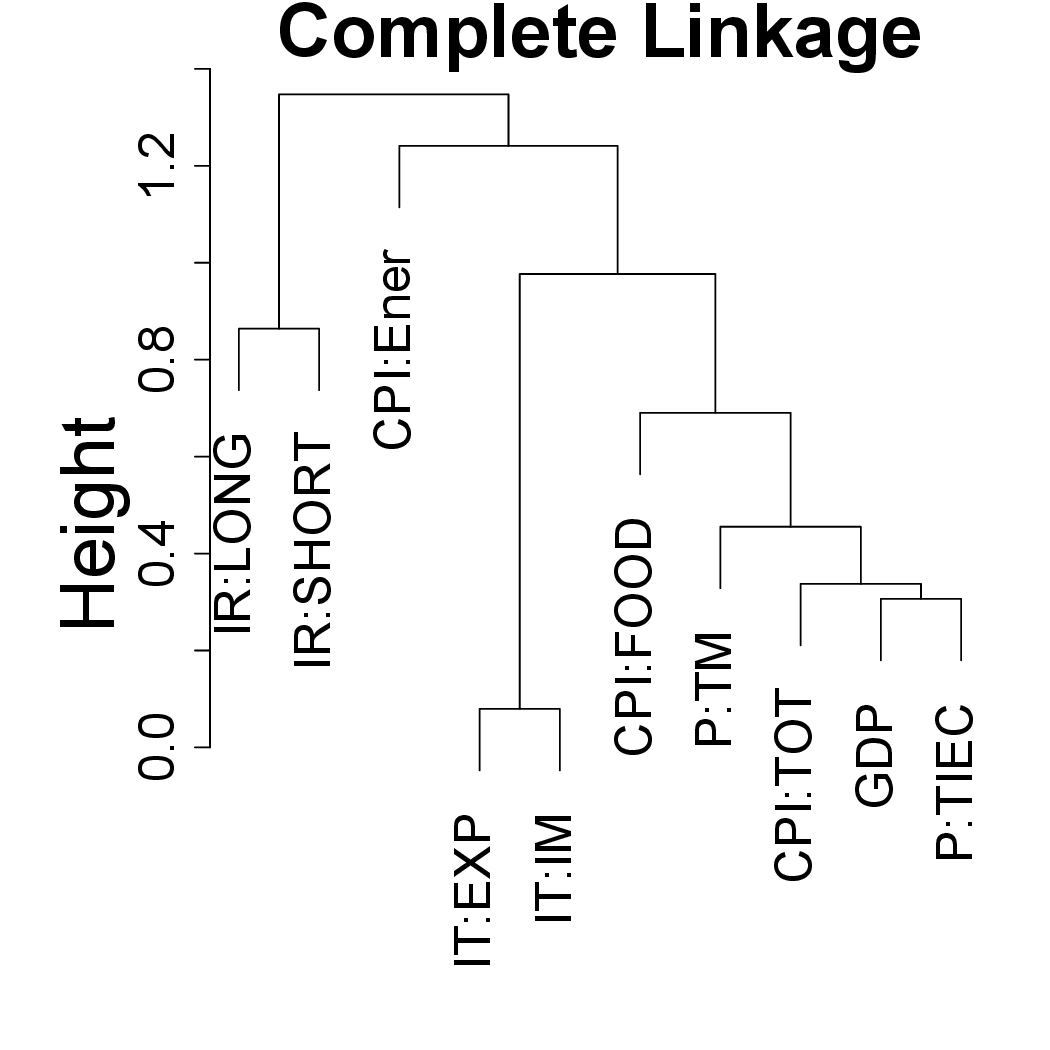}}
\quad
\subfigure[Step-App.]
{\includegraphics[height=1.3in,width=1.8in]{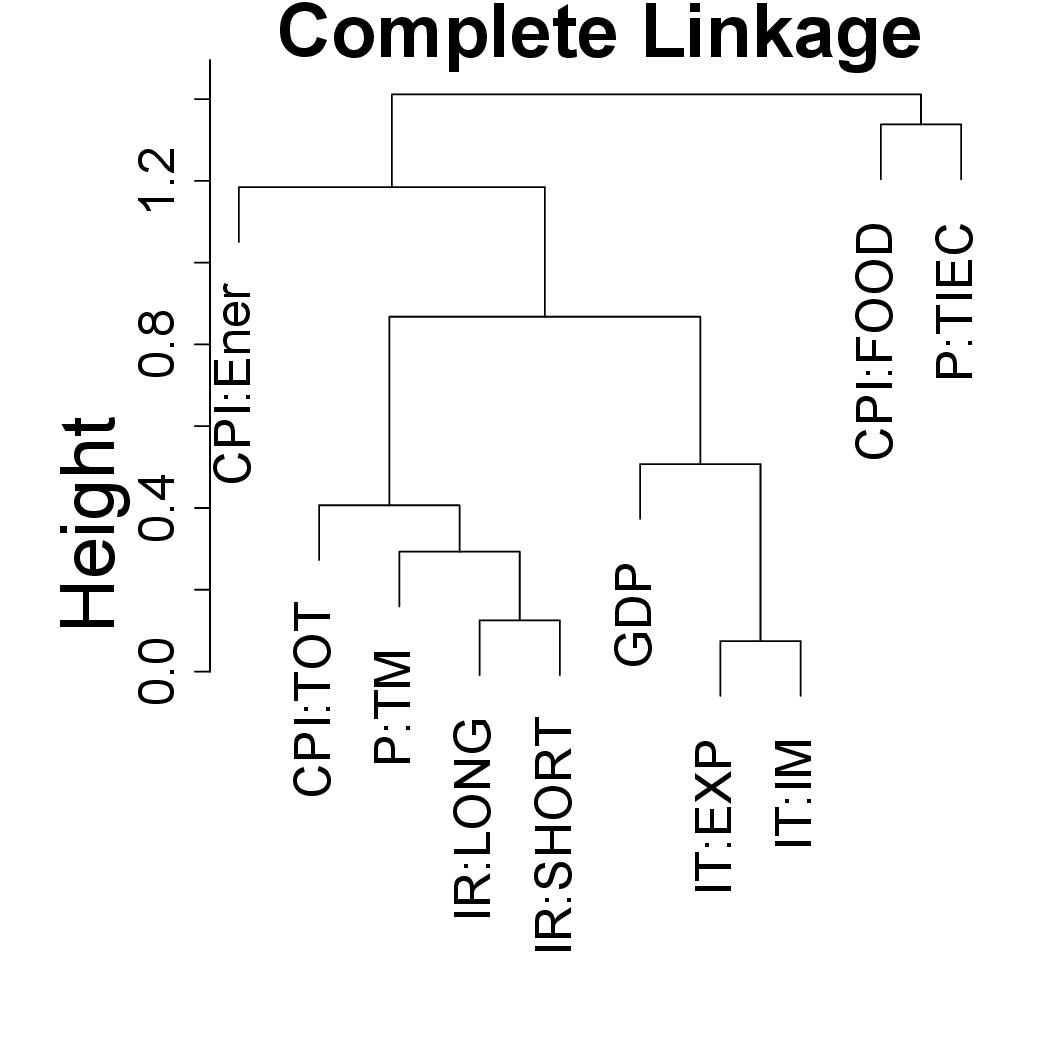}}
\quad
\subfigure[Q-MLE.]
{\includegraphics[height=1.3in,width=1.8in]{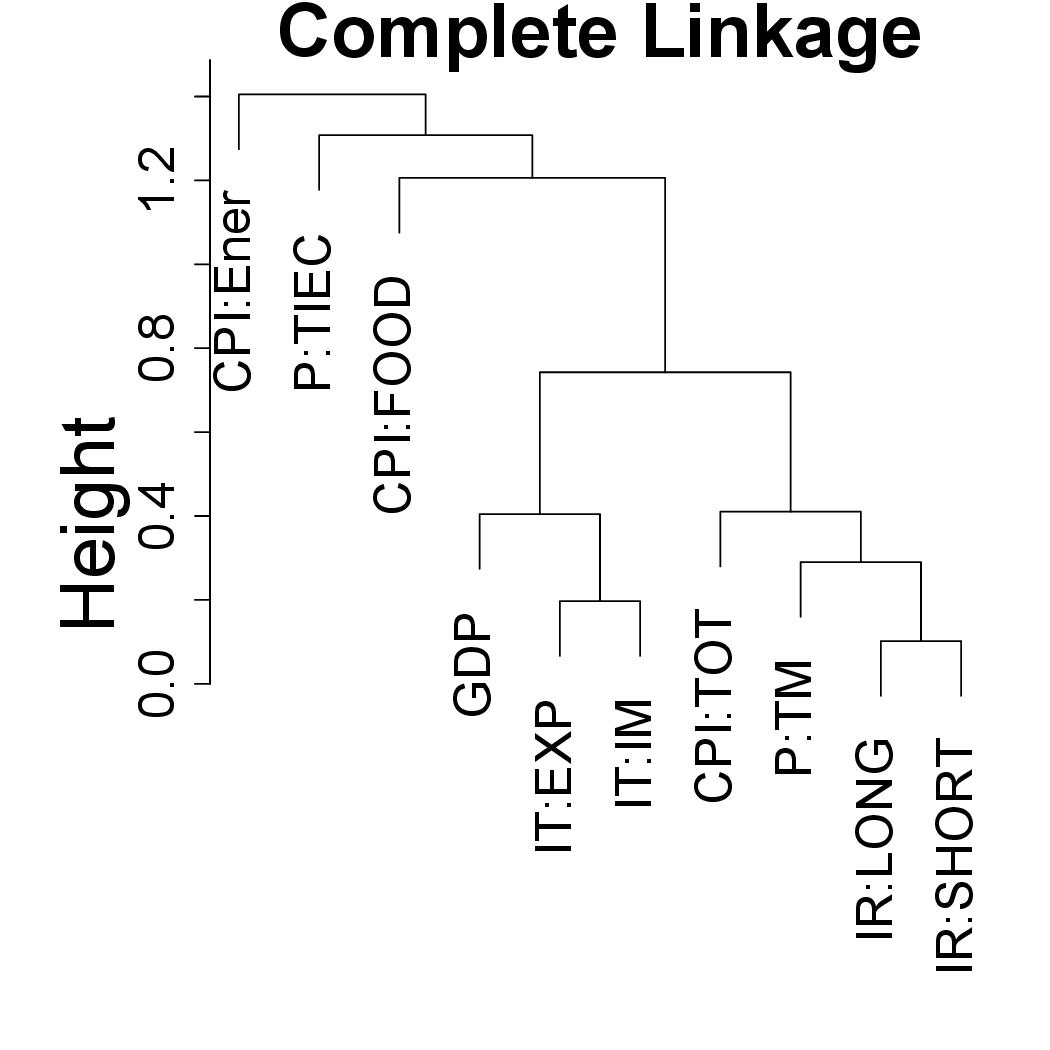}}
\quad
\subfigure[sPCA.]
{\includegraphics[height=1.3in,width=1.8in]{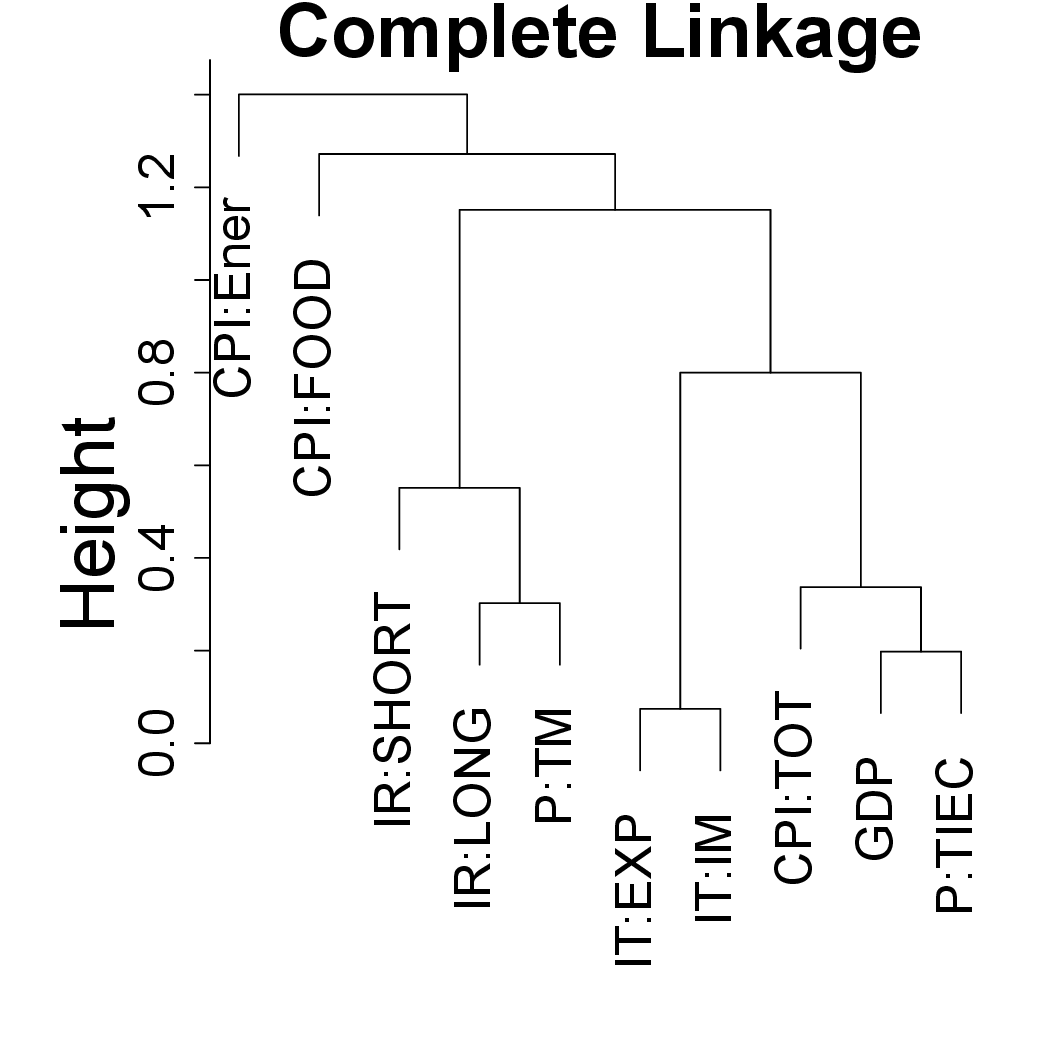}}
\caption{Hierarchical clustering results from the estimated column-wise loading matrix for different methods of estimation applied to the multinational, macroeconomic indices data.}
\label{hclustering_C}
\end{figure}

\textbf{\textit{Clustering}} We choose factor numbers of $k_1=3$ and $k_2=4$.
For interpretability, the varimax rotation is applied to loading matrix estimates, and then the rotated loadings are multiplied by $30$ and truncated to integer values.
The hierarchical clustering results based on the rotated row-wise and column-wise loading matrix estimates are shown in Figures \ref{hclustering_R} and \ref{hclustering_C} (2w-PCA is omitted for space saving), respectively.
For row-wise loadings (countries), the results of $\alpha$-PCA, 2w-PCA, and sPCA are identical, but are different from those of autoPCA, proPCA, Step-App, and Q-MLE.
All methods cluster the United States of America (USA), Canada (CAN) and New Zealand (NZL) into one group.
For column-wise loadings (indices), proPCA gives the most close clustering to the true index groups, followed closely by $\alpha$-PCA, 2w-PCA, and sPCA.
The tables of final loading matrix estimates are given in Appendix I of the online supplement.

\begin{figure}[H]
\centering
\subfigure[True: row]
{\includegraphics[height=1.5in,width=1.5in]{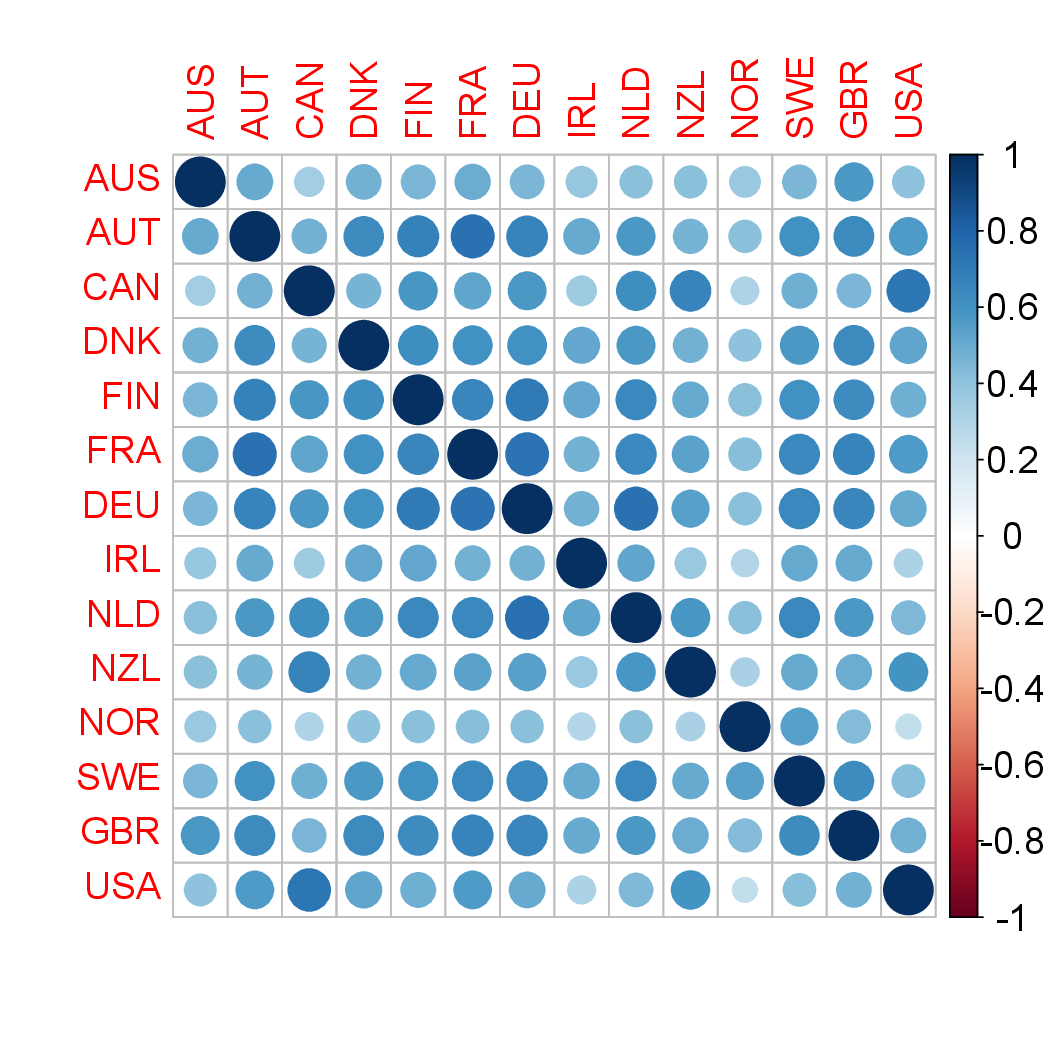}}
\subfigure[autoPCA: row]
{\includegraphics[height=1.5in,width=1.5in]{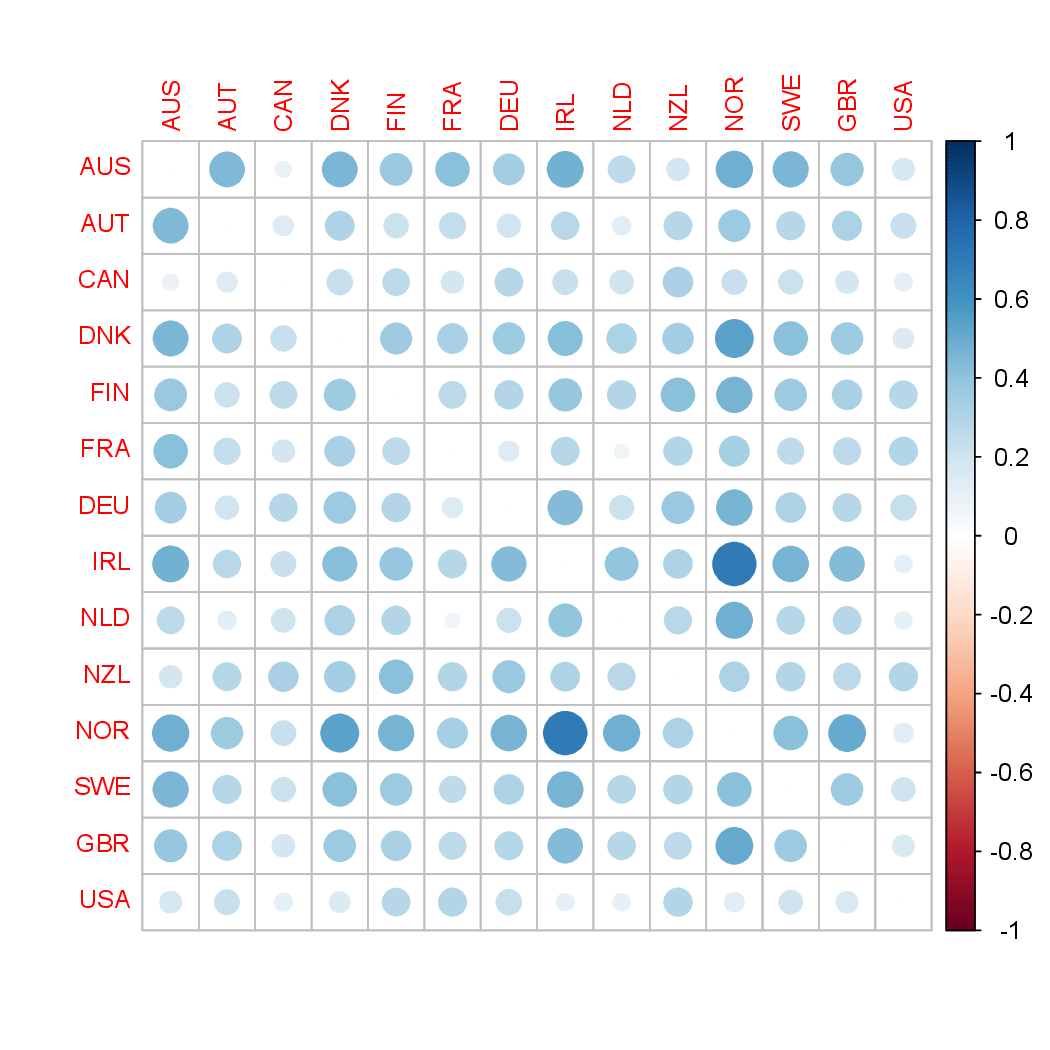}}
\subfigure[$\alpha$-PCA: row]
{\includegraphics[height=1.5in,width=1.5in]{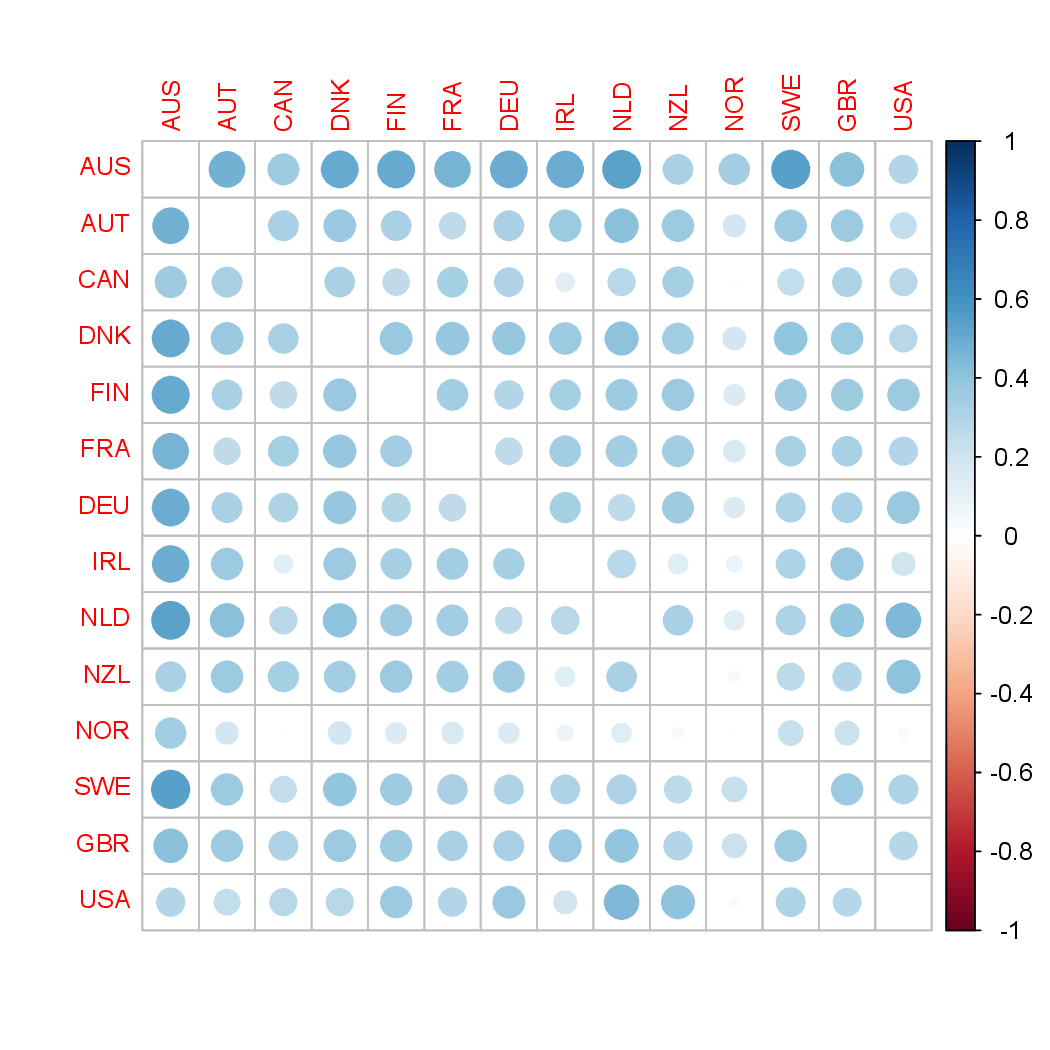}}
\subfigure[proPCA: row]
{\includegraphics[height=1.5in,width=1.5in]{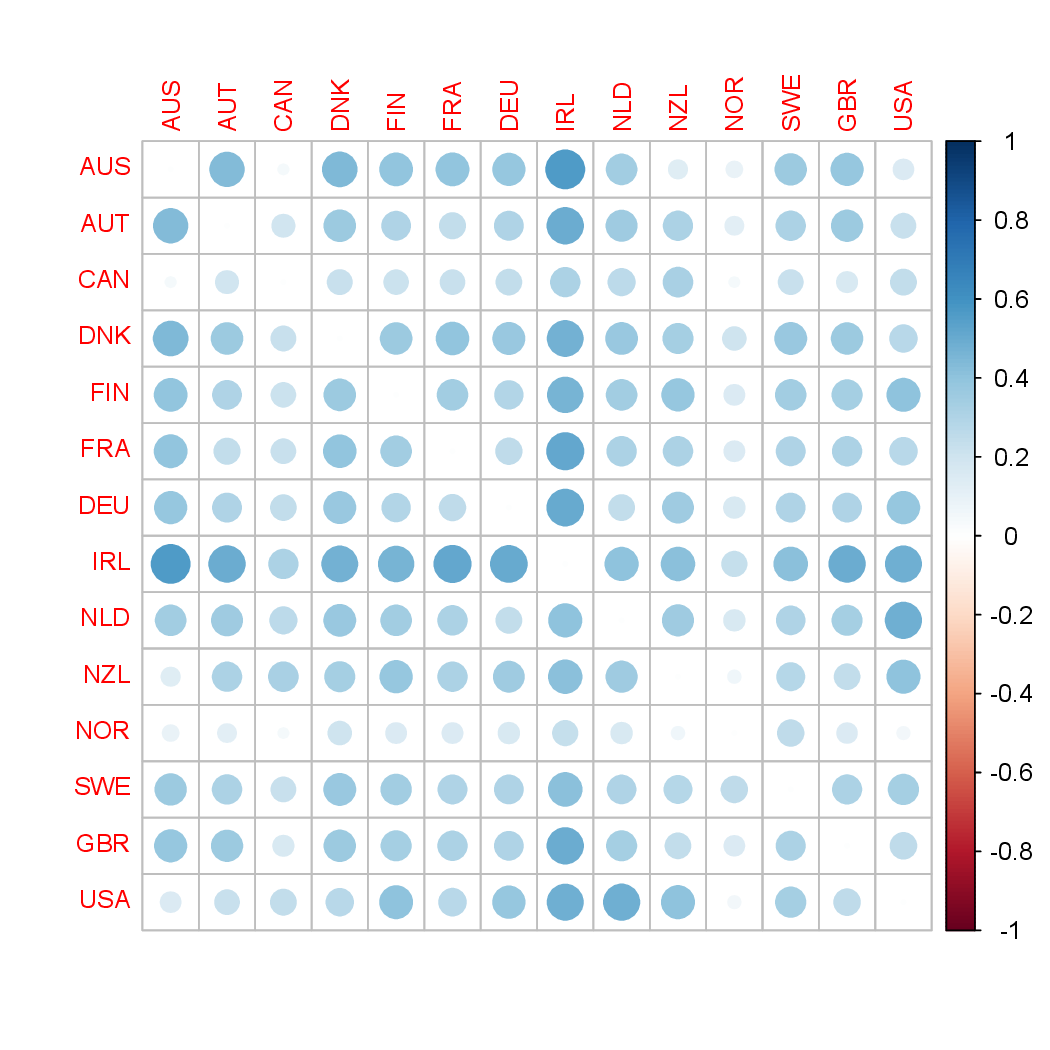}}
\subfigure[True: column]
{\includegraphics[height=1.5in,width=1.5in]{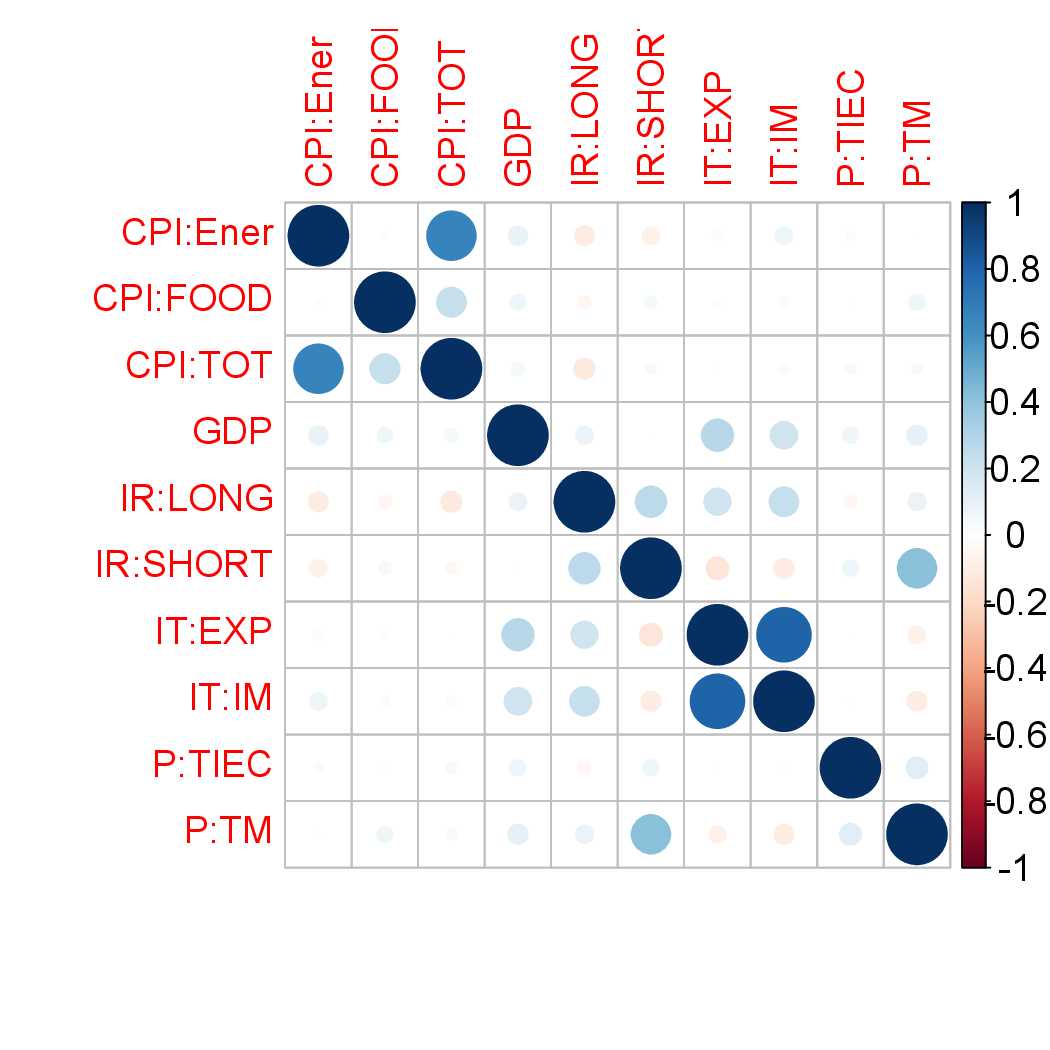}}
\subfigure[autoPCA: column]
{\includegraphics[height=1.5in,width=1.5in]{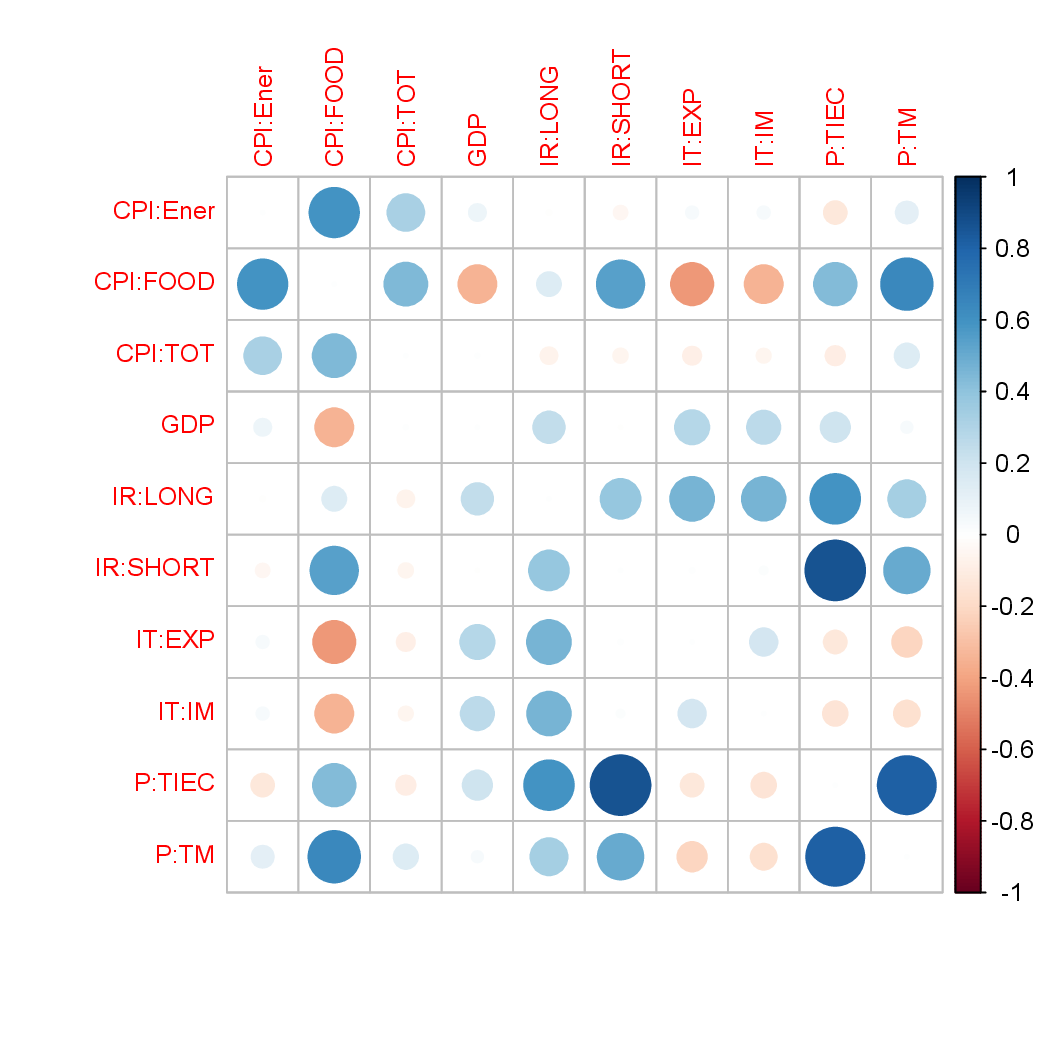}}
\subfigure[$\alpha$-PCA: column]
{\includegraphics[height=1.5in,width=1.5in]{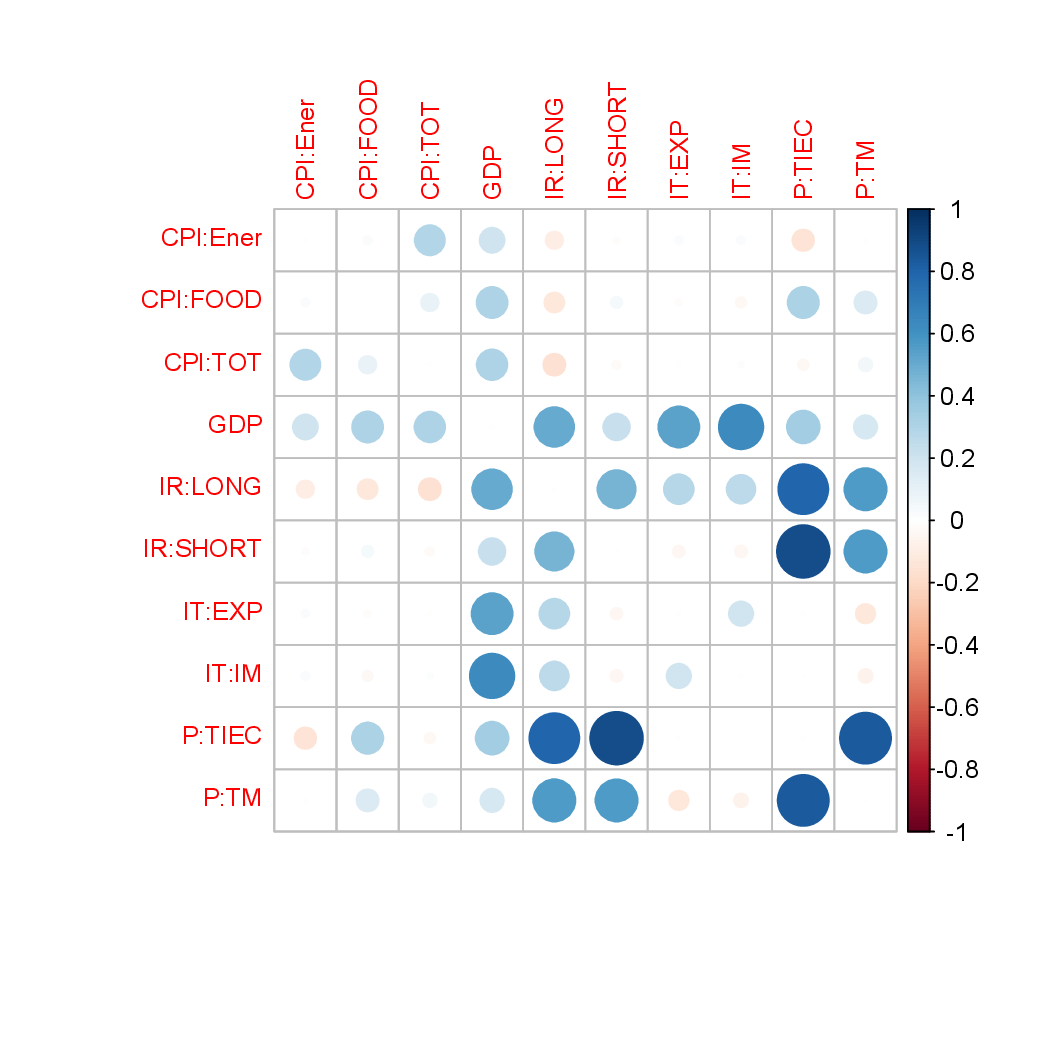}}
\subfigure[proPCA: column]
{\includegraphics[height=1.5in,width=1.5in]{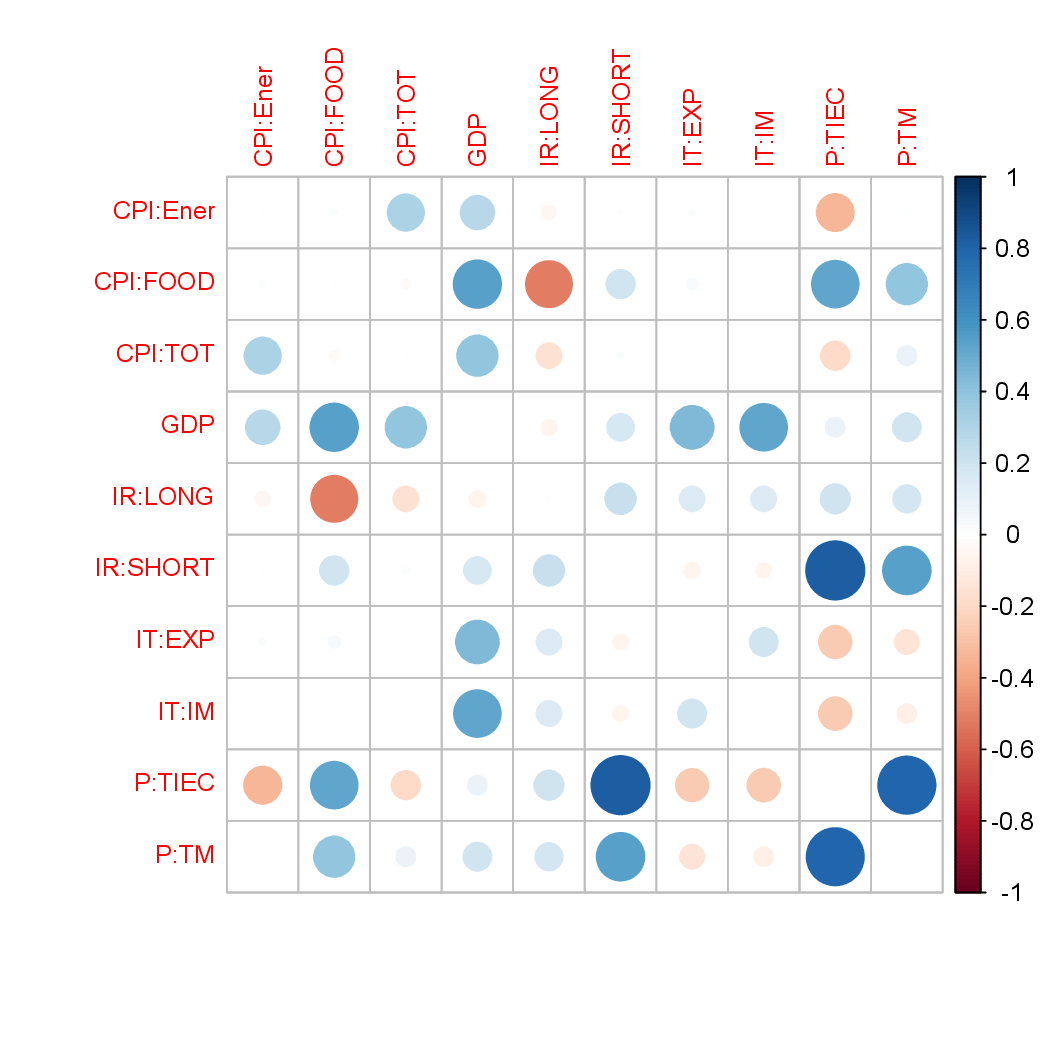}}
\subfigure[True: interaction]
{\includegraphics[height=1.5in,width=1.5in]{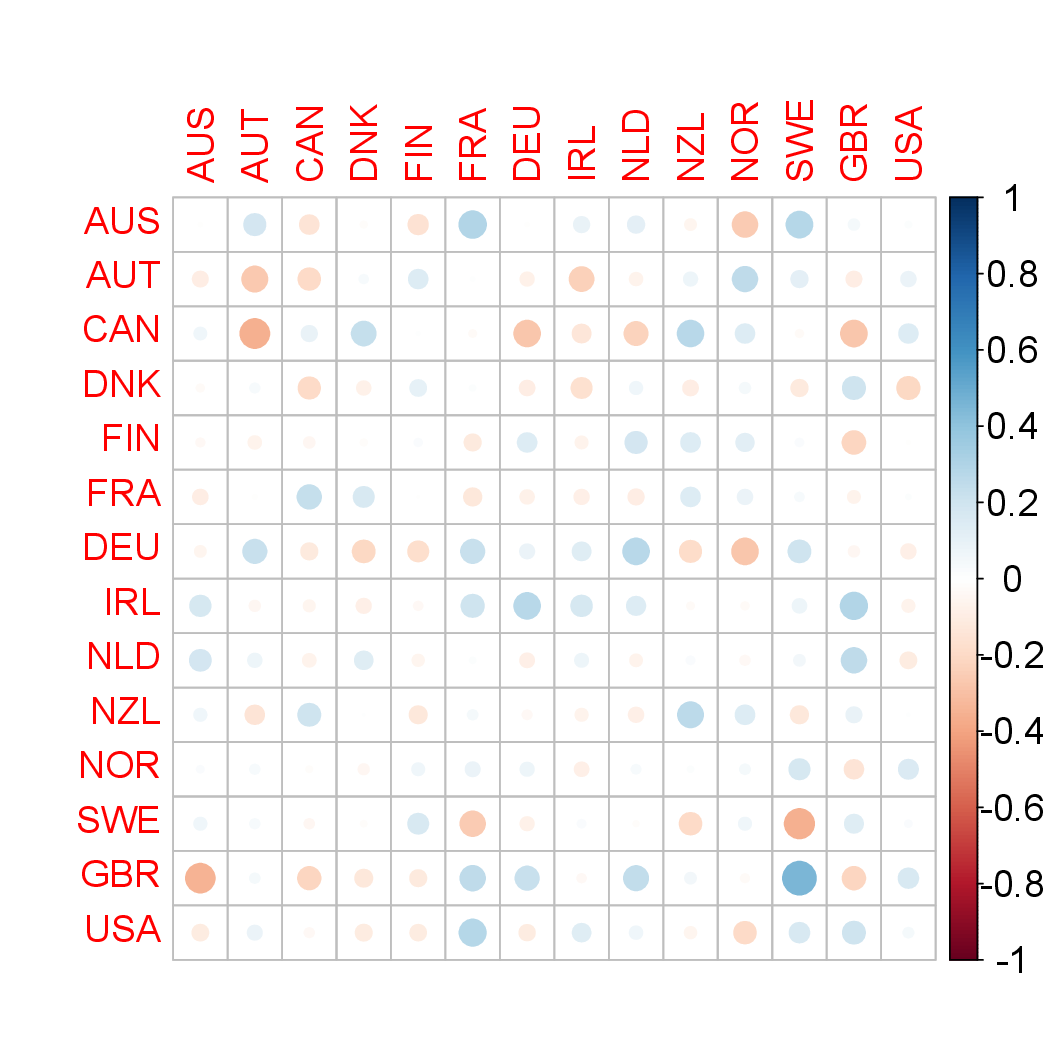}}
\subfigure[autoPCA: interaction]
{\includegraphics[height=1.5in,width=1.5in]{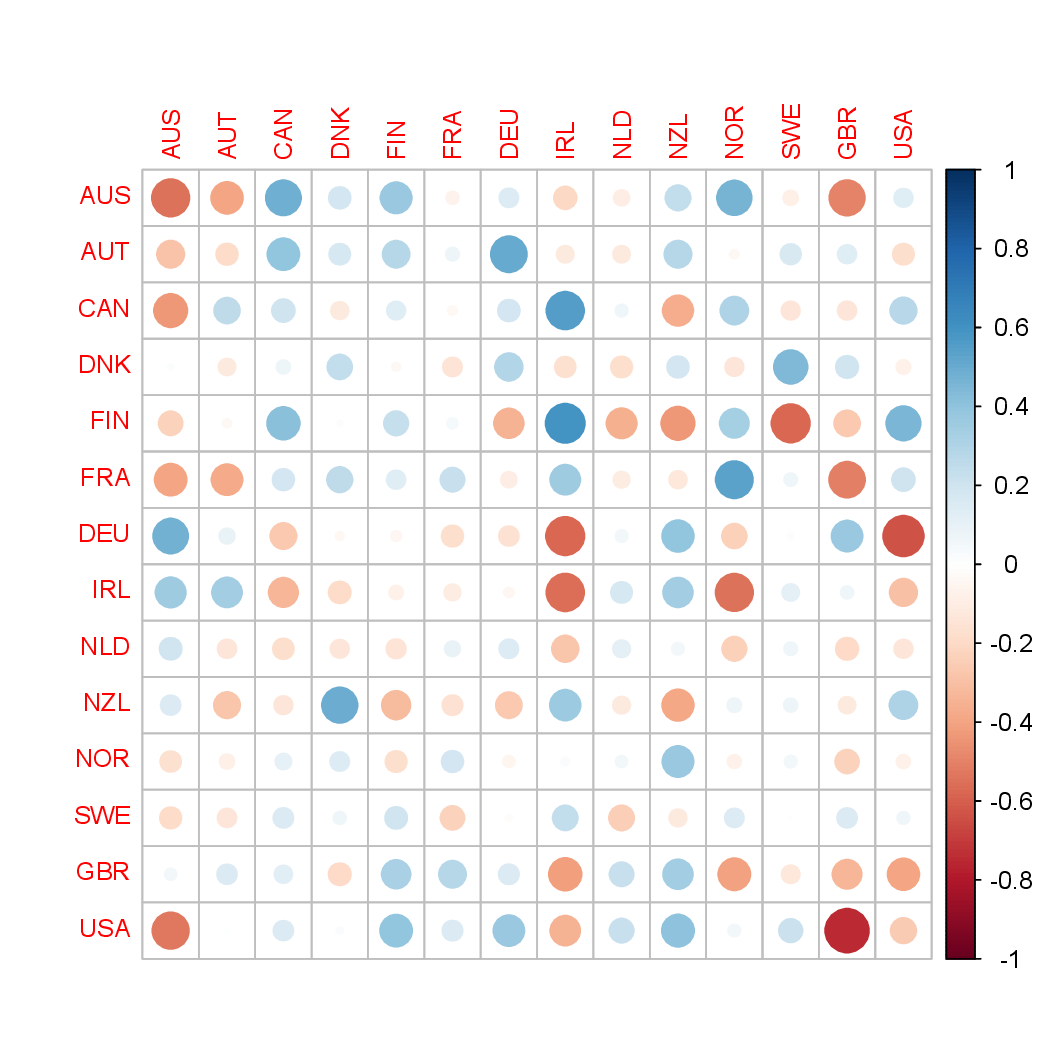}}
\subfigure[$\alpha$-PCA: interaction]
{\includegraphics[height=1.5in,width=1.5in]{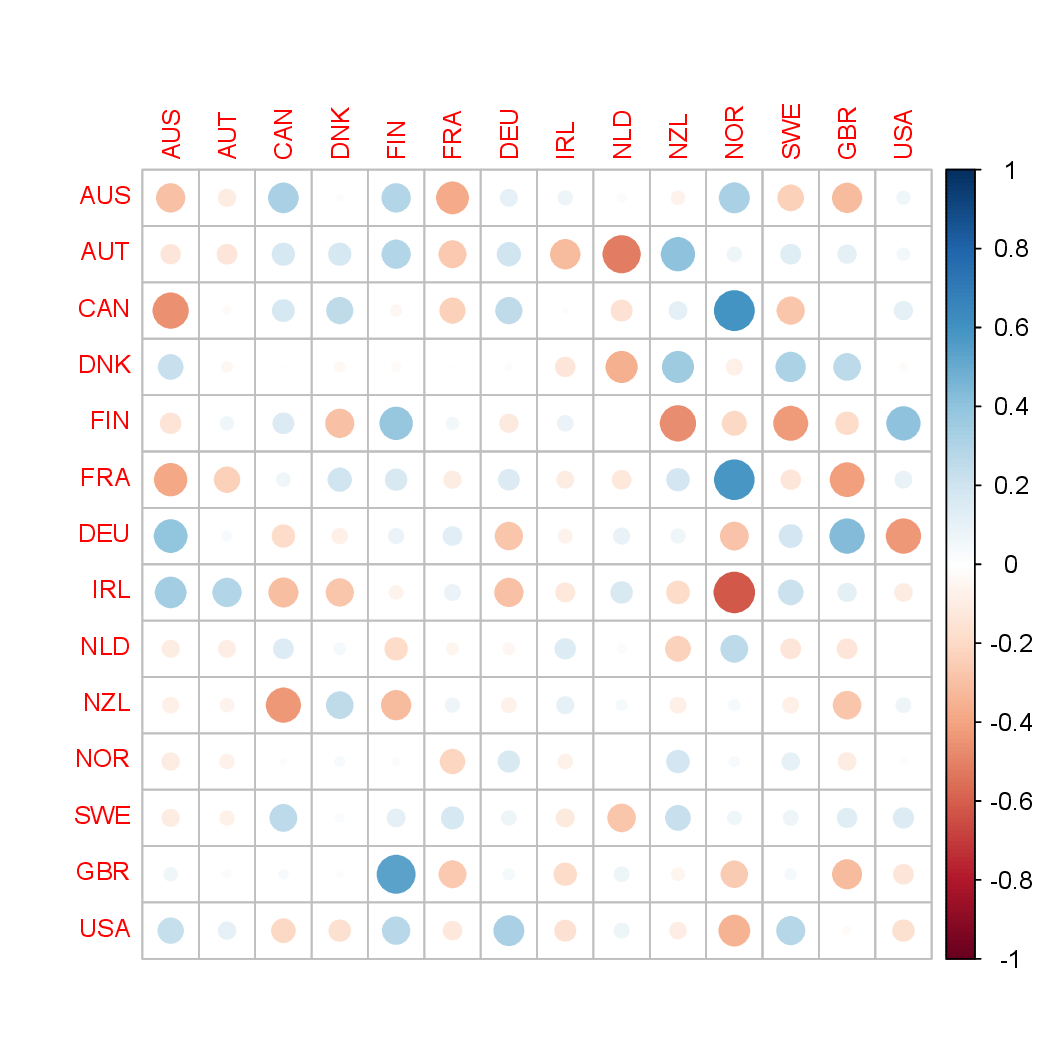}}
\subfigure[proPCA: interaction]
{\includegraphics[height=1.5in,width=1.5in]{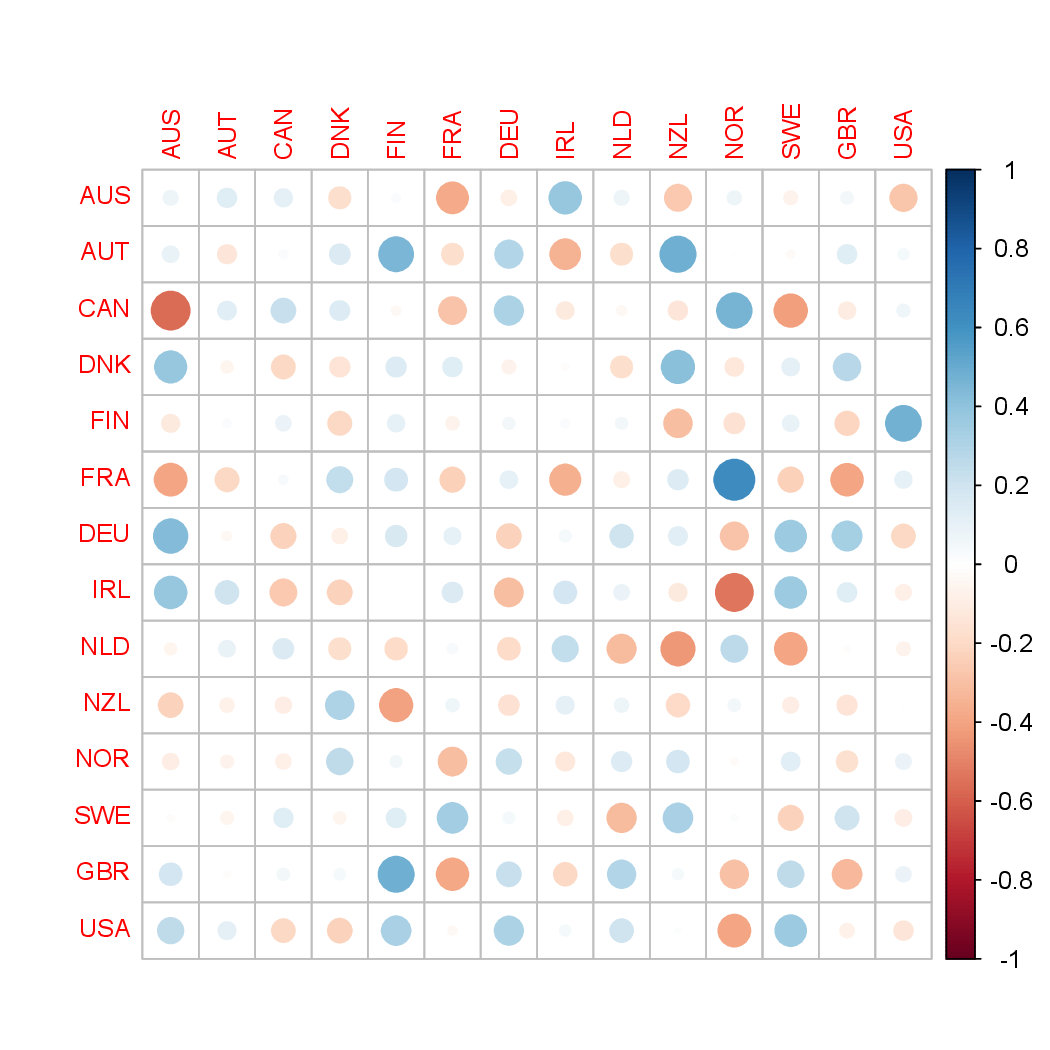}}
\caption{Observed correlations (the $1$st column) and differences between estimated and observed correlations for the multinational, macroeconomic indices data under BiMFaM (the $2$nd to the $4$th columns).}
\label{corr_BiMFaM}
\end{figure}

\begin{figure}[H]
\centering
\subfigure[2w-PCA: row]
{\includegraphics[height=1.5in,width=1.5in]{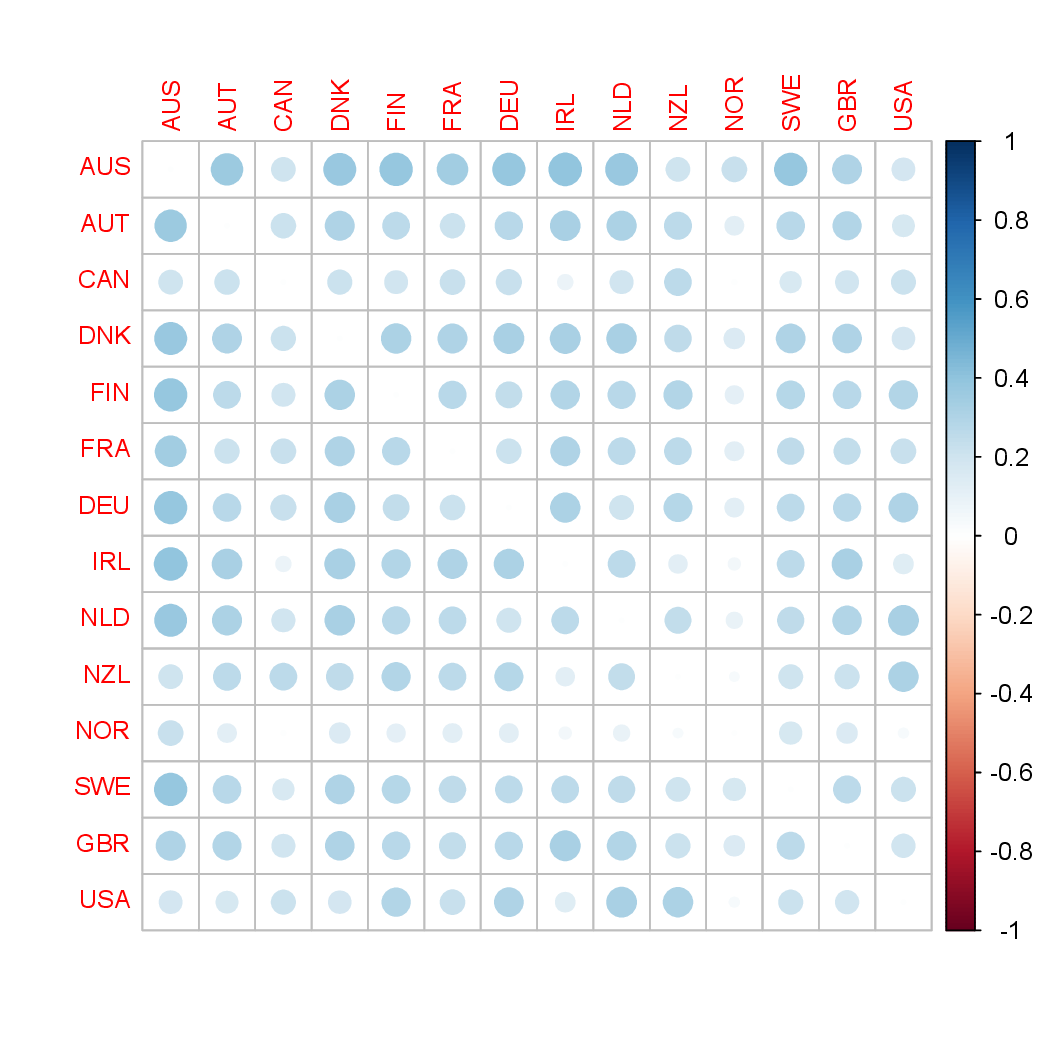}}
\subfigure[Step-App: row]
{\includegraphics[height=1.5in,width=1.5in]{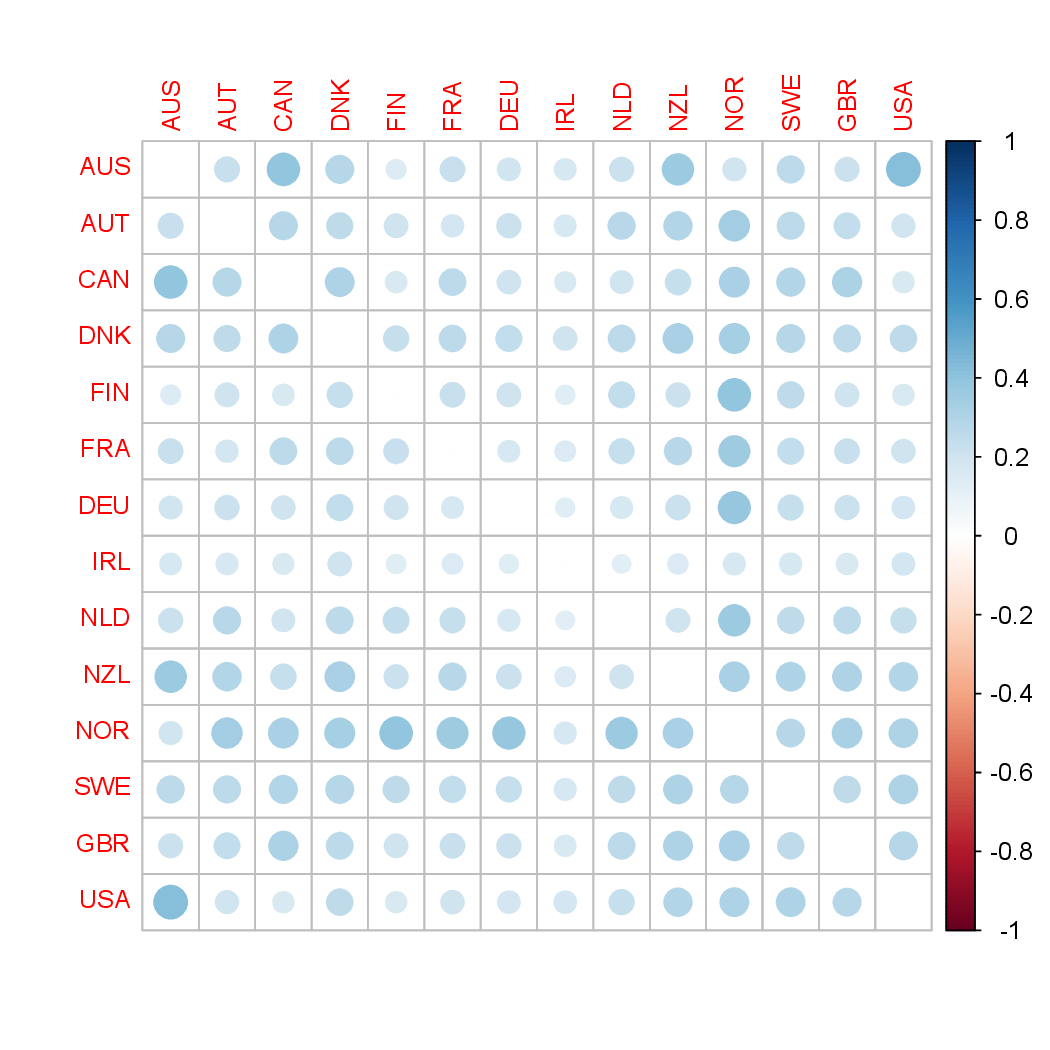}}
\subfigure[Q-MLE: row]
{\includegraphics[height=1.5in,width=1.5in]{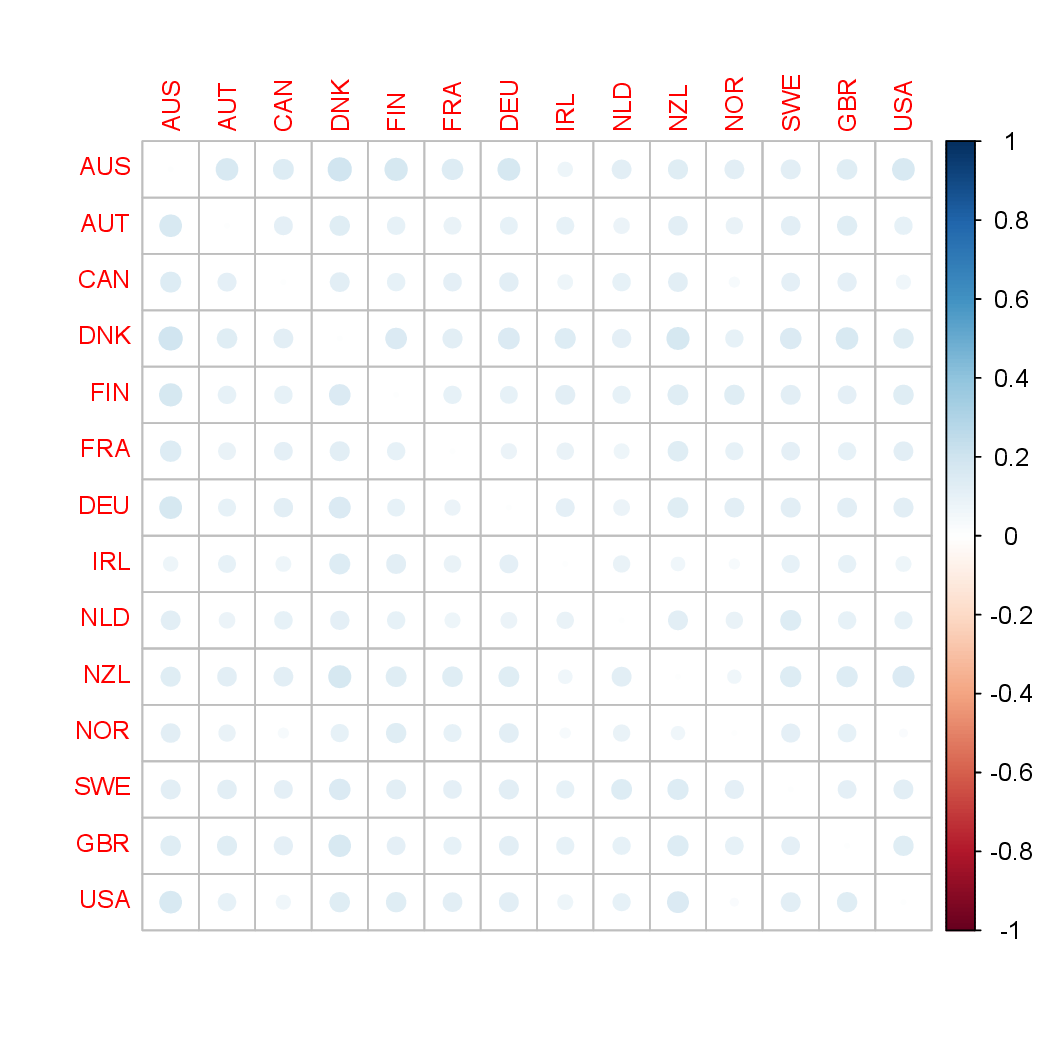}}
\subfigure[sPCA: row]
{\includegraphics[height=1.5in,width=1.5in]{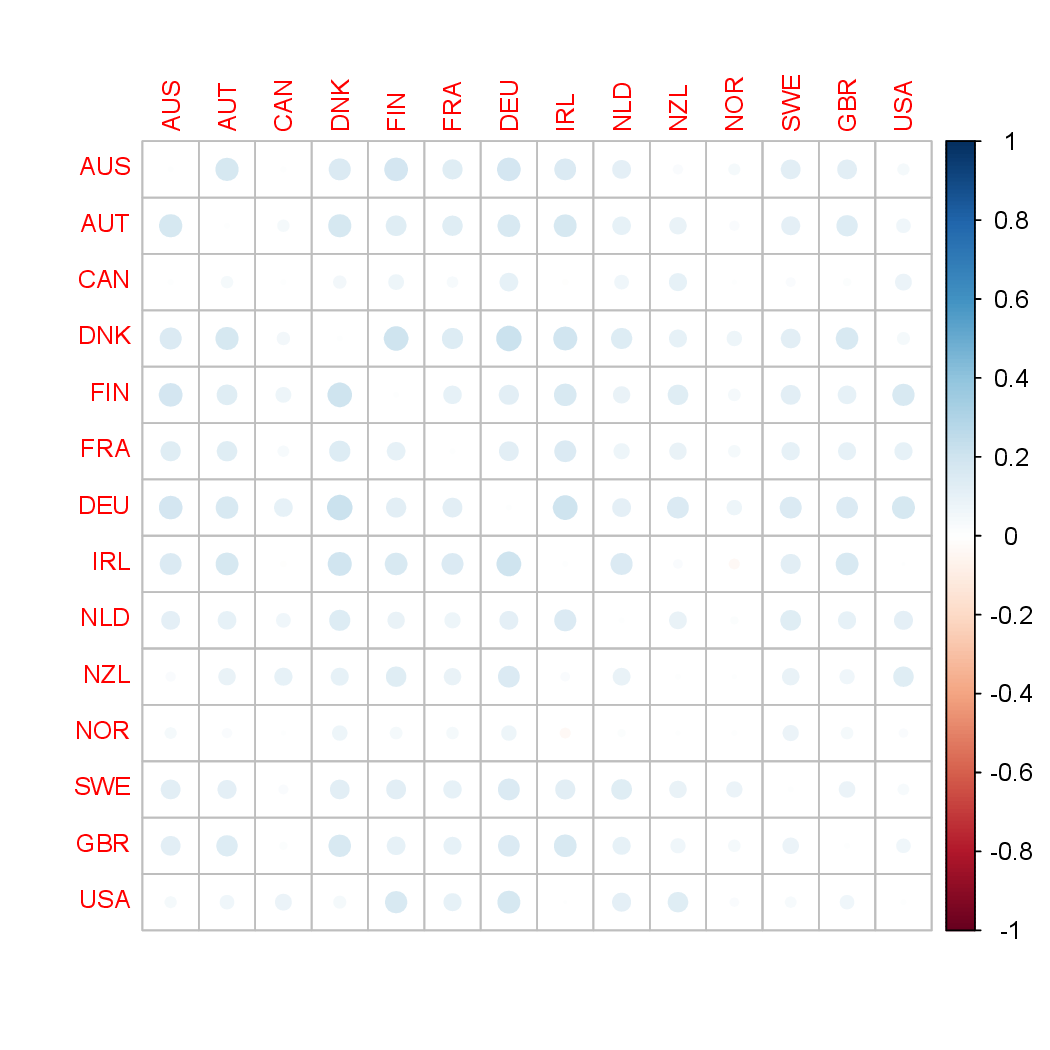}}
\subfigure[2w-PCA: column]
{\includegraphics[height=1.5in,width=1.5in]{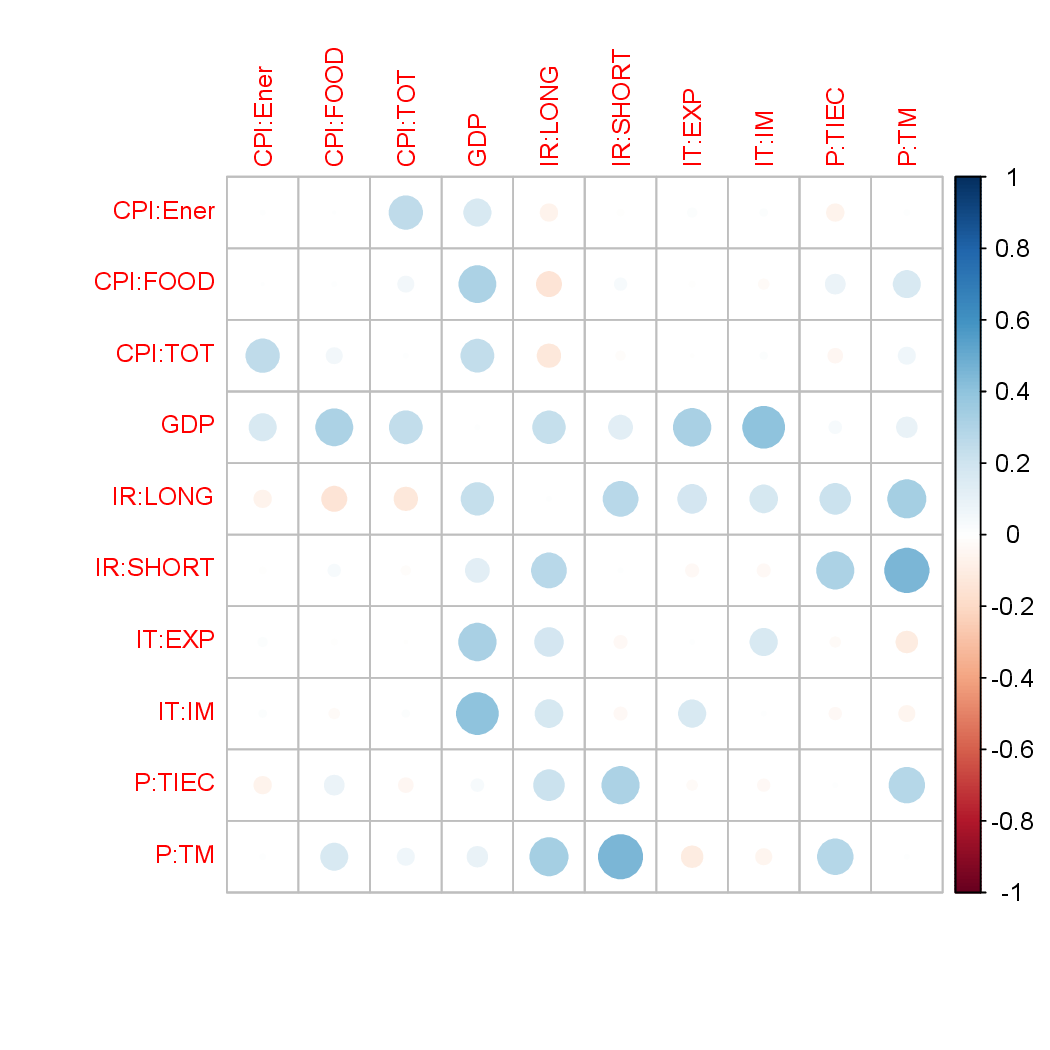}}
\subfigure[Step-App: column]
{\includegraphics[height=1.5in,width=1.5in]{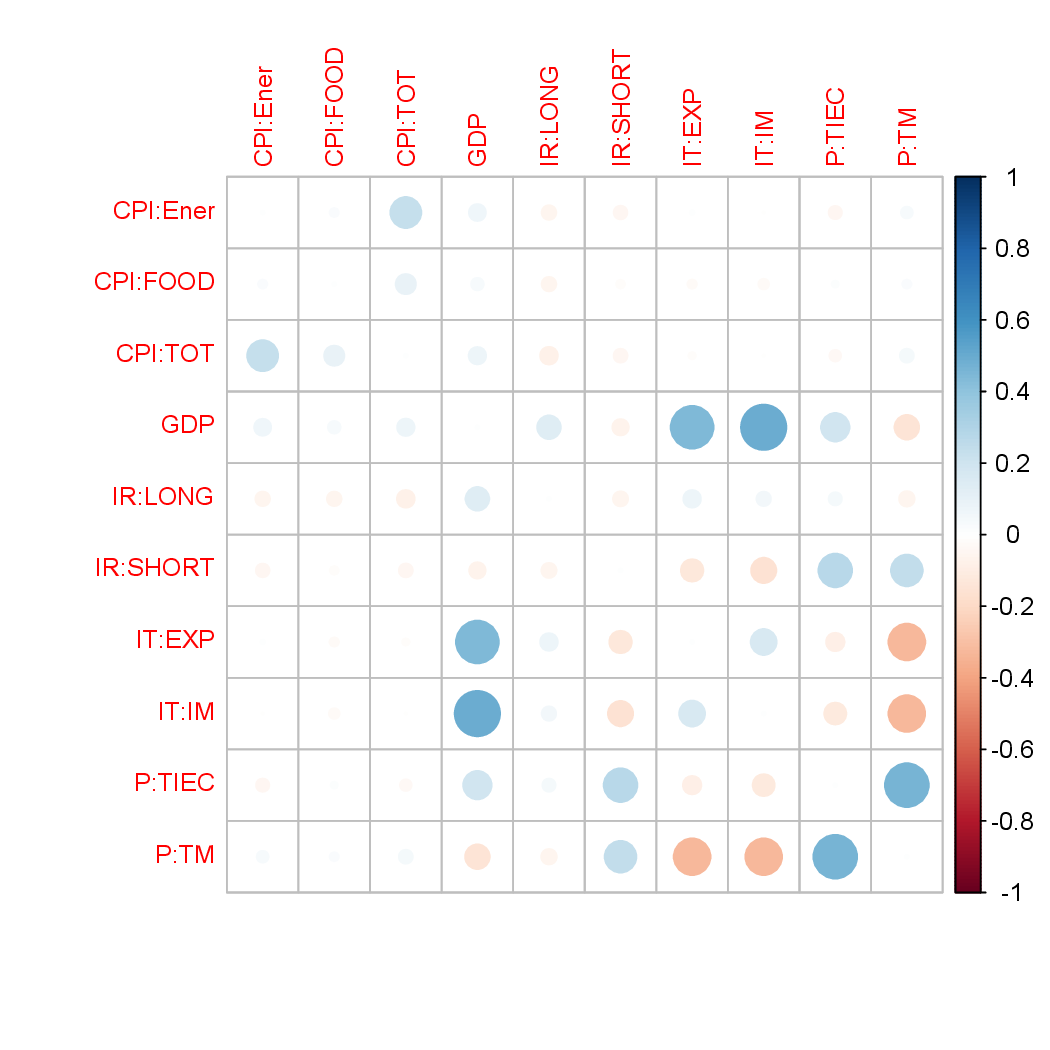}}
\subfigure[Q-MLE: column]
{\includegraphics[height=1.5in,width=1.5in]{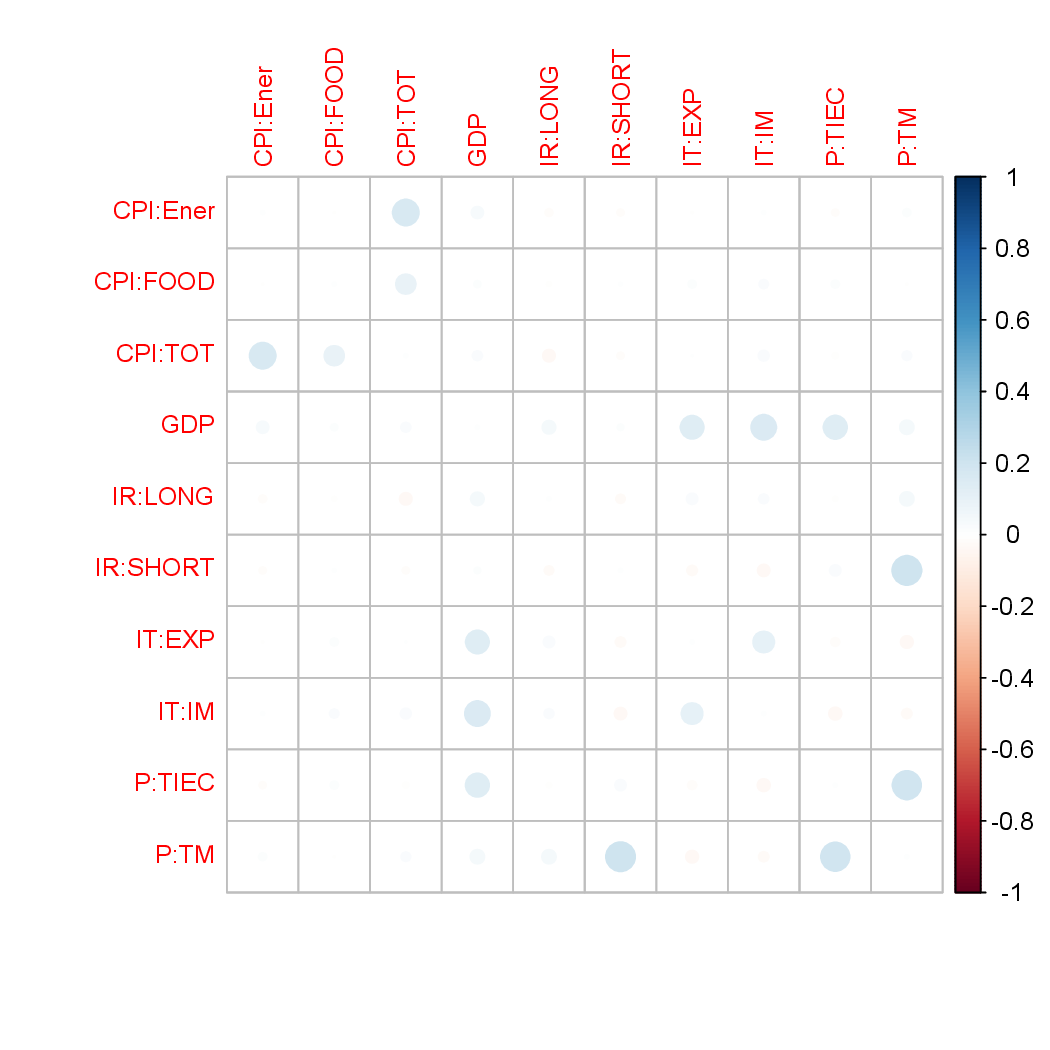}}
\subfigure[sPCA: column]
{\includegraphics[height=1.5in,width=1.5in]{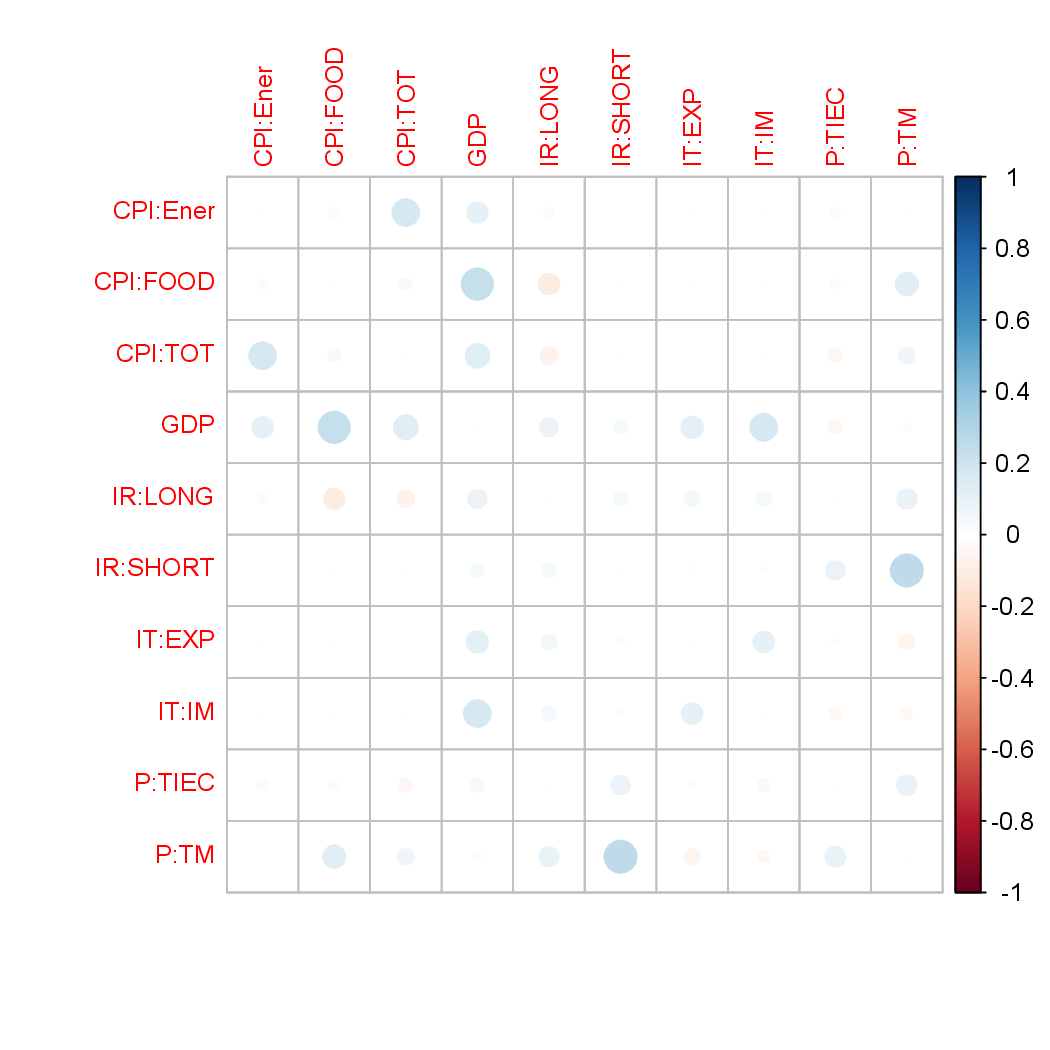}}
\subfigure[2w-PCA: interaction]
{\includegraphics[height=1.5in,width=1.5in]{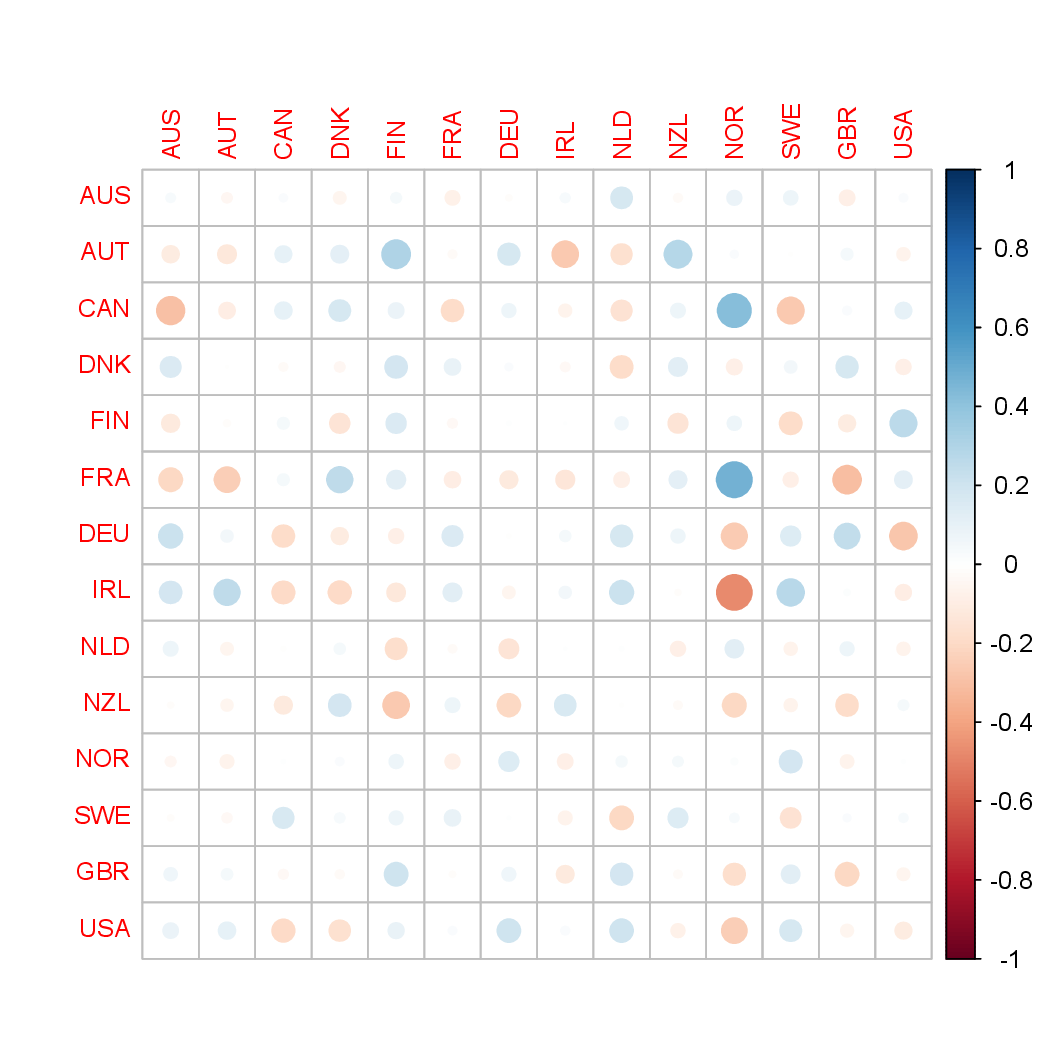}}
{\captionsetup{font=footnotesize}
\subfigure[Step-App:interaction]
{\includegraphics[height=1.5in,width=1.5in]{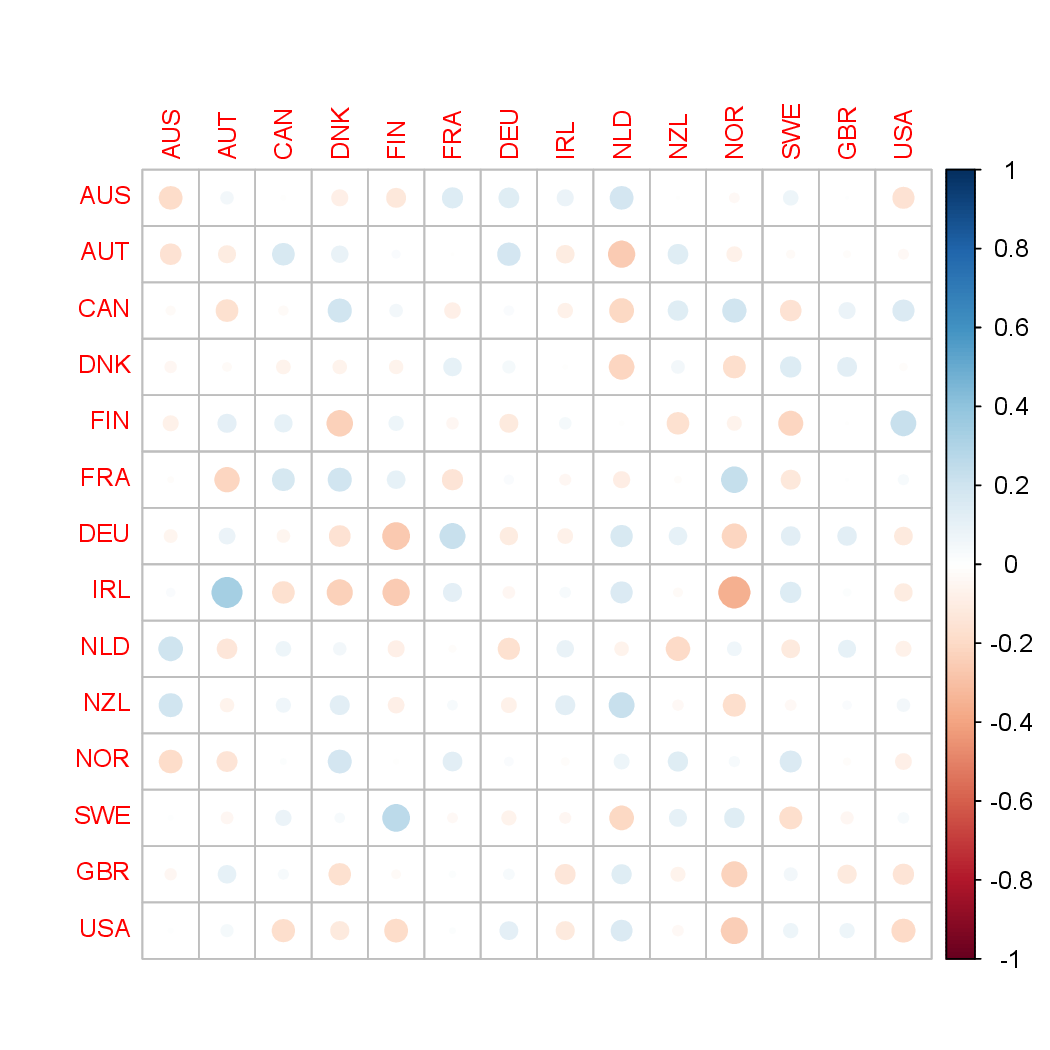}}}
\subfigure[Q-MLE: interaction]
{\includegraphics[height=1.5in,width=1.5in]{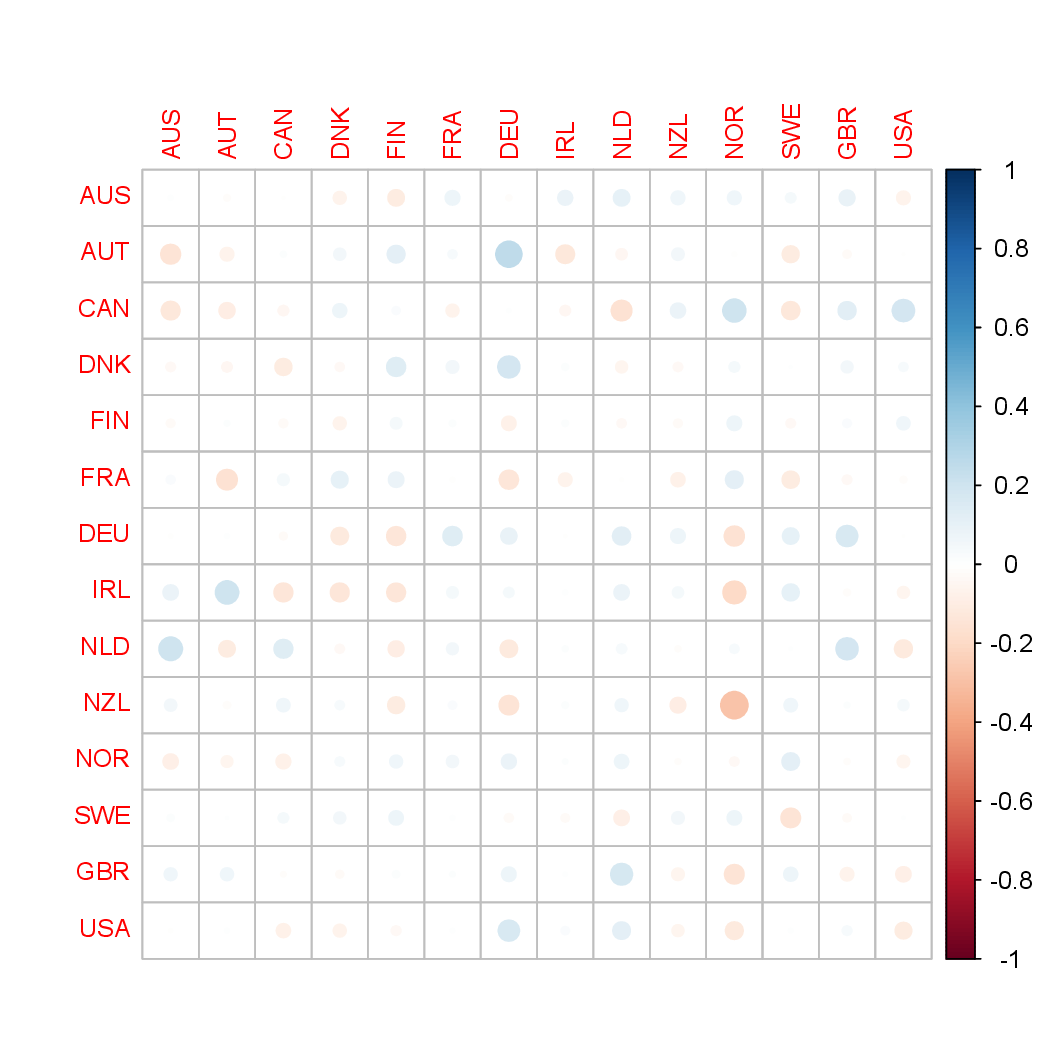}}
\subfigure[sPCA: interaction]
{\includegraphics[height=1.5in,width=1.5in]{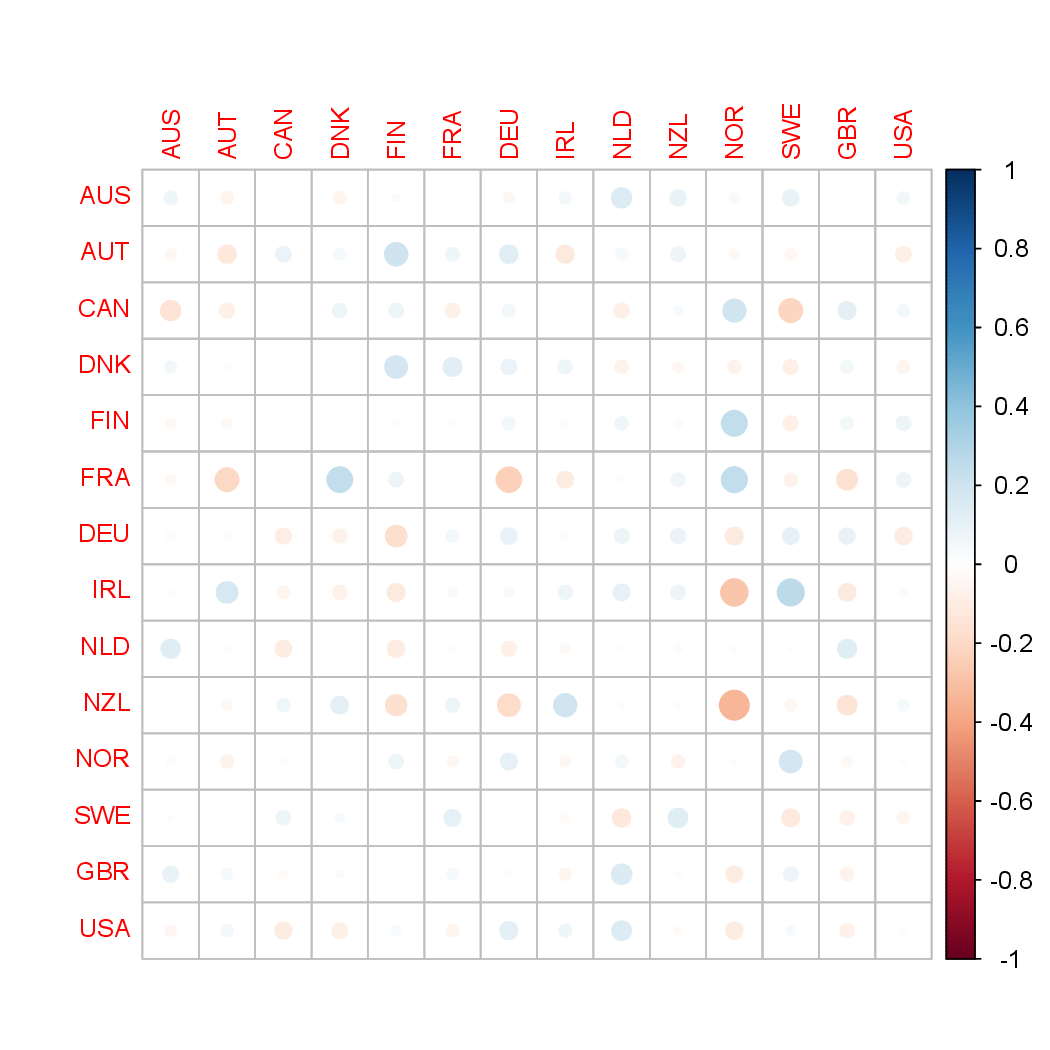}}
\caption{Differences between estimated and observed correlations for the multinational, macroeconomic indices data under 2w-DFM (the $1$st to the $3$rd columns) and RaDFaM (the $4$th column)}
\label{corr_2w&RaDFaM}
\end{figure}

\textit{\textbf{Reconstruction comparison in terms of correlation effects}}
To compare the reconstruction performance of BiMFaM, 2w-DFM, and RaDFaM, we calculate the difference between the row-wise, column-wise, and interactive correlation from the estimated signal part and the observations.
The results in Figures \ref{corr_BiMFaM}-\ref{corr_2w&RaDFaM} show that the recovered row-wise, column-wise, and interactive correlation for sPCA and Q-MLE are closer (much lighter color) to the observed ones (the $1$st column in Figure \ref{corr_BiMFaM}).
This result provides evidence that mode-wise latent factors influence significantly on the reconstruction performance.

\textit{\textbf{Reconstruction comparison in terms of the RMSE}} We use 10 fold cross-validation to evaluate the difference between the observed and estimated matrices in terms of the RMSE for different values of factor numbers.
The results in Table \ref{OECD} demonstrate that sPCA performs the best; Q-MLE is out of work for the $2$nd and the $4$th factor numbers, which may attribute to the small sample size and matrix dimensions.

\begin{table}[H]
\caption{\label{OECD}RMSEs under different choices of $(k_1,\, k_2)$ for the multinational macroeconomic indices data.}
\centering
\begin{tabular}{*{6}{c}}
\hline\hline
Model & Method /$(k_1,k_2)$ & $(2,2)$ & $(3,4)$ & $(5,6)$ & $(8,8)$ \\
\hline\hline
BiMFaM & autoPCA      & 0.3232 & 0.2736 & 0.2272 & 0.1743\\
       & $\alpha$-PCA & 0.3075 & 0.2596 & 0.2142 & 0.1576\\
       & proPCA       & 0.3103 & 0.2608 & 0.2068 & 0.1578\\
       \hline
2w-DFM & 2w-PCA       & 0.2886 & 0.3029 & 0.3149 & 0.3380\\
       & Step-App     & 0.2440 & 0.2677 & 0.3009 & 0.3263\\
       &   Q-MLE      & 0.2443 & -      & 0.2941 & -\\
       \hline
RaDFaM & sPCA         &\textbf{0.1979} & \textbf{0.1480} & \textbf{0.0880} & \textbf{0.0429}\\
\hline\hline
\end{tabular}
\end{table}

\section{Discussion and Future Work}
\label{sec:conc}

The Bayesian networks of the three models are provided in Appendix A of the online supplement, which visually show their modeling strategies.
Another insight maybe the block diagonal thought under BiMFaM, which may result in a special form of RaDFaM \citep{HeKongTrapaniYu2023JOE-one}.

We end this paper with a suggestion for future work.
Since, as we illustrated in Subsection \ref{subsec:subspace},
BiMFaM has a signal part in the form of the $2$nd-order Tucker decomposition, it can be extended naturally  to the higher-order Tucker tensor factor model \citep{ChenYangZhang2022JASA-factor, ZhangLiLiuGuo2022arXiv-Tucker}.
Similarly, we can rewrite expression \eqref{RaDFaM_rank} of RaDFaM as
$$
\sum\limits_{i=1}^{l}\mathbf{U}_{i}\mathbf{V}_{i}^{\top}
= \sum\limits_{i=1}^{l}\mathbf{U}_{i}\circ\mathbf{V}_{i},
$$
which is a special $2$nd-order CANDECOMP/PARAFAC (CP) decomposition \citep[Section 3]{KoldaBader2009SIAM-tensor}.
Due to the strong signal strength of RaDFaM, we may extend the hierarchical spirit to model a new tensor decomposition, and hence an induced tensor factor model.
Specifically, for any $D$th-order tensor $\mathcal{X}\in\mR^{p_1\times\ldots\times p_D}$, we postulate the model below
\beqrs
\left\{\begin{array}{l}
\mathcal{X}=\sum\limits_{i=1}^{l}\mathbf{U}_{1,i}\circ\ldots\circ\mathbf{U}_{D,i},\\
\mathbf{U}_{d,i}=\mathbf{R}_{d}\mathbf{A}_{d,i\cdot}+\boldsymbol{\xi}_{d,i},~d\in[D].
\end{array}\right.
\eeqrs
One may consider developing PCA-type estimators or building EM-variant algorithms to find the maximum likelihood estimators.

\bigskip
\begin{center}
{\large\bf SUPPLEMENTARY MATERIAL}
\end{center}

\begin{description}

\item[Title:] Supplementary Material for ``Modeling and Learning on High-Dimensional Matrix-Variate Sequences"

\item[Overview:] The illustration of the Bayesian networks of the three models is provided in \textbf{Appendix A}.
The proofs of Propositions 1-2 are provided in \textbf{Appendices B-C}, respectively.
The proofs of Theorems 1-4 with the necessary lemmas are provided in \textbf{Appendices D-G}, respectively.
Additional results for simulations and real data analysis are provided in \textbf{Appendices H and I}, respectively.

\end{description}

\bibliographystyle{chicago}
\bibliography{RaDFaM}

\end{document}